\documentclass[%
reprint,
superscriptaddress,
longbibliography,
nofootinbib,
noeprint,
amsmath,amssymb,
aps
]{revtex4-2}
\usepackage[utf8]{inputenc}
\usepackage{bbm}
\usepackage{bm}
\usepackage[left=2cm,right=2cm,top=2cm,bottom=2cm]{geometry}
\usepackage{enumitem}
\usepackage{amsmath,amsfonts,amssymb,amsthm,graphics,graphicx}
\usepackage{hyperref}

\usepackage{chngcntr}

\theoremstyle{definition}
\usepackage [autostyle, english = american]{csquotes}
\usepackage[caption=false]{subfig}
\graphicspath{{figures/}}
\MakeOuterQuote{"}
\newcommand{\ket}[1]{|{#1}\rangle}
\newcommand{\bra}[1]{\langle {#1} |}

\newcommand{\pc}{0.089}

\def\equationautorefname~#1\null{Eq. (#1)\null}

\newcommand{\appref}[1]{\hyperref[#1]{App.~\ref*{#1}}}
\renewcommand{\vec}[1]{\mathbf{#1}}

\newcommand{\gs}[1]{#1}
\newcommand{\eff}{\mathrm{eff}}
\newcommand{\comment}[1]{}
\newcommand{\CNOT}{\mathrm{CNOT}}
\newcommand{\lstate}[1]{\ket{\overline{{#1}}}}
\usepackage{tikzit}
\usepackage{subfiles}

\newcommand{\finalS}{\langle \mathcal{S}_T \rangle}
\newcommand{\GSC}[1]{\ensuremath{\mathrm{GS_{#1}}}}
\usepackage{hyperref}
\newcommand{\pauliplus}{\oplus}

\newcommand{\coset}[1]{\mathbf{A}_{t}^{(L,{{{#1}}})}}
\makeatletter 
    
\renewcommand\onecolumngrid{
\do@columngrid{one}{\@ne}%
\def\set@footnotewidth{\onecolumngrid}
\def\footnoterule{\kern-6pt\hrule width 1.5in\kern6pt}%
}

\renewcommand\twocolumngrid{
        \def\footnoterule{
        \dimen@\skip\footins\divide\dimen@\thr@@
        \kern-\dimen@\hrule width.5in\kern\dimen@}
        \do@columngrid{mlt}{\tw@}
}%

\makeatother    
\begin{document}

\tikzstyle{stabilizer}=[fill={rgb,255: red,191; green,191; blue,191}, draw={rgb,255: red,191; green,191; blue,191}, shape=rectangle]
\tikzstyle{error}=[fill={rgb,255: red,255; green,242; blue,0}, draw=red, shape=circle, minimum width = 0.1 cm]
\tikzstyle{physical}=[fill={rgb,255: red,128; green,128; blue,128}, draw={rgb,255: red,128; green,128; blue,128}, shape=circle, minimum width=0.1 cm]
\tikzstyle{U}=[fill=white, draw=black, shape=circle, minimum height=0.6 cm, minimum width=0.6 cm, line width=1 pt, tikzit fill=magenta]
\tikzstyle{cnot}=[fill=white, draw=black, shape=rectangle, minimum height=0.7 cm, minimum width=0.7 cm, line width=1 pt]
\tikzstyle{X-input}=[fill=red, draw=red, shape=rectangle]
\tikzstyle{Z-input}=[fill=blue, draw=blue, shape=rectangle]
\tikzstyle{CNOT}=[fill=white, draw=blue, shape=circle, minimum width=0.7 cm, minimum height=0.7 cm, line width=1 pt, tikzit fill=cyan]
\tikzstyle{NOTC}=[fill=white, draw=red, shape=circle, minimum height=0.7 cm, minimum width=0.7 cm, line width=1 pt, tikzit fill={rgb,255: red,255; green,191; blue,191}]
\tikzstyle{x spider}=[fill={rgb,255: red,232; green,165; blue,165}, draw=black, shape=circle]
\tikzstyle{z spider}=[fill={rgb,255: red,216; green,248; blue,216}, draw=black, shape=circle]
\tikzstyle{bigU}=[fill=white, draw=black, shape=circle, minimum width=0.7 cm, minimum height=0.7 cm, line width=1 pt]
\tikzstyle{trace}=[fill=white, draw=black, shape=rectangle]
\tikzstyle{bulkerror}=[fill={rgb,255: red,0; green,255; blue,255}, draw=red, shape=circle]

\tikzstyle{gray}=[-, draw={rgb,255: red,128; green,128; blue,128}, line width=1 pt]
\tikzstyle{stack}=[-, draw={rgb,255: red,140; green,175; blue,255}]
\tikzstyle{grayout}=[-, fill={rgb,255: red,255; green,191; blue,191}, opacity=0.6, draw={rgb,255: red,128; green,128; blue,128}, dashed]
\tikzstyle{thick}=[-, line width=1.5 pt, draw=magenta]
\tikzstyle{black}=[-, line width=1 pt, draw=black]

\title{Observation of a Fault Tolerance Threshold with Concatenated Codes}
\author{Grace M. Sommers}
\address{Physics Department, Princeton University, Princeton, NJ 08544}
\author{Michael Foss-Feig}
\affiliation{Quantinuum, 303 S. Technology Ct., Broomfield, CO 80021, USA}
\author{David Hayes}
\affiliation{Quantinuum, 303 S. Technology Ct., Broomfield, CO 80021, USA}
\author{David A. Huse}
\address{Physics Department, Princeton University, Princeton, NJ 08544}
\author{Michael J. Gullans}
\address{Joint Center for Quantum Information and Computer Science,
NIST/University of Maryland, College Park, Maryland 20742, USA}
\date{\today}

\begin{abstract} 
 We introduce a fault-tolerant protocol for code concatenation {of a generalized Shor code} using a butterfly network architecture with high noise thresholds and low ancilla overhead to allow implementation on current devices.  
We develop a probability passing decoder using tensor networks that applies Bayesian updates to the marginal error probabilities after each layer of checks, achieving a state preparation threshold of $e_c \approx \pc$ for erasure errors, and $\approx 0.015$ for unheralded noise. 
We implement our state preparation protocol on ion-trap hardware with added noise to demonstrate the threshold behavior in a real quantum device.  We further theoretically test the performance of our scheme as a quantum memory and for universal quantum computation through the preparation of low-noise magic states for state distillation and $T$-gate injection.
\end{abstract}
\maketitle

\section{Introduction}
Recent years have seen dramatic experimental advances in quantum error correction and fault tolerance, ranging from demonstrations of fault-tolerant gadgets including repeated rounds of quantum error correction \cite{egan2021fault-tolerant-4e9,Postler2024,Huang2023,silva2024demonstration-a9a,krinner2021realizing-d73,ai2023suppressing-df2}, fault-tolerant algorithms with many logical qubits  \cite{bluvstein2023logical-448}, and sub-threshold scaling in topological codes \cite{collaborators2025quantum-3c7}.  Interestingly, though, fault-tolerant threshold behavior has not been experimentally observed in concatenated codes, despite their foundational role in fault-tolerant quantum computing~\cite{shor1996fault,Knill1996,aharonov1997,Kitaev1997,Knill1998,reichardt2005,Aliferis2006}. 

{At the same time, recent years have seen steady \textit{theoretical} progress in the study of concatenated codes, from decoding algorithms~\cite{Poulin2006,Evans2012,Yadavalli2024,Pato2024}, to ancilla preparation~\cite{Reichardt2004,Steane2004,Paetznick2013} and associated overhead analysis \cite{Steane2003,Knill2005,Cross2009,Suchara2013,Chamberland2017}, to the construction of universal gate sets without magic state distillation~\cite{Jochym-OConnor2014,Chamberland2016,Yoder2016,Nikahd2017,Lin2020}.} More recently, there is an increasing interest in concatenated codes from a \textit{practical} quantum computing perspective now that optimized schemes~\cite{yamasaki2024time-efficient-b61,yoshida2024concatenate-611} have brought their performance (measured asymptotically by noise threshold and/or encoding rate) to a level on par with low-density parity check (LDPC) codes like the surface code \cite{Fowler2012fi} or high-rate codes on non-local geometries \cite{Gottesman2013ug,bravyi2024high-threshold-623}.  Moreover, from a statistical mechanics perspective,  concatenated codes allow novel realizations of spin-glass phases and transitions in tree-like geometries \cite{Poulin2006,Yadavalli2024,Sommers2024tree}.  These developments motivate an investigation of the threshold behavior of concatenated codes on current quantum devices.


In this work, we introduce a  family of efficient fault-tolerant protocols for code concatenation with high thresholds and low ancilla overheads to enable their realization on quantum hardware, at the expense of increased decoding complexity.
{We begin in~\autoref{sect:concatenate} by introducing the concatenated code family, {which consists of two-qubit repetition codes in alternating bases,} and designing a state preparation circuit which prepares the logical states $\lstate{0}$ and $\lstate{+}$ with zero ancilla overhead. In~\autoref{sect:state-prep}, we evaluate the state preparation gadget's threshold performance under both heralded and unheralded noise in  classical simulations and on the Quantinuum System Model H2, a trapped-ion quantum processor with all-to-all connectivity and mid-circuit measurement~\cite{quantinuum2023}. The state preparation gadget is then incorporated into a quantum memory test in~\autoref{sect:memory}, theoretically demonstrating the survival of logical information through many rounds of Steane error correction~\cite{Steane1997}. Finally, in~\autoref{sect:universal} we analytically examine a protocol for preparing logical magic states below the distillation threshold \cite{Bravyi2005}, establishing a path toward universal quantum computation in our low-overhead and high-threshold concatenation scheme. We conclude in~\autoref{sect:conclude}. The appendices and Supplemental Material~\cite{supp-ref} contain a detailed exposition of the error models and decoding methods. The tensor-network-based approximate decoder introduced therein, which combines methods from belief propagation and weight enumerator tensor networks \cite{Cao2022lego,Cao2024,Cao2024expansion}, may be of independent interest.}

\section{Single- and multitree encodings}\label{sect:concatenate}
\begin{figure*}
\includegraphics[width=\linewidth]{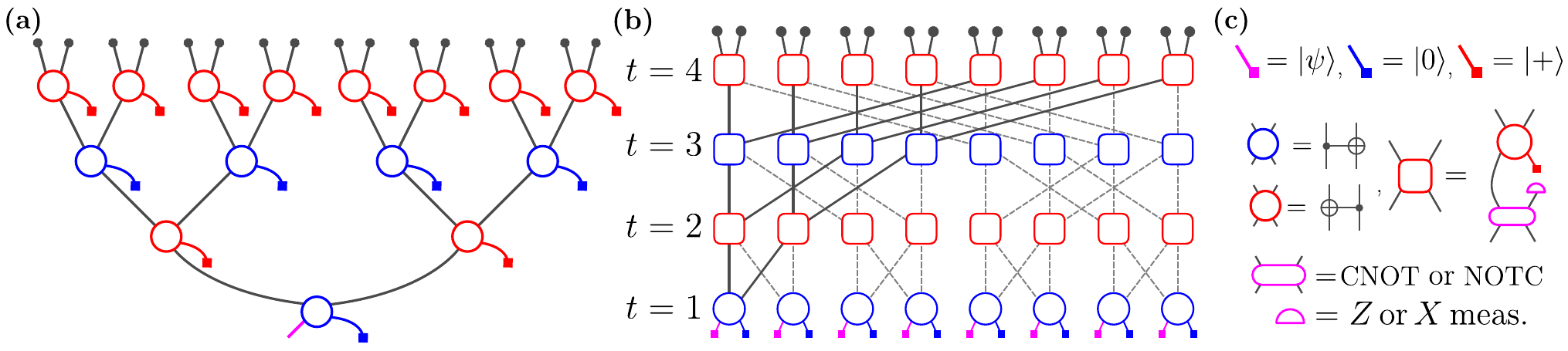}
\caption{ (a) Example depth-4 encoding circuit of a \GSC{o} code. In each layer, the left qubit entering a gate comes from the previous layer, while the right qubit is in a fresh pure state. {One logical qubit (purple leg at root) is encoded into 16 physical qubits (leaves of tree).} (b) Multitree (butterfly network) state preparation circuit for $\lstate{0}$ (purple legs all $\ket{0}$) or $\lstate{+}$ (purple legs all $\ket{+}$).  Thick solid lines trace the tree that encodes the state on the ``system,'' {corresponding to the branches of the singletree in (a)}, {while dashed gray lines indicate qubits used as ancillas}. (c) Legend for (a) and (b). Blue and red circles denote CNOT {(left qubit as control)} and NOTC {(right qubit as control)}, respectively. Each square node of the butterfly network consists of: (1) a ``check gate'' (CNOT for $\lstate{0}$, NOTC for $\lstate{+}$), (2) a measurement of the ancilla qubit in the corresponding basis ($Z$ or $X$), {used to read off the syndrome in that layer}, (3) resetting the ancilla to a fresh stabilizer state, (4) an encoding gate {(CNOT for blue square nodes, NOTC for red square nodes)}. Each of these steps can incur errors, {as described in the main text.} \label{fig:model}}
\end{figure*}
The encoding circuit of a self-concatenated code, with a base code on $b$ physical qubits, can be written as a 
$b$-ary tree, whose depth $t$ is the number of generations~\cite{Yadavalli2024,Sommers2024tree}.  In a previous work~\cite{Sommers2024tree}, we investigated the phase diagrams of concatenated codes generated by binary ($b=2$) trees; 
although the base code necessarily has distance one, the appropriate choice of encoding isometry can generate codes with distance growing exponentially in $t$. One of these is the binary tree version of a generalized Shor (GS) code~\cite{Shor1995,Nguyen2021}, which alternatively concatenates a 2-qubit repetition code in the $Z$ basis and $X$ basis. {That is, as shown in~\autoref{fig:model}a~\cite{zx}, the encoding isometry alternates between
\begin{equation}\label{eq:copy}
\CNOT_{12}\ket{0}_2 = \ket{00}\bra{0} + \ket{11}\bra{1}
\end{equation}
and
\begin{equation}\label{eq:delocal}
\CNOT_{21}\ket{+}_2 = \ket{++}\bra{+} + \ket{--}\bra{-}.
\end{equation}
In the ensuing discussion, we will drop subscripts and denote $\mathrm{NOTC} \equiv \CNOT_{21}$. Where a distinction is meaningful, we use \GSC{o} (\GSC{e}) to denote the code family where \autoref{eq:copy} occurs on odd (even) layers, mainly employing the former encoding order.}

The $b=2$ GS code family has {the parameters $[[n=2^t, k=1,d=2^{\lfloor t/2 \rfloor}]]$}. While there exist other binary tree codes with superior distance scaling and performance under Pauli errors, the GS code is well-suited to fault-tolerant protocols because it is a CSS code, in which the logical operators and stabilizer generators can all be chosen to be all $Z$'s or all $X$'s~\cite{Calderbank1996,Steane1996}. {CSS codes are automatically equipped with a transversal CNOT gate, and Steane error correction reduces the task of fault-tolerant syndrome measurement to the task of preparing ancillas encoded in $\lstate{0}$ and $\lstate{+}$~\cite{Steane1997}.}

{The binary tree encoding circuit offers one way to prepare a given logical state $\lstate{\psi}$: feeding the single-qubit  state $\ket{\psi}$ into the root of the tree, the output on the leaves is $\lstate{\psi}$. When $\ket{\psi}$ is a stabilizer state, stabilized by the Pauli operator $P$, we can think of this as setting the ``logical leg'' to be an additional stabilizer, e.g., to prepare the state $\lstate{0}$, the root is stabilized by $Z$~\cite{root}.} 

However, {the encoding circuit inevitably suffers from noise}, and errors near the root of the tree can accumulate and propagate through the circuit. Consequently, if we only read out the syndrome in the final layer, at any finite noise rate $p$, the logical failure probability $P_F(p,t)$ is monotone-increasing with tree depth $t$. Prior work~\cite{Yadavalli2024,Sommers2024tree} has demonstrated a threshold $p_{single}$ below which \textit{some} quantum information survives to infinite depth, i.e. $\lim_{t\rightarrow\infty} P_F(p,t) < 1/2$. To make the logical failure rate converge to zero as $t\rightarrow \infty$, we could then perform state distillation~\cite{Bravyi2005}, consuming many noisy states to produce one less noisy state.  

The protocol we present here, for preparing the logical states $\lstate{0}$ and $\lstate{+}$ used in Steane error correction, improves upon that approach by introducing checks between each layer of encoding, rather than just at the end. This method achieves a higher threshold $p_c > p_{single}$ and requires zero qubit overhead, {in the sense that the circuit preparing a depth $T$ encoded state requires only $2^T$ total qubits.} 

{While interleaving checks in the encoding circuit is not a new idea---they are implicit in the hierarchical structure of many Steane EC-style ancilla preparation circuits~\cite{Paetznick2013}, and a similar approach underlies the Knill-style Bell state preparation protocol in Ref.~\cite{Knill2005}---they take an especially simple form for our code family}, turning the singletree into a multitree known as a \textit{butterfly network} (\autoref{fig:model}b)~\cite{Leighton1992}. {The same network geometry appears in the fast Fourier transform~\cite{cooley1965algorithm,weinstein1969} and in the encoding circuit of a quantum polar code~\cite{Ferris2014}.} At level $t$, each depth-$t$ state $S$ is checked against a depth-$t$ ancilla $A$, {yielding partial information about the errors on the system up to that layer}. The ``checking'' procedure for state preparation of $\lstate{0}$ in the \GSC{o} code family consists of the following steps (\autoref{fig:model}c):
\begin{enumerate}
\item Perform a transversal CNOT between $S$ and $A$.
\item Measure each qubit of $A$ in the $Z$ basis. From these measurement outcomes, we can infer the $Z$ syndrome of the joint system $S\otimes A$.
\item After odd (even) layers, reset each qubit of $A$ to $\ket{+}$ ($\ket{0}$), and feed it in as a fresh stabilizer to the next level of encoding $S$, {including the root stabilizers}. 
\end{enumerate}
To prepare $\lstate{+}$, we fix all the root stabilizers to be $X$, replace the transversal CNOTs with transversal NOTCs, and measure in the $X$ basis.
\begin{figure}[t]
\centering
\includegraphics[width=0.9\linewidth]{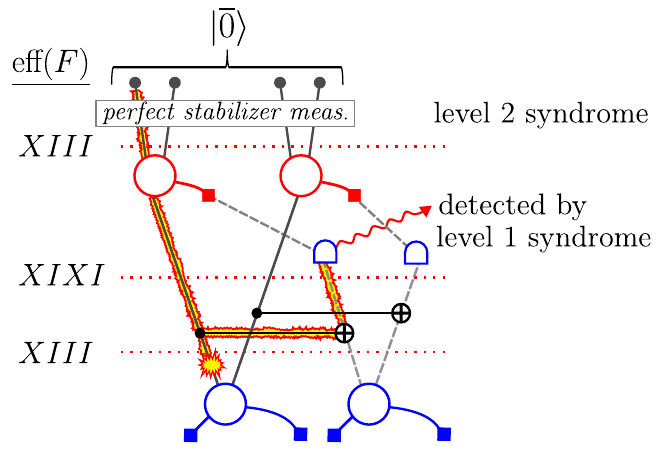}
\caption{{Depth-2 multitree circuit for $\lstate{0}$, with a round of perfect stabilizer measurements appended to evaluate its performance. Highlighted wires show how a fault (yellow starburst) propagates through the circuit. The first round of checks (level 1 syndrome) distinguishes this fault from a bit flip on the second system qubit, which would propagate to a logically inequivalent error with the same level 2 syndrome. }\label{fig:error-prop}}
\end{figure}
\begin{figure*}[t]
\centering
\subfloat[Erasure errors]{
\includegraphics[height=0.17\textheight]{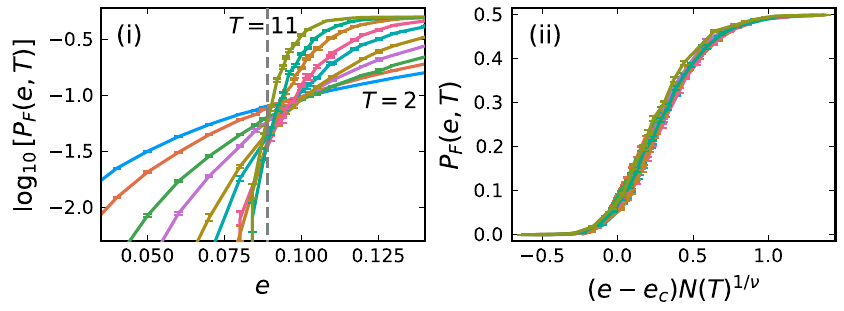}
\label{fig:heralded}}\hfill
\subfloat[Unheralded errors]{
\includegraphics[height=0.17\textheight]{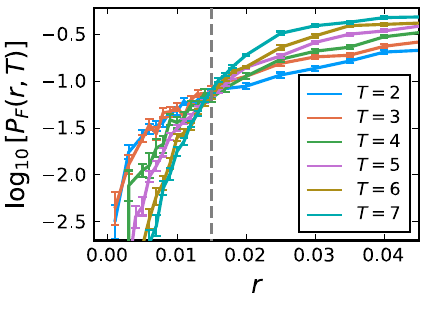}
\label{fig:unheralded}}
\caption{Logical failure probability of state preparation in classical simulations. (a) Erasure errors at rate $e$, optimal decoder. {Left panel (i) shows system sizes $T=2$ to $T=11$, plotted on a log scale. Gray dashed line marks the threshold $e_c \approx \pc$. Right panel (ii) is a scaling collapse to the form~\autoref{eq:scaling} with $e_c = \pc, \nu = 2.85$, plotted on a linear scale with depths $T=3$-11. (b) Unheralded bit/phase flips at rate $r$, decoded via stacked probability passing. Gray dashed line marks the threshold $r_c \approx 0.015$.} In all panels, the plotted quantity is averaged over $\lstate{0}$ and $\lstate{+}$. In this paper, all logarithms are base 10, {and uncertainties are determined as the standard error of the mean across independent samples or the one sigma confidence interval across Bernoulli trials}. {We use a linear scale for the horizontal axis as our focus is on the near-threshold behavior, not the behavior near $p=0$.}}
\end{figure*}

Therefore, at level $t$ of preparing a depth-$T$ state, there are $2^{T-t}$ identically prepared depth-$t$ states. After the $T$th level of encoding, there is only one state left. As is standard in fault tolerance, to analyze the performance of this gadget in isolation, we conclude the circuit by perfectly measuring all the stabilizers, \textit{except} the root stabilizer. {The decoder then uses the syndromes measured in levels $1,2,...,T$, which collectively form the ``spacetime syndrome,'' to (1) find a Pauli error $F$ in spacetime that produces this syndrome, (2) decide whether to apply $\eff(F)$ or $\eff(F) L$. Here $\eff(F)$ is the fault propagated to the last layer of the circuit~\cite{Delfosse2023}, and $L$ is the logical operator that flips the root stabilizer ($L=\overline{X}$ for $\lstate{0}$, $L=\overline{Z}$ for $\lstate{+}$). As a minimal example, \autoref{fig:error-prop} shows a fault with $\eff(F)=X_1$. This fault would be decoded incorrectly if we only measured the syndrome in the final layer, but is correctly decoded using the full spacetime syndrome.}

In previous overhead analysis of fault-tolerant protocols for concatenated codes~\cite{Chamberland2017}, each round of stabilizer checks consists of both $X$-type and $Z$-type checks. In contrast, our protocol only performs one type of check, enabling zero qubit overhead {as each ancilla is eventually reset and fed into the growing system}. In the context of state preparation, this is justified because the decoder succeeds as long as it correctly fixes the sign of the root stabilizer (returns it to $+1$). Therefore at each level, we need only measure the syndrome that ``matches'' the root stabilizer, which for $\lstate{0}$ is all $Z$'s (and thus immune to phase flips) and for $\lstate{+}$ is all $X$'s (and thus immune to bit flips). This strategy is modified when using the prepared states for Steane error correction, discussed below.

\section{State preparation thresholds}\label{sect:state-prep}
\subsection{Theoretical demonstration}
To test the protocol introduced in the previous section, we first consider a theoretical noise model with Pauli errors in three different locations: (1) on each stabilizer input (that is, on all the input legs before $t=1$, and after every qubit reset in subsequent steps), (2) after each two-qubit gate, and (3) before each measurement on the ancilla (see \appref{app:models})~\cite{errors}. 

\subsubsection{Erasure errors}
{We first consider a model in which all errors are \textit{heralded}: each error raises a flag on the noisy qubit, while unflagged qubits are guaranteed to have suffered no errors. The most common heralded error channel is the erasure channel, in which the flagged qubit is replaced with a fully mixed state: 
\begin{equation}\label{eq:erasures}
\mathcal{E}_e(\rho) = (1-e)\rho \otimes \ket{0} \bra{0} + e \frac{\mathbbm{1}}{2} \otimes \ket{1}\bra{1}
\end{equation}
where $e$ is the erasure rate and $\ket{0}, \ket{1}$ is the state of a classical register. For CSS codes, the erasure channel is equivalent to the composition of a heralded bit flip channel 
\begin{equation}\label{eq:herald-X}
\mathcal{E}_{X,e}(\rho) = (1-e)\rho \otimes \ket{0} \bra{0} + \frac{e}{2} (\rho + X \rho X^\dag)   \otimes \ket{1}\bra{1}
\end{equation}
with a heralded phase flip channel $\mathcal{E}_{Z,e}$.} 

From the circuit architecture, we construct the stabilizer generators of the spacetime code~\cite{Bacon2017,Gottesman2022,Delfosse2023}, wherein each possible error location is assigned a qubit, and each bit of syndrome information is associated to a stabilizer generator in spacetime. Under fully heralded noise, optimal decoding of this spacetime code can be done in polynomial time (see~\autoref{app:erasures} of the Supplemental Material~\cite{supp-ref}). For i.i.d. erasures at rate $e$, the threshold for both $\lstate{+}$ and $\lstate{0}$ is $e_c\approx \pc$, as shown in~\autoref{fig:heralded}(i). The crossing between system sizes drifts to smaller $e$ and smaller $P_F(e)$ with increasing $T$. A fit to the one-parameter scaling form 
\begin{equation}\label{eq:scaling}
P_F(e, T) = f((e-e_c) N(T)^{1/\nu}),
\end{equation}
with $f$ an unknown scaling function, $e_c = \pc$, $\nu = 2.85$, and $N(T) \sim T 2^T$ the number of spacetime qubits, is shown in~\autoref{fig:heralded}(ii)~\cite{spacetime}.

\subsubsection{Stacked probability passing} 
Next we consider unheralded noise, taken to be independent bit and phase flips at rate $r$:
\begin{equation}\label{eq:unheralded}
\mathcal{N}_r(\rho) = (\mathcal{N}_{X,r} \circ \mathcal{N}_{Z,r})(\rho)
\end{equation}
where
\begin{subequations}
\begin{align}
\mathcal{N}_{X,r} &= (1-r) \rho + r X \rho X^\dag \label{eq:unherald-X}\\
\mathcal{N}_{Z,r} &= (1-r) \rho + r Z \rho Z^\dag. 
\end{align}
\end{subequations}
The lack of correlations between $X$ and $Z$ errors means that, for CSS codes, we can decode the $Z$ and $X$ syndromes independently.

Without perfect heralding, optimal decoding via the spacetime code is no longer feasible. Instead, we use an approximate ``stacked probability passing'' decoder, detailed in~\appref{app:decoding} and the Supplemental Material~\cite{supp-ref}. 
Under this noise model, the stacked probability passing decoder has an estimated threshold of $r_c \approx 0.015$, {as shown in~\autoref{fig:unheralded}.} 

To benchmark the approximate decoder, we also use it to decode erasure errors, comparing its performance to that of the optimal decoder. For that error model, the approximate decoder achieves a similar threshold and subthreshold performance to the optimal decoder, though further systematic improvements could be made~\cite{supp-ref}.

\subsection{Hardware implementation} 
Having demonstrated the effectiveness of our circuit layout and decoding method in theory, we now test the protocol on quantum hardware.

We implement state-preparation circuits using up to 32 qubits (depth $T=5$) on the Quantinuum System Model H2. The dominant error sources native to the device are two-qubit gate errors and measurement errors, each occurring at a rate $\lesssim 0.0025$. Modeled as independent
bit and phase flips, this noise rate is well
below $r_c$. To tune through a transition, we stochastically introduce gates in two different ways: (1) incoherent noise and (2) coherent rotations. In the final layer of circuit, each qubit in the system is destructively measured in the basis of the root stabilizer; decoding is successful if the decoder’s guess of the logical state, using the heralding record and syndrome, matches the measured state.

We choose both types of added errors to be heralded, so the threshold is at large enough added noise rate to clearly distinguish the subthreshold and above-threshold phases. 

\subsubsection{Added incoherent errors}
\begin{figure}[t]
\includegraphics[width=\linewidth]{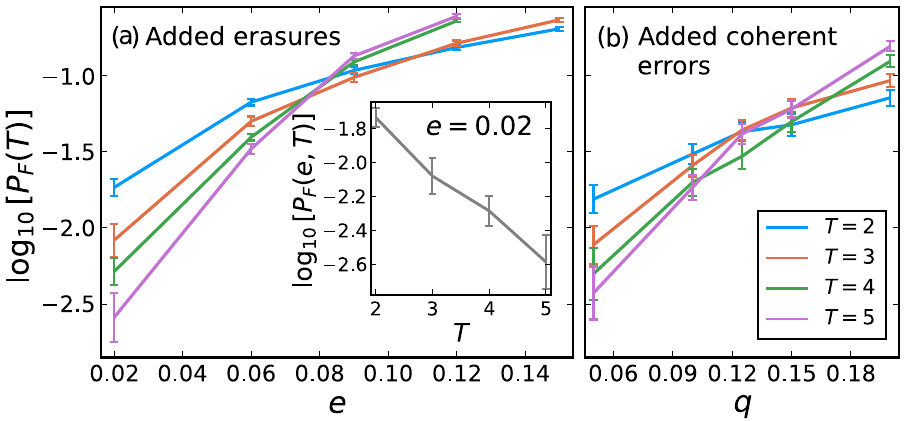}
\caption{Logarithm of the failure probability on Quantinuum System Model H2, native noise + added heralded bit/phase flips at rate $e$ (\autoref{eq:herald-X}) (a) and  heralded coherent errors (\autoref{eq:coherent}) at rate $q$ (b). The inset to (a) shows the subthreshold scaling at added noise rate $e=0.02$. In all panels, the plotted quantity is averaged over $\lstate{0}$ and $\lstate{+}$.\label{fig:qtuum}}
\end{figure}
The incoherent error model is heralded bit flips [\autoref{eq:herald-X}] (for $\lstate{0}$) and phase flips (for $\lstate{+}$) at rate $e$ after each two-qubit gate and before each measurement. Even at these modest system sizes,~\autoref{fig:qtuum}(a)  shows evidence of a threshold between $e=0.06$ and $e=0.09$.  In the subthreshold phase, the logical failure probability is exponentially suppressed in $T$, as shown in the inset.

{We stress that while the form of the added noise is artificial, the exponential suppression of the logical failure probability at low rates of added noise already demonstrates the success of the protocol in handling the noise native to the device. Injecting noise of our choosing increases the logical failure rate, allowing the trend with system size to be resolved using fewer shots.}

\subsubsection{Added coherent errors}

We next benchmark the state preparation protocol in the presence of added coherent errors at rate $q$. That is, we introduce coherent non-Clifford rotations (with perfect heralding to boost the threshold): {if a flag is raised (with probability $q$), then a coherent rotation by $\theta$ about the $X$ (or $Z$) axis on the Bloch sphere is applied:
\begin{equation}\label{eq:coherent}
\mathcal{C}_{X,q,\theta}(\rho) = (1-q)\rho \otimes \ket{0}\bra{0} + q e^{i\theta X/2} \rho e^{-i\theta X/2} \otimes \ket{1}\bra{1}.
\end{equation}
We set the rotation angle $\theta = \pi/4$ so that the rotation is maximally non-Clifford.}

Our motivation for considering this second model is that while incoherent noise is typically assumed when developing decoders, real devices also suffer systematic small rotations~\cite{Bravyi2018}. Part of the difficulty in studying coherent noise is the phenomenon of quantum interference, which makes syndrome sampling classically inefficient in general and gives rise to complex Boltzmann weights in the associated statistical mechanics mapping~\cite{Venn2023}. With quantum hardware, the inefficiency of syndrome sampling is circumvented as the syndromes are generated for us by the circuit. {Though the heralding is unphysical, it serves only to boost the threshold and does not alter the implementation of the decoder.}

Although the added errors are coherent, our approximate decoder still models the noise as incoherent by taking the Pauli twirl,
\begin{align}\label{eq:twirl}
\tilde{\mathcal{C}}_{X,q,\theta}(\rho) &= q \left[\cos^2(\theta/2) \rho + \sin^2(\theta/2) X \rho X^\dag\right] \otimes \ket{1}\bra{1} \notag \\ &+ (1-q)\rho \otimes \ket{0}\bra{0}.
\end{align}

At small $q$, this decoder achieves modest suppression of the failure probability with increasing $T$, with an apparent threshold around $q\approx 0.125$ [\autoref{fig:qtuum}(b)]. 

\subsubsection{Future directions}
{A potential optimization, enabled by the hardware's capacity for mid-circuit measurement and feedback, is to adapt the checking procedure in real time based on syndrome information from earlier rounds. While a rewiring protocol for a fully heralded error model does not yield asymptotic improvements for preparation of $\lstate{0}$ and $\lstate{+}$, adaptive rewiring may be an asset for other tasks~\cite{supp-ref}.}

{Successful preparation of $\lstate{0}$ and $\lstate{+}$ only demonstrates the code's \textit{classical} performance, since the measurements are performed in a single basis. As a more stringent test of the code performance, one could prepare a logical Bell pair, which would require defense against both types of noise simultaneously. In an experiment, the error-corrected correlation between measurement outcomes would serve as a measure of entanglement in the logical state~\cite{Audenaert2006}. However, the logical Bell state uses twice as many qubits for a given $T$ as $\lstate{0}$ and $\lstate{+}$, thus limiting the number of generations accessible on current hardware. We therefore opt not to pursue this avenue, instead turning to a \textit{theoretical} demonstration of the code's use as a quantum memory.}

\section{Quantum memory}\label{sect:memory}
\begin{figure*}[t]
\subfloat[]{
\includegraphics[width=0.25\linewidth]{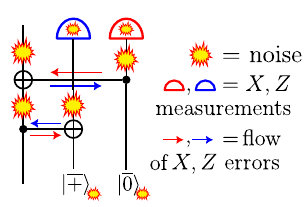}\label{fig:steane-circuit}}
\subfloat[]{
\includegraphics[width=0.32\linewidth]{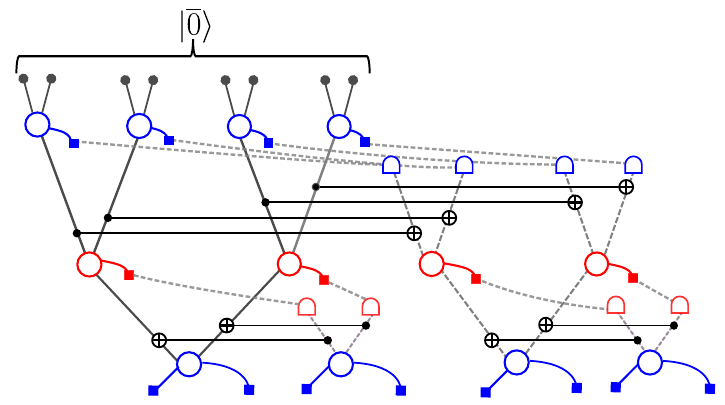}\label{fig:ancilla-circuit}}
\subfloat[]{
\includegraphics[width=0.42\linewidth]{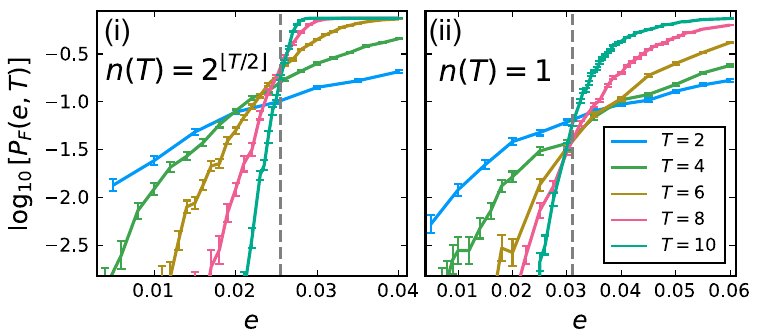}}
\caption{{(a) Steane syndrome measurement. Leftmost wire is an unknown logical state on the system. Yellow starbursts show the noise locations modeled in this work. Arrows indicate the flow of Pauli errors. (b) Depth-3 preparation circuit for an ancilla encoded in $\lstate{0}$, for use in the gadget shown in (a). This preparation circuit includes $X$ checks after the first layer and $Z$ checks after the second layer. Legend is the same as in~\autoref{fig:model}c.} (c) Logical failure probability after (i) $2^{\lfloor T/2 \rfloor}$ rounds and (ii) 1 round of Steane syndrome measurement, under a bulk erasure model with optimal decoding. Apparent thresholds, {achieving the best collapse to the scaling form~\autoref{eq:scaling},} are indicated by dashed gray lines. {Uncertainties are determined as the standard error of the sample mean.} \label{fig:memory}}
\end{figure*}

For the quantum memory demonstration, we use the encoded logical states as ancillas in a Steane syndrome extraction gadget. A review of Steane error correction and details of our implementation are provided in~\autoref{app:memory} of the Supplemental Material. For the present discussion, it suffices to recall {(see~\autoref{fig:steane-circuit})} that one round of Steane error correction uses two ancillas, encoded in the states $\lstate{0}$ and $\lstate{+}$ and coupled to the system via transversal CNOT gates, to perform $X$ and $Z$ checks on a CSS code 
\textit{without} learning the logical state~\cite{Steane1997}.  These ancillas are prepared via a modification of the state preparation protocol discussed above. {The state preparation circuit in~\autoref{fig:model}b defends against only one type of errors; for example, the circuit for $\lstate{0}$ uses $Z$ checks in each layer, allowing phase flip errors to accumulate since they do not affect the logical state. In contrast, when $\lstate{0}$ is used in the Steane EC gadget, we must defend against both bit flip errors (which otherwise will leak onto the system during the transversal CNOT) and phase flip errors (since $\lstate{0}$ is used to to learn the $X$ syndrome). To address this issue while maintaining the butterfly network structure of the circuit, we simply alternate between $X$ and $Z$ checks in the multitree, as shown for a depth-3 ancilla in~\autoref{fig:ancilla-circuit}.}

Suppose that an unknown logical state has been perfectly encoded in a depth $T$ tree code. We can think of this state as being entangled with a reference qubit. We then use $n\geq 1$ rounds of the syndrome extraction gadget to read out the syndrome. Since the ancillas used in the syndrome extraction are noisy, and the system can suffer additional errors after each layer of gates, the decoder's task is to predict the correction operator that returns the system to its initial (unknown) state, based on the syndrome information read out from the ancilla measurements and from the perfect stabilizer measurements at the end of the circuit.

When the errors are heralded, we infer the performance of an optimal decoder by computing the mutual information between the final state of the system $S$ and the initial reference $R$. Perfect decoding is possible when the mutual information is maximal, $I(S:R)=2$. Otherwise, for each lost bit of mutual information, the recovery probability degrades by a factor of 2, such that the logical failure probability of the optimal decoder for a given $I(S:R)$ is
\begin{equation}\label{eq:mutual}
P_F(e) = 1 - \frac{1}{2^{2-I(S:R)}}.
\end{equation}
To test the survival of logical information after many rounds of syndrome extraction, we subject a depth $T$ logical state to $n=d(T)=2^{\lfloor T/2 \rfloor}$ rounds of noisy syndrome measurement, followed by a layer of perfect stabilizer measurements. To test the performance of a single Steane error correction gadget in isolation, we also examine the logical failure probability for $n=1$ round. The thresholds for the two circuits are $e_c \approx 0.0255$, $e_c \approx 0.031$, respectively, as shown in~\autoref{fig:memory}. Both transitions are broadened with an exponent $\nu \approx 2.75$, with ``system size'' $N(T) \sim T 2^{T} n(T)$. 

The approximate probability passing decoder described in~\autoref{app:memory-ppd} of the Supplemental Material has a much lower threshold, and inferior subthreshold scaling. Therefore, a transition under unheralded errors is only visible if we restrict to $n=1$ rounds of checks. In this case, we attain a pseudothreshold of $r_c\approx 0.003$ below which the failure probability does not increase with depth. We leave improvements of the decoder for unheralded and partially heralded noise to future work. 

\section{Universal gates}\label{sect:universal}
{Finally, we theoretically examine how to construct a fault-tolerant universal gate set for the GS code family. For this purpose, it suffices to implement the generators of the Clifford group---CNOT, Hadamard, and the phase gate $S$---supplemented with a logical $T$ gate.}

{As a CSS code, the GS code automatically comes equipped with a transversal CNOT, a fact exploited above in the state preparation and Steane EC gadgets. We propose to implement a logical Hadamard using a transversal gate + code switching~\cite{Stephens2008,Paetznick2013switch,Anderson2014,Kubica2015,Bombin2016,Vuillot2019,pogorelov2025}, discussed below. To implement the remaining gates at the logical level, we rely on gate injection/gate teleportation~\cite{Bravyi2005,Gottesman2024}: (1) prepare an ancilla in a logical resource state, (2) couple the ancilla to the system via transversal CNOT, (3) measure the ancilla, and (4) conditioned on the measurement outcome, apply an (easier-to-implement) gate on the system.}

\subsection{Magic state distillation}
We first consider state preparation of logical $T$ states, which can be used for $T$-gate injection~\cite{Bravyi2005}. Specifically, we prepare the state~\cite{T}.
{\begin{equation}
    \lstate{T} \equiv e^{-i \pi \overline{Z}/8} \lstate{+}.
\end{equation}}
Since this state does not point along the $X$ or $Z$ axis {of the logical Bloch sphere}, it cannot be prepared using the multitree circuit in~\autoref{fig:model}b. Instead, we prepare the state using the singletree circuit {(\autoref{fig:model}a with $\ket{T} = \exp(-i\pi Z/8) \ket{+}$ fed into the root)}, followed by state distillation. 

\begin{figure}[t]
\centering
\includegraphics[width=\linewidth]{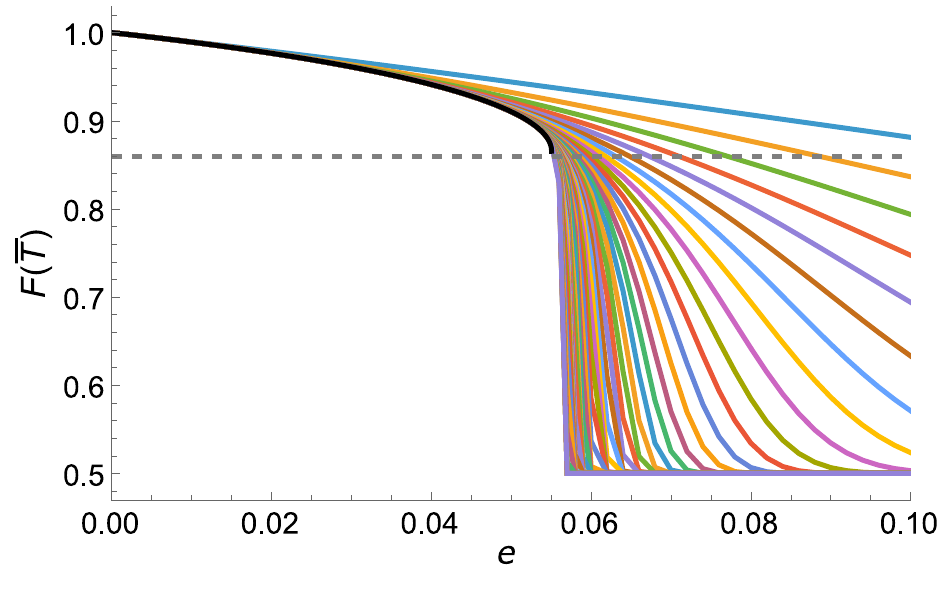}
\caption{Fidelity to the logical $T$ state (\autoref{eq:T}) after decoding erasure errors on the branches of the singletree encoding circuit. Colored curves are for even depths up to 100, using the  \GSC{e} encoding order~\cite{even}. Black curve shows the fixed-point fidelity. Dashed curve shows the threshold for magic state distillation from Ref.~\cite{Bravyi2005}.\label{fig:T-fidelity}}
\end{figure}

Let $\mathcal{N}$ denote the noisy encoding isometry (from the state $\ket{T}$ on the root to a noisy encoded state on the leaves) and $\mathcal{R}$ the optimal recovery channel. If the noise + decoder implements a logical $P_\alpha$ with probability $\mathbbm{P}_\alpha$, then the logical $T$ state fidelity is 
\begin{equation}\label{eq:T}
F(\overline{T}) \equiv \bra{\overline{T}} \mathcal{R} \circ \mathcal{N} (\ket{T} \bra{T}) \lstate{T} = 1 - \mathbbm{P}_Z  - (\mathbbm{P}_X + \mathbbm{P}_Y)/2.
\end{equation}
For models of bulk errors parameterized by $p$, with perfect syndrome measurement in the final layer, there exists a ``coding phase'' in which $F(\overline{T})$ converges, in the limit of infinite depth, to a fixed point $F^*(p) >1/2$ for $p<p_{single}$. If the noise is composed of single-qubit independent unheralded bit and phase flips, the threshold can be determined numerically to high precision as $r_{single}=0.0066 \pm 0.0004$, whereas under erasure errors, an exact calculation yields $e_{single}=0.05505...$~\cite{Sommers2024tree}. {\autoref{fig:T-fidelity} shows the fidelity as a function of erasure rate $e$, for even depths up to 100. The fixed-point fidelity within the coding phase is marked as a black curve.}

Optimal decoding of the singletree is followed by distillation. To converge to zero failure probability, the fidelity of the inputs to the distillation protocol must exceed a certain value, which equals $0.859$ in the 15-to-1 scheme introduced in Ref.~\cite{Bravyi2005}. An analytic calculation for erasure errors finds that for all $e\leq e_{single}$, $F^*(e) > 0.859$, so distillation is successful throughout the entire coding phase~\cite{threshold}. Additional details are provided in~\autoref{app:T} of the Supplemental Material~\cite{supp-ref}.

{\subsection{Logical Clifford gates}
The standard $T$-gate injection circuit uses a logical phase gate $S$, conditioned on the measurement outcome on the ancilla~\cite{Gottesman2024}. In turn, we can also implement $S$ fault-tolerantly also using gate injection, with the resource state $\lstate{i} = (\lstate{0} + i\lstate{1})/\sqrt{2}$, the unique $+1$ eigenstate of $\overline{Y}$. To prepare $\lstate{i}$, we once again use the singletree encoding circuit, now with the logical leg fixed to be a $-Y$ stabilizer, followed by a distillation procedure that produces one less noisy $\lstate{i}$ from 7 noisy $\lstate{-i}$'s~\cite{Cross2009}. The classically conditioned gate in step 4 of the gate injection circuit is just a Pauli gate, which admits a transversal implementation.}

{To complete our universal gate set, we need a fault-tolerant Hadamard gate. Applying $H$ transversally does not quite implement a logical Hadamard: it also exchanges the $X$-type stabilizers with $Z$-type stabilizers, switching between the \GSC{o} and \GSC{e} code family. We can account for this by simply keeping track of which of the two codes each logical qubit is in. The only complication is that before performing a gate between a \GSC{o}-encoded qubit and a \GSC{e}-encoded qubit, we must 
\textit{code switch} one of the two qubits, to put them in the same code.} 

{The code switching protocol works as follows. Suppose we want to code switch from $\GSC{e}(t)$ to $\GSC{o}(t)$. To do so, we need to project onto the codespace of \GSC{o}, while preserving the logical state~\cite{pogorelov2025}. Projecting onto the codespace is simple: just measure the syndrome  using \GSC{o}-encoded ancillas, and fix up the signs of the wrong-signed stabilizers by applying Pauli corrections. As long as these Pauli corrections commute with the logical operators, the logical state is preserved. To meet this condition, we first permute the physical qubits, mapping (for even $t$) the time-evolved operators $(X_{\mathrm{e}}(t),Z_{\mathrm{e}}(t))$, which are logical representatives of the $\GSC{e}(t)$ code, onto the operators $(X_{\mathrm{o}}(t),Z_{\mathrm{o}}(t))$ of $\GSC{o}(t)$. Then, the permuted code $\widetilde{\mathrm{GS}}_{\mathrm{e}}(t)$ and the target code $\GSC{o}(t)$ are different gauge fixings of a common subsystem code~\cite{Anderson2014}, so the Pauli corrections can be implemented using stabilizers of $\widetilde{\mathrm{GS}}_{\mathrm{e}}$, which preserve the logical operators. Further detail is provided in~\autoref{app:code-switch} of the Supplemental Material~\cite{supp-ref}.}

\section{Discussion}\label{sect:conclude}
In this work we have taken steps toward universal fault-tolerant quantum computation using a family of concatenated CSS codes whose base code consists of 2 qubits. We constructed optimal and approximate decoders for a zero-ancilla-overhead state preparation protocol, and demonstrated thresholds on trapped ion hardware. The state preparation circuit was then adapted for use in a quantum memory test. Finally, we showed how the bare (single tree) circuit, followed by state distillation, can be used to prepare resource states such as the logical $T$ state, an essential ingredient for performing universal logical gates. 

Concatenated codes with Steane error correction, and LDPC codes such as the surface code, are two paths toward fault-tolerant quantum computing. {One advantage of the former approach is that clever choices of concatenation enable fault-tolerant implementation of universal gates without gate injection, thus circumventing magic state distillation (or cultivation~\cite{Gidney2024}) entirely~\cite{Jochym-OConnor2014,Chamberland2016,Yoder2016,Nikahd2017,Chamberland2017,Lin2020}. While Ref.~\cite{Chamberland2017} found that such distillation-free concatenated codes still require more resources to achieve the same logical error rate as surface codes with magic state distillation, their approach is dominated by the overhead associated with the preparation and verification of ancilla states for Steane error correction, which we apply more sparingly. In particular, our state preparation circuit has zero qubit overhead, in contrast with many previously introduced protocols that rely on post-selection, incurring significant time and qubit overhead~\cite{Steane1997,Knill1998,Reichardt2004,Steane2003}. The GS code used for our demonstration \textit{does} require magic states to perform universal gates, but the high-level strategy outlined in this work---multitree state preparation with qubit reuse and no postselection, made possible by a tensor-network-based decoder---applies to general concatenated CSS codes.} {The simplest way to generalize the multitree circuit to base codes with $b>2$ would be to replace~\autoref{fig:model}b with a radix-$b$ butterfly network, meaning that each layer of concatenation would be followed by checking a system block against $b-1$ ancilla blocks, whose qubits are then reset and fed into the next layer. A fault occurring in the last layer can only spread to at most $b$ qubits, which is a bounded error in a code with macroscopic distance ($t\rightarrow \infty$, $b$ fixed).} 

{The appeal of the GS code lies in the simplicity of one concatenation level, which involves just a single two-qubit gate in contrast to the larger base codes studied in earlier works. With large base codes, analysis is often limited to one or two concatenation levels, yet the ``pseudothreshold'' inferred from just one concatenation level can differ significantly from the asymptotic threshold~\cite{Svore2005,Svore2006,Chamberland2016}. With our small base code, we can access multiple generations on the hardware, demonstrating a suppression of the logical failure probability in the subthreshold phase, and many more generations in classical simulations, enabling a finite-size scaling analysis.}

{It is worth noting that while our multitree state preparation avoids the \textit{qubit} overhead involved in a protocol that applies postselection or distillation of several singletrees, it incurs a \textit{decoder} overhead due to the relative complexity of the stacked probability passing decoder vs. the singletree decoder.} Decoding is more efficient, and the threshold much higher, when the errors are  heralded. While perfect heralding of errors is an idealized scenario, qubits on a range of experimental platforms can be designed so that \textit{most} errors are erasures, via a technique known as erasure conversion \cite{Wu2022,Sahay2023,Kang2023,Kubica2023,Teoh2023,Scholl2023,Ma2023,Levine2024, Chou2024,Koottandavida2024,Holland2024,Zhang2025}. A preliminary investigation indicates that the threshold of the state preparation gadget increases sharply as the heralding rate approaches 1, consistent with the findings of Ref.~\cite{Wu2022} for the surface code.

{Finally, as noted above, the butterfly network geometry appears in several other contexts, and an avenue for future study is to consider how extensions in those contexts translate into generalizations of the multitree state preparation circuit. For example, mixed-radix algorithms for the fast Fourier transform when the input size is not a power of 2~\cite{1162042} could be useful for preparing codes with different base codes in different layers. The encoding circuit of a quantum polar code is also a butterfly network, with each node corresponding to a CNOT gate rather than the ``check and reset'' black box in~\autoref{fig:model}c. As polar codes are a special case of branching MERA codes~\cite{Ferris2014}, it may be interesting to explore the circuit that results when each node of the branching MERA graph is replaced with a check and reset.}
\section*{Acknowledgments} {We thank Zhi-Yuan Wei for pointing out the connection to quantum polar codes and branching MERA.}
This research was supported in part by NSF QLCI grant OMA-2120757, including an Institute for Robust Quantum Simulation (RQS) seed grant.  We gratefully acknowledge the Quantum Computing User Program (QCUP) at Oak Ridge National Lab for providing the credits for running on the Quantinuum device. Numerical work was completed using computational resources managed and supported by Princeton Research Computing, a consortium of groups including the Princeton Institute for Computational Science and Engineering (PICSciE) and the Office of Information Technology's High Performance Computing Center and Visualization Laboratory at Princeton University. {The open-source \texttt{QuantumClifford.jl} package was used to simulate Clifford circuits with erasure errors~\cite{QuantumClifford}.} Any mention of commercial products in this paper is for information only; it does not imply recommendation or endorsement by NIST, nor does it imply that the equipment identified are necessarily the best available for the purpose.
\section*{Code availability}
The central code base for this project is available at Ref.~\cite{mycode}. 
\appendix
\section{Error Models}\label{app:models}
In this appendix, we
summarize the various error models considered in the work.

\subsection{Types of errors}
Three types of noise channels are considered in this work: erasure errors [\autoref{eq:erasures}], which in a CSS code are equivalent to independent heralded bit and phase flips, $\mathcal{E}_{X,Z}$ [\autoref{eq:herald-X}]; independent unheralded bit and phase flips $\mathcal{N}_{X,Z}$ [\autoref{eq:unheralded}];  and heralded coherent errors $\mathcal{C}_{X,Z}$ [\autoref{eq:coherent}]. To emphasize the difference between the channels $\mathcal{E},\mathcal{N}, \mathcal{C}$, we parameterize them by different letters: $r$, $e$, and $q$, respectively~\cite{channels}. When not referring to a specific channel, we parameterize the noise rate by $p$.

In the state preparation gadget, noise can be treated classical, because the logical state $\lstate{0}$ ($\lstate{+}$) is immune to $Z$ ($X$) errors. Thus, we need only consider $ \mathcal{E}_{X},\mathcal{N}_{X}, \mathcal{C}_{X,\pi/4}$ in preparing $\lstate{0}$ and vice versa for $\lstate{+}$. 

\subsection{Error locations}
All the error models considered in this work consist of independent and identically distributed single-qubit errors, at three different types of locations:
\begin{enumerate}
\item On each stabilizer input, at rate $p_{i}$
\item On the two output wires of each two-qubit gate, at rate $p_g$
\item Before each measurement (i.e. random bit flip of measurement outcome), at rate $p_m$.
\end{enumerate}

In classical simulations (\autoref{fig:heralded} and~\autoref{fig:memory}), we take $p_i = p_g = p_m = p$, though gate errors have the largest effect on the performance due to the larger number of available error locations and the potential for errors to spread in subsequent layers.

In the experiment, we first infer the native unheralded noise rates $r_i, r_g, r_m$ from the reported rate of two-qubit gate errors and state preparation and measurement (SPAM) errors. {At the time of the experiment, $p_{SPAM} \approx 0.0025$ and is dominated by measurement errors, so we take $r_i=0, r_m=0.0025$. The probability of a nontrivial two-qubit Pauli operator following a two-qubit gate was $p_{2qb} \lesssim 0.002$. Our decoder models these errors as independent $X$ and $Z$ errors on each qubit (\autoref{eq:unheralded}), i.e. $1-p_{2qb} = (1-r_g)^4$, yielding $r_g \approx 0.0005$.} 

Then we introduce either (1) heralded bit or phase flips at rate $e$ or (2) heralded coherent errors at rate $q$, after each gate. For (1), we also add heralded bit flips of the measurement outcomes in post-processing. 

In the logical $T$ state preparation, we consider noiseless measurements on the leaves and single-qubit erasure errors on every branch of the tree, including the root.

\section{Decoding Methods}\label{app:decoding}
Consider a faulty circuit (for state preparation or syndrome extraction) with several noisy measurement layers. To assess the performance of this gadget in isolation, we include a layer of perfect stabilizer measurements at the end of the circuit. The decoding task then consists of the following steps: 

\begin{enumerate}
    \item Given the syndrome information collected in each layer of measurements, find a Pauli operator $E$ that, when applied to the system qubits at the end of the circuit, returns us to the code space.
    \item Determine which of the logically inequivalent classes is more likely. 
    \begin{itemize}
        \item For state preparation, there are only two logical classes to consider, $E L$ and $E$, where $L$ is the logical operator that flips the logical state being prepared ($\overline{X}$ for $\lstate{0}$, $\overline{Z}$ for $\lstate{+}$).
        \item For Steane syndrome extraction, there are four logical classes, but if $X$ and $Z$ errors occur independently, the decoder independently chooses from $\{\overline{I}, \overline{X}\}$, $\{\overline{I},\overline{Z}\}$. 
        \end{itemize}
    \item Apply the operator $E L^*$ where $L^*$ is the most likely logical class in step (2). Decoding fails if the actual fault that occurred propagates to an error that differs from $E L^*$ by a nontrivial logical operator.
    \end{enumerate}
An optimal maximum likelihood decoder computes the probability of each logical class in step (2) exactly. A suboptimal decoder approximates the likelihood of each logical class, for example by using a simplified error model (as when we use the Pauli twirl approximation for coherent errors, or model the native noise as independent bit and phase flips) or by discarding some syndrome information.

\subsection{Erasure Errors}
When the noise is perfectly heralded, we can decode optimally. In theory, this is achieved by mapping the faulty circuit to a spacetime code, which is a subsystem code described further in~\autoref{app:erasures} of the Supplemental Material. In practice, the logical failure probability of this decoder can be determined without implementing the actual decoder, by simulating the noisy circuit evolution in a stabilizer tableau. The simulation is efficient because the erasure operation---tracing out a qubit---can be done in the stabilizer formalism, with worst-case complexity $O(N^2)$. In the state preparation protocol, each root stabilizer is set to $Z$ or $X$, and the entropy of the output state, $S(\rho_S)$, is computed after the final round of perfect stabilizer measurements. The decoder succeeds with probability $1/2^{S(\rho_S)}$, where $S(\rho_S)$ is either 0 or 1, so in a given shot $i$, $P_F(i) = S(\rho_S(i))/2$. This quantity is then averaged over $N_{shots}$ shots, separately for $\lstate{0}$ and $\lstate{+}$, to obtain $P_F(\lstate{0}), P_F(\lstate{+})$. In~\autoref{fig:heralded}(ii), we plot
\begin{equation}
    P_F = \frac{1}{2} \left[P_F(\lstate{0}) + P_F(\lstate{+})\right]
\end{equation}
with error bars
\begin{equation}\label{eq:error-pfail}
\sqrt{\left[\mathrm{SE}(P_F(\lstate{0}))^2 + \mathrm{SE}(P_F(\lstate{0}))^2\right]/2}
\end{equation}
where $\mathrm{SE}(y) = \sqrt{\frac{\sum{(y_i-\overline{y})^2}}{N_{shots} (N_{shots}-1)}}$ is the standard error of the sample mean of $y$. In~\autoref{fig:heralded}(i) and supplemental figures, to better discern the subthreshold performance, we plot the average of the log failure probability,
\begin{equation}\label{eq:ave-log}
\log_{10}[P_F] = \frac{1}{2} \left[\log_{10}(P_F(\lstate{0})) + \log_{10}(P_F(\lstate{+}))\right].
\end{equation}

In the quantum memory demonstration, the input state is a logical Bell pair $(\lstate{0}_S \ket{0}_R + \lstate{1}_S \ket{1}_R)/\sqrt{2}$, and the decoder succeeds with probability $1/2^{2-I(S:R)}$ where $I(S:R)$ is the mutual information between the system and reference (c.f.~\autoref{eq:mutual}).

\subsection{Probability passing}
In the general case, the evaluation of logical class probabilities can be implemented via the contraction of a tensor network. 

In the singletree setting used to prepare logical $T$ states, with measurements only performed in the final layer, this tensor network has the same tree structure as the circuit itself (\autoref{fig:model}a). The tree has an open leg at the root, and an open leg sticking out of each possible error location in the bulk. Since faults at different locations occur independently, a vector of error probabilities can be contracted locally on each bulk leg, yielding a tree tensor network whose output (on the root leg) is the vector of logical class probabilities.

In the fault-tolerant multitree setting, each layer of ancilla measurements gives us partial information about the errors occurring on the system. With this information, we can update the distribution of errors on the system tree via Bayes' rule. Optimal decoding would entail updating the full distribution exactly, introducing correlations between errors at different locations. The resulting tensor network is no longer efficiently contractible due to the large entanglement. 

We therefore instead implement an approximate decoder that uses the idea of \textit{probability passing.}  A simple approximation would ignore all correlations by applying the Bayesian update marginally on each error location. For state preparation, this method achieves a pseudothreshold below which the logical failure probability is roughly independent of system size. To achieve a subthreshold phase in which the logical failure probability decreases with system size, our ``stacked probability passing decoder'' improves upon this crude approximation by keeping the correlations that develop from ancillas that interact directly with the system. The tensor network implementation of this algorithm is detailed in~\autoref{app:ppd} of the Supplemental Material.

{Each shot of the state preparation circuit yields a sample $y_i \in \{0,1\}$ where $y=0$ denotes success, $y=1$ denotes failure. The average failure probability for preparing state $\lstate{\psi} = \lstate{0}, \lstate{+}$ is then
\begin{equation}\label{eq:pfail-bern}
    P_F(\lstate{\psi}) = \frac{\sum_{i} y_i}{N_{shots}} 
\end{equation}
with estimated one sigma confidence interval $\sigma \approx \sqrt{P_F(1-P_F)/N_{shots}}$. The uncertainty on $\log_{10}[P_F(\lstate{\psi})]$ and its average across $\lstate{0}, \lstate{+}$ is determined through error propagation.}

%

\clearpage
\onecolumngrid
\setcounter{figure}{0}
\setcounter{section}{0}
\setcounter{page}{1}
\let\oldthefigure\thefigure
\renewcommand{\thefigure}{S\oldthefigure}
\setcounter{equation}{0}
\counterwithout{equation}{section}
\renewcommand{\theequation}{S\arabic{equation}}
\newpage

\title{Supplemental Information: Observation of a Fault Tolerance Threshold with Concatenated Codes}
\maketitle
\onecolumngrid
The Supplemental Material is organized as follows.
\begin{itemize}
    \item \autoref{app:intro} breaks down the steps for decoding the multitree state preparation circuit and defines the spacetime code.
    \item \autoref{app:erasures} elaborates on the methods for decoding fully heralded errors in the state preparation circuit. Results on subthreshold scaling, variations on the basic noise model, and dynamical rewiring are also presented.
    \item Sections~\ref{app:tensor-networks}-\ref{app:experiment} present the background, methods, and results of decoding unheralded and partially heralded errors.
    \begin{itemize}
    \item \autoref{app:tensor-networks} introduces the tensor network notation used for decoding both the state preparation gadget and the syndrome extraction gadget. As a warmup, it reviews the optimal decoder for the singletree, which is relevant to the preparation of logical $T$ states.
    \item \autoref{app:ppd} describes the stacked probability passing decoder.
    \item \autoref{app:experiment} contains additional details on the two state preparation experiments.
    \end{itemize}
    \item \autoref{app:memory} and \autoref{app:memory-ppd} expand on the discussion of the quantum memory demonstration, presenting results on the optimal decoder and an approximate probability passing decoder, respectively.
    \item {\autoref{app:code-switch} elaborates on the code switching protocol between \GSC{o} and \GSC{e} codes.}
\end{itemize}

\section{Decoding for state preparation: General Theory}\label{app:intro}

\subsection{From quantum to classical}\label{app:classical}
Acting on the state $\lstate{0}$ or $\lstate{+}$, there are only two logically inequivalent classes to consider:
\begin{equation}\label{eq:0state}
    \overline{I} \lstate{0} = \overline{Z} \lstate{0} = \lstate{0}, \quad \overline{X} \lstate{0} = i \overline{Y} \lstate{0} = \lstate{1}
\end{equation}
and similarly for $\lstate{+}$. Without loss of generality, we will focus on $\lstate{0}$ in the ensuing discussion.

\autoref{eq:0state} says that on a logical level, we need only distinguish between two classes, $E_s$ and $E_s \overline{X}$. If, in addition, $X$ and $Z$ errors occur independently on the \textit{physical} level, then the entire decoding procedure can be treated classically, in the following sense.

Consider the stabilizer group of the state $\lstate{0}$ at depth $t$. This group is generated by a set of $Z$-type stabilizer generators coming from the stabilizer inputs on odd layers, as well as the ``root stabilizer'' $\overline{Z}$; and a set of $X$-type stabilizer generators, coming from inputs on even layers. The $Z$-type stabilizers are sensitive only to bit flips, which we will refer to as ``matching-type'' errors because they can flip the root stabilizer, whereas the $X$-type stabilizers are sensitive only to phase flips, ``opposite-type'' errors. 
Since the fault-tolerant gadget includes a round of perfect measurements in which we measure all the stabilizers except the root stabilizer, we can unambiguously fix up the signs on the $X$-type stabilizers in this final round. Moreover, if bit flips and phase flips occur independently\footnote{This includes erasure errors, since erasing a qubit corresponds to applying a bit flip and phase flip each with probability 1/2.}, performing $X$ checks tells us only about the phase flip errors, which have no effect on the relative likelihood of the two logical classes. This means that we gain nothing by measuring the $X$ syndrome in earlier levels, justifying our choice to only perform $Z$-type checks in the first $T-1$ levels. 

Under the transversal CNOT between system $S$ and ancilla $A$ involved in each round of $Z$-type checks, $X$ errors propagate from $S$ to $A$, while $Z$ errors leak from $A$ to $S$. One might think that the latter type of spreading is dangerous. However, since all $Z$ errors are perfectly corrected by the final layer of perfect stabilizer measurements, we can ignore this backwards flow of errors. Indeed, this argument shows independent bit and phase flips in state preparation of $\lstate{0}$ using circuit~\autoref{fig:model}b is entirely equivalent to a model with pure bit flip errors.

In light of this discussion, the errors introduced into the circuits on Quantinuum hardware are only of one type (\autoref{eq:herald-X} and~\autoref{eq:coherent}).

\subsection{Formulation of the problem}\label{app:steps}
Let's break down the decoding task outlined in~\appref{app:decoding} even further. Recall that the ``effect'' of a fault $F$, denoted $\eff(F)$, is the Pauli error on the final time slice that results from propagating $F$ to the end of the circuit.  Consider state preparation of a depth $T$ logical state. Since each time step of the circuit in~\autoref{fig:model}b contains several sublayers with fault locations, we will use $t=1,...,T$ to denote the time step and $\tau=1,...,\tau_{max}$ to denote the spacetime layer.

Any decoder consists of the following steps:
\begin{enumerate}
\item Let $\vec{s} = (s_1,...,s_T)$ denote the vector of syndromes from all $T$ levels of the circuit, such that $\tilde{s} = (s_1,...,s_{T-1})$ are syndromes from the ancilla measurements, and $s_T$ is from the final round of perfect measurements (where the root stabilizer is not measured). Also let $\finalS$ denote the final stabilizer group.
\item Given $\vec{s}$, find a fault operator $F_{\vec{s}}$ (a Pauli operator defined on the spacetime qubits) that produces that syndrome, i.e. $\sigma(F_{\vec{s}})=\vec{s}$. Given the stabilizer generators of the spacetime code, this could be done by solving a system of linear equations. In what follows, we will use $\tilde{\sigma}(F)$ to denote the syndrome from all but the last level, i.e. $\tilde{\sigma}(F_{\vec{s}}) = \tilde{\vec{s}}$. 

We can always break up $F_{\vec{s}}$ into two parts: $F_{\vec{s}} = G_{\tilde{\vec{s}}} \eta_{\tau_{max}}(P_{s_T})$. Following the notation of Ref.~\cite{Delfosse2023}, $\eta_{\tau_{max}}(P)$ denotes a fault operator composed of applying the Pauli operator $P$ to the final time slice, with identity on all previous layers. $P_{s_T}$ is a Pauli that has the matching syndrome $s_T$ on the last level, so that $\sigma(\eta_{\tau_{max}}(P{s_T})) = (\vec{0}, s_T)$. Meanwhile, $G_{\tilde{\vec{s}}}$ is chosen such that $\sigma(G_{\tilde{\vec{s}}}) = (\tilde{\vec{s}}, 0)$.  Therefore, $\eff(G_{\tilde{\vec{s}}}) \in \finalS$, $\eff(\eta_{\tau_{max}}(P_{s_T})) = P_{s_T}$, and  $\eff(F_{\vec{s}}) \in P_{s_T} \finalS$.
\item Our task is to choose a correction operator from the most likely logical class of $F_{\vec{s}}$, given that our noise model induces a prior probability distribution on the faults $\mathbb{P}(F)$. As noted above, for state preparation there are only two classes: one is the identity coset of $F_{\vec{s}}$ in the gauge group $\mathcal{G}$ of the spacetime code discussed below, the other is the logical $X$ coset, which differs by $\overline{X}$, the logical $X$ operator of the final stabilizer group. Let $I(F_{\vec{s}}), X(F_{\vec{s}})$ denote the (not necessarily normalized) weights of these classes.

\item Choose the logical class with the greater probability. Let $L^*(F_{\vec{s}})=I$ if $I(F_{\vec{s}})>X(F_{\vec{s}})$, and $L^* = \overline{X}$ otherwise. Then we can correct the faulty final state by applying $P_{s_T} L^*$. Note that this correction operator lives on the final time slice, so its syndrome is $(\vec{0}, s_T)$, not the syndrome $(\tilde{\vec{s}},s_T)$ that was actually observed. But since we only care about correcting the final state, this is perfectly fine.

The decoder succeeds for a given fault $F$ if $\eff(F) \in \finalS P_{s_T} L^*(F_{\vec{s}})$. Thus the recovery probability, averaged over all possible fault operators, is
\begin{align}\label{eq:recovery-ft}
\sum_{F} \mathbb{P}(F) \mathrm{prob}(F \, \mathrm{is \, decoded \, successfully}) &= \sum_{\tilde{\vec{s}}, s_T} \sum_F \mathbb{P}(F) \delta[\tilde{\sigma}(F) = \tilde{\vec{s}}] \delta[\eff(F) \in \finalS P_{s_T} L^*(F_{\vec{s}})].  
\end{align}
\comment{
\begin{align}\label{eq:recovery-ft}
\sum_{F} \mathbb{P}(F) \mathrm{prob}(F \, \mathrm{is \, decoded \, successfully}) &= \sum_{\vec{s}} \sum_{F: \tilde{\sigma}(F) = \tilde{\vec{s}}} \mathbb{P}(F) \delta[ e(F) \in \finalS P_{s_T} L^*(F_{\vec{s}})] \notag \\
&=\sum_{\tilde{\vec{s}}, s_T} \sum_{F:  e(F) \in \finalS P_{s_T} L^*(F_{\vec{s}})} \mathbb{P}(F) \delta[\tilde{\sigma}(F) = \tilde{\vec{s}}].  
\end{align}
}
\end{enumerate}

\subsection{Spacetime code formalism}
One way to evaluate and interpret the logical class probabilities is by mapping the faulty circuit onto a \textit{spacetime code}~\cite{Bacon2017,Gottesman2022,Delfosse2023}. Each possible error location in the circuit is a qubit of the spacetime code. Each independent bit of syndrome information obtained from the measurements corresponds to a stabilizer generator supported on the spacetime qubits. In the present setting, these stabilizer generators come in two groups.

The first group are the stabilizers associated with the syndrome $\tilde{\vec{s}}$. Consider the $Z$ checks performed in time step $t < T$. There are $2^{T-t-1}$ pairs of $S$ and $A$ blocks. When we couple $A$ to $S$ via a transversal CNOT and measure all of the ancilla qubits in the $Z$ basis, we are learning the $Z$ syndrome of a code on $S\otimes A$. The stabilizer group of this $2^{t+1}$-qubit code is generated by
\begin{equation}\label{eq:block-s}
g_1 \otimes g_1, g_2 \otimes g_2, ..., g_{n} \otimes g_{n(t)}, \overline{Z} \otimes \overline{Z}
\end{equation}
where $g_1,...,g_{n(t)}$ are the $Z$-type stabilizers, associated with the fresh qubits fed in on the right leg of the CNOT gate in odd layers, and $\overline{Z}$ is the root stabilizer, time-evolved from the $Z$ operator fed in on the left leg of the bottom CNOT gate. If $g_i$ originates from a stabilizer input on leg $j_i$ in layer $\tau_i$, then the corresponding stabilizer generator in the spacetime code is a Pauli string supported on spacetime qubits from layers $\tau_i$ up to $\tau$. The support on an intermediate layer $\tau'$ is the Pauli obtained by propagating the pair of $Z$ stabilizer inputs $Z_{j_i,S}, Z_{j_i,A}$ from $\tau_i$ to $\tau'$~\cite{Delfosse2023}. Explicitly, if $P(\tau\rightarrow \tau')$ denotes the Pauli obtained by propagating $\eta_\tau(P)$ up to time $\tau'$, then let\footnote{Such operators are dubbed ``spackle operators'' in Ref.~\cite{Bacon2017}, a term which may have meaning to home improvement experts.}
\begin{equation}
    \overrightarrow{P}_{\tau_1,\tau_2} = \prod_{\tau'=\tau_1}^{\tau_2} \eta_{\tau'}(P(\tau_1 \rightarrow \tau'))
\end{equation}
Then the spacetime stabilizers are $\left(\overrightarrow{Z_{j_i,S}\otimes Z_{j_i,A}}\right)_{\tau_i,\tau}$ for $i=1,...,n(t)$.

The second group are the stabilizers read out on the system in the final layer, with syndrome $s_T$. Now we learn both the $X$ and $Z$ syndrome, but again we can disregard the $X$ stabilizers since we are focusing on bit flip errors. For the \GSC{o} code family, the $Z$ part of the syndrome contains $n(T) = (4^{\lfloor (T-1)/2\rfloor}-1)/3$ bits, each of which corresponds to a spacetime stabilizer propagated from its input to time $T$, $\left(\overrightarrow{Z_{j_i}}\right)_{\tau_i,\tau_{max}}$.

For state preparation, the spacetime code has only half of a logical. To form a canonical generating set of stabilizers and logicals, we define the logical representative $Z_L = \left(\overrightarrow{Z_1}\right)_{1,\tau_{max}}$ where $1$ is the logical leg.

The spacetime code is a subsystem code, whose gauge group relates operators that propagate to the same error. In particular, for each two-qubit gate $U$ acting on qubits $i$ and $j$ between layers $\tau$ and $\tau+1$, then $\overrightarrow{P}_{\tau,\tau+1}$ are in the gauge group for $P=X_i, X_j, Z_i, Z_j$. Our choice of $Z_L$ is made so that it is a ``bare'' logical operator of this subsystem code, meaning that it commutes with the entire gauge group~\cite{Poulin2005,Vuillot2019,Albert2023}.

\subsection{Alternative Conventions for the GS Code}

As defined in~\autoref{fig:model}a, the single tree encoding of the \GSC{o} code alternates between the ``encoding gate'' CNOT with $Z$ inputs (\autoref{eq:copy}) and NOTC with $X$ inputs (\autoref{eq:delocal}). An alternative form, used in some of our numerics and the experimental implementation, defines the isometry in \textit{every} layer as
\begin{equation}\label{eq:had-isometry}
    (H\otimes H) \CNOT_{12} \ket{0}_2.
\end{equation}
Pushing the Hadamard gates through to the end of the circuit, we recover the alternating isometries in the main text.

If~\autoref{eq:had-isometry} is used, then the root stabilizer alternates between being all $X$'s and all $Z$'s, so the matching-type errors alternate between bit flips and phase flips in odd and even layers. This technicality means that, in practice, the added noise in the {trapped-ion demonstration} consists of alternating $Z$ and $X$ rotations. We stress, however, that the placement of Hadamards has no effect on the performance, provided that the underlying noise is symmetric between $X$ and $Z$. If, for example, the noise channel were biased toward bit flip errors, then we should place the Hadamard gates in such a way that the root stabilizer is always composed of all $X$'s. Since all of our noise models are unbiased\footnote{That is, the channels~\autoref{eq:unheralded} and~\autoref{eq:erasures} are symmetric under $X\leftrightarrow Z$, \textit{before} we use the argument from~\autoref{app:classical} to disregard the opposite-type errors in the state preparation circuit.}, we will keep with the convention in the main text for ease of presentation. {Unless otherwise noted, we present results for the \GSC{o} version of the code.}

\subsection{Asymmetry between $\lstate{0}$ and $\lstate{+}$}
Despite the symmetry between $X$ and $Z$ errors at the physical level, we empirically find that, for a given error rate and tree depth, the logical failure probability for preparing $\lstate{+}$ {in the \GSC{o} encoding} is significantly higher then that of $\lstate{0}$. {A plausible origin for this trend is that (regardless of whether we push Hadamards through or not), in the first layer of checks, the syndrome for each pair of $S$ and $A$ blocks has 2 bits for $\lstate{0}$, but only 1 for $\lstate{+}$. This gives $\lstate{0}$ a ``head start'' which persists to large depths.}

The two logical states also exhibit opposite odd-even effects with $T$. For that reason, in the main text we averaged over $\lstate{0}$ and $\lstate{+}$. The data are shown separately in~\autoref{fig:heralded-sep},~\autoref{fig:ppd-sep},~\autoref{fig:qtuum-sep}, and~\autoref{fig:coherent-sep} of this Supplement. 

\section{State preparation with erasure errors}\label{app:erasures}

\subsection{Failure probability from the spacetime code}\label{app:rM}
At the heart of the decoding task laid out in~\autoref{app:steps} of this Supplement is the calculation (exact or approximate) of the the logical class probabilities $I(F_{\vec{s}}), X(F_{\vec{s}})$. This computation is easy for erasure errors, because every fault that contributes to a logical class is either equally likely (if it is supported on the flagged spacetime locations and produces the observed syndrome) or has zero conditional probability (is inconsistent with the heralding information or measurement outcomes). Moreover, the total probability of a logical class is either zero or uniform. For state preparation, that means either $I(F_\vec{s}) = X(F_\vec{s})$, or only one is nonzero. Thus, the following polynomial-time decoder is optimal:
\begin{enumerate}
\item Given spacetime syndrome $\vec{s}$ and heralded locations $\mathcal{E}$, find $F_{\vec{s}}$ such that $\sigma(F_{\vec{s}}) = \vec{s}$ and $F_\vec{s}$ is contained in $\mathcal{E}$. $F_\vec{s}$ can be taken as a Pauli $X$ string whose support is a vector $\vec{b}$, where $\vec{b}$ is obtained from a system of $|\vec{s}|$ equations in $|\mathcal{E}|$ boolean variables.
\item By definition, $I(F_\vec{s})$ is nonzero since $F_\vec{s}$ is consistent with the syndrome and erasure pattern. So $\eff(F_\vec{s})$ is as good a choice as any for the correction operator.
\end{enumerate}

The failure probability of the optimal erasure decoder for a (subsystem) code can therefore be obtained from a canonical set of stabilizers and logicals as follows~\cite{Gullans21}. Sample an erasure pattern $\mathcal{E}$. For each erased site $\vec{r}$ (i.e. a qubit of the spacetime code), we form an error basis $X_{\vec{r}}, Z_{\vec{r}}$. Then each stabilizer generator and logical corresponds to a row of a binary matrix $M$, and each error in the basis corresponds to a column, where $M_{ij}=1$ if the $i$th generator anticommutes with the $j$th error. The stabilizer rows form the submatrix $M_S$.

The failure probability for a given erasure pattern is then
\begin{equation}\label{eq:rM}
P_F = 1 - \frac{1}{2^{r_M}}, \quad \mathrm{where} \, r_M = \mathrm{rank}(M)-\mathrm{rank}(M_S).
\end{equation}
In the present case, $r_M \in \{0,1\}$: either we can decode perfectly, or else a logical $X$ error is undetectable, meaning there is some error supported on the erased spacetime region that has a trivial syndrome but anticommutes with $Z_L$.

In practice, we find that a more efficient method for extracting the logical failure probability is to simulate the noisy circuit and compute the entropy of the system after the final round of perfect stabilizer measurements. The entropy of the final state, $S(\rho)$, is either 0 or 1: it corresponds directly to the quantity $r_M$ defined above. For some of the data collection, we used a recursive implementation of the circuit by which, in the course of simulating a depth $T$ circuit, we also determine $S(\rho)$ for $2^{T-t-1}$ depth-$t$ circuits, thus obtaining many more samples at smaller depths.

\subsection{Subthreshold scaling}
Given a stabilizer code family with distance $d(n)$ for system size $n$, well below the code capacity threshold, the failure probability follows the ansatz
\begin{equation}\label{eq:subthreshold}
P_F(p, n) \propto \exp[-c(p) d(n)].
\end{equation}

The GS code has a code distance of $d(T) = 2^{\lfloor T/2 \rfloor} \approx n^{1/2}$, so if noise is applied at the end of the singletree encoding circuit (on the leaves of~\autoref{fig:model}a), then $\log[P_F(p,T)]$ decreases linearly with $x=2^{T/2}$. In contrast, when noise is applied in the bulk of the singletree, the subthreshold behavior is replaced with
\begin{equation}
    P_F(p, T) = a(p) - b(p) 2^{-c(p) T},
\end{equation}
i.e., as $T\rightarrow \infty$, the system converges to a fixed point with finite failure probability $a(p)$, and the convergence to this fixed point is only a simple exponential in $t$, rather than doubly exponential~\cite{Sommers2024tree}. 

\begin{figure}[hbtp]
\subfloat[]{
\includegraphics[height=0.195\textheight]{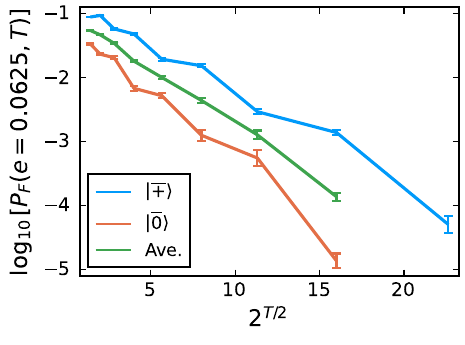}\label{fig:subthreshold-state-prep}}\hfill
\subfloat[]{
\includegraphics[height=0.19\textheight]{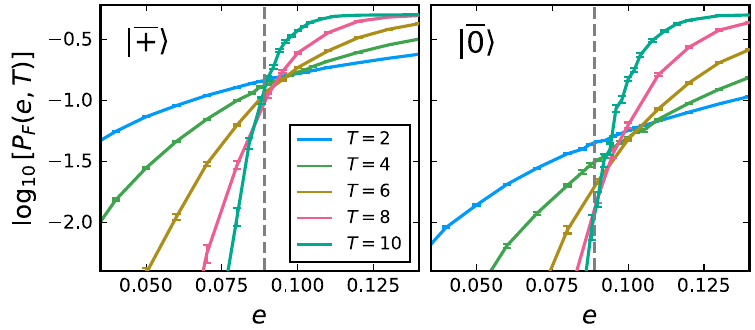}\label{fig:heralded-sep}}
\caption{Logarithm of the logical failure probability for state preparation of $\lstate{+}$ and $\lstate{0}$, at erasure rate $e_i=e_g=e_m=e$. (a) Subthreshold scaling at $e=0.0625$, for $\lstate{+}$ (blue), $\lstate{0}$ (orange), and the average between them (green). For $\lstate{+}$, the number of shots ranges from 5000 at $T=3$ to $100\,000$ at $T=9$. For $\lstate{0}$, the number of shots ranges from 1000 at $T=3$ to $520\,000$ at $T=8$. {The average across logical states is defined as in~\autoref{eq:ave-log}, taking the logarithm first.} (b) Log failure probability at even depths $T$. At least 1000 shots are averaged over for each data point at each $T$, with many more shots obtained at smaller $T$ by virtue of the data collection process. {Error bars denote the standard error of the mean.}}
\end{figure}

With the butterfly network state preparation circuit, under erasure errors, the scaling form of~\autoref{eq:subthreshold} is restored. At erasure rate $e$ well below $e_c=\pc$, a 3-parameter fit to $\log(P_F(T)) = 2^{a T} b + c$ yields $a \approx  1/2$. The linear trend in $\log(P_F(e,T))$ as a function of $2^{T/2}$ is shown in~\autoref{fig:subthreshold-state-prep} for $e=0.0625$ and depths up to $T=8$. While the log failure probabilities for $\lstate{0}$ and $\lstate{+}$ each show opposite odd-even trends, their average smoothly decreases.

In the main text, we presented (the logarithm of) the logical failure probability averaged over $\lstate{+}$ and $\lstate{0}$(\autoref{fig:heralded}).~\autoref{fig:heralded-sep} shows the logical failure probabilities for $\lstate{0}$ and $\lstate{+}$ separately, again on a logarithmic scale to highlight the subthreshold performance.

\subsection{Different noise models}
The basic noise model described in~\appref{app:models} has three types of error locations: those associated with state preparation/``input errors'', errors after two-qubit gates, and measurement errors. These three error sources have different densities in the circuit. The total number of possible input error locations is the smallest---$(\lfloor T/2 \rfloor +1)2^{T-1}$ for preparing $\lstate{+}$, $\lfloor{(T+3)/2}\rfloor 2^{T-1}$ for preparing $\lstate{0}$---because the input errors on alternating levels (where the opposite-type stabilizers are fed in) have no effect at all on the ability to decode at the end. There are approximately twice as many possible erasure locations associated with measurements---$(T-1) 2^{T-1}$---and 8$\times$ as many associated with gates: $(2T-1) 2^T$.

In the main text, we took the rate of erasure errors to be the same across all three types: $e_{i}=e_{g}=e_m$. To examine the relative effect of these three error sources on the threshold, we now consider three edge cases: $e_i = 0$ (keeping $e_g = e_m = e$), $e_m = 0$, and $e_{g} = 0$.

The erasure rate $e$ is then defined as the total density of erasures:
\begin{equation}\label{eq:e}
e = \frac{e_i n_i + e_g n_g + e_m n_m}{n_{i} + n_{g} + n_{m}}
\end{equation}

Compared to the homogeneous case, the threshold stays roughly the same for $e_i = 0$ and $e_m = 0$. On the other hand, the failure probability is strongly suppressed in the absence of gate errors, even after accounting for their higher density by defining $e$ as in~\autoref{eq:e}. This suggests that gate errors are more potent, presumably because they can propagate in a more dangerous fashion.

We also explored several models in which the total number of erasures is fixed, either per layer, per error type, or per subtree of the butterfly network. Recall from~\autoref{fig:heralded} that the logical failure probability with an i.i.d. erasure model obeys a scaling collapse (\autoref{eq:scaling}) with $\nu \approx 2.85$.\footnote{The inferred exponent is smaller, $\nu \approx 2.25$, if we use $N(T) = 2^T$, the number of physical qubits, rather than $N(T) = T 2^T$.} For all models with i.i.d. heralded errors, we generically expect, and all our data are consistent with, $\nu \geq 2$, because the quenched randomness in the number of errors results in a $\sqrt{N}$ broadening of the transition. In random stabilizer codes subject to end-of-circuit noise, changing from i.i.d. erasures to a fixed-fraction model significantly alters the critical behavior, sharpening the scaling function near the threshold from $f((e-e_c)N)^{1/2})$ to $g((e-e_c)N)$~\cite{Gullans21}. Can we similarly modify the scaling of our fault tolerance transitions by reducing the randomness in the error model? In all variations of the multitree state preparation circuit we have tried, the answer is no: we find $\nu\gtrsim 2$, with no significant sharpening of the $P_F(e,T)$ curves.

\subsection{Dynamical rewiring}
The state preparation protocol depicted in~\autoref{fig:model}b  involves performing checks on $2^{T-t-1}$ pairs of $S$ and $A$ blocks in step $t=1,...,T-1$. But the pairing of blocks in each step is arbitrary, as is the designation of $S$ and $A$ within each pair. If we know (based on prior rounds of checks and/or heralded error locations) that one member of the pair is less noisy than the other, we might want to keep the less noisy block as the system -- or we might gain an even greater advantage by not performing checks between the pair at all, and simply discarding the noisier block. The idea that the designation of system vs. ancilla and the decision of whether or not to perform a check can be optimized based on information learned from earlier rounds motivates the ``dynamical rewiring'' protocol discussed in this subsection.

\begin{figure}[hbtp]
\subfloat[]{
\includegraphics[width=0.3\linewidth]{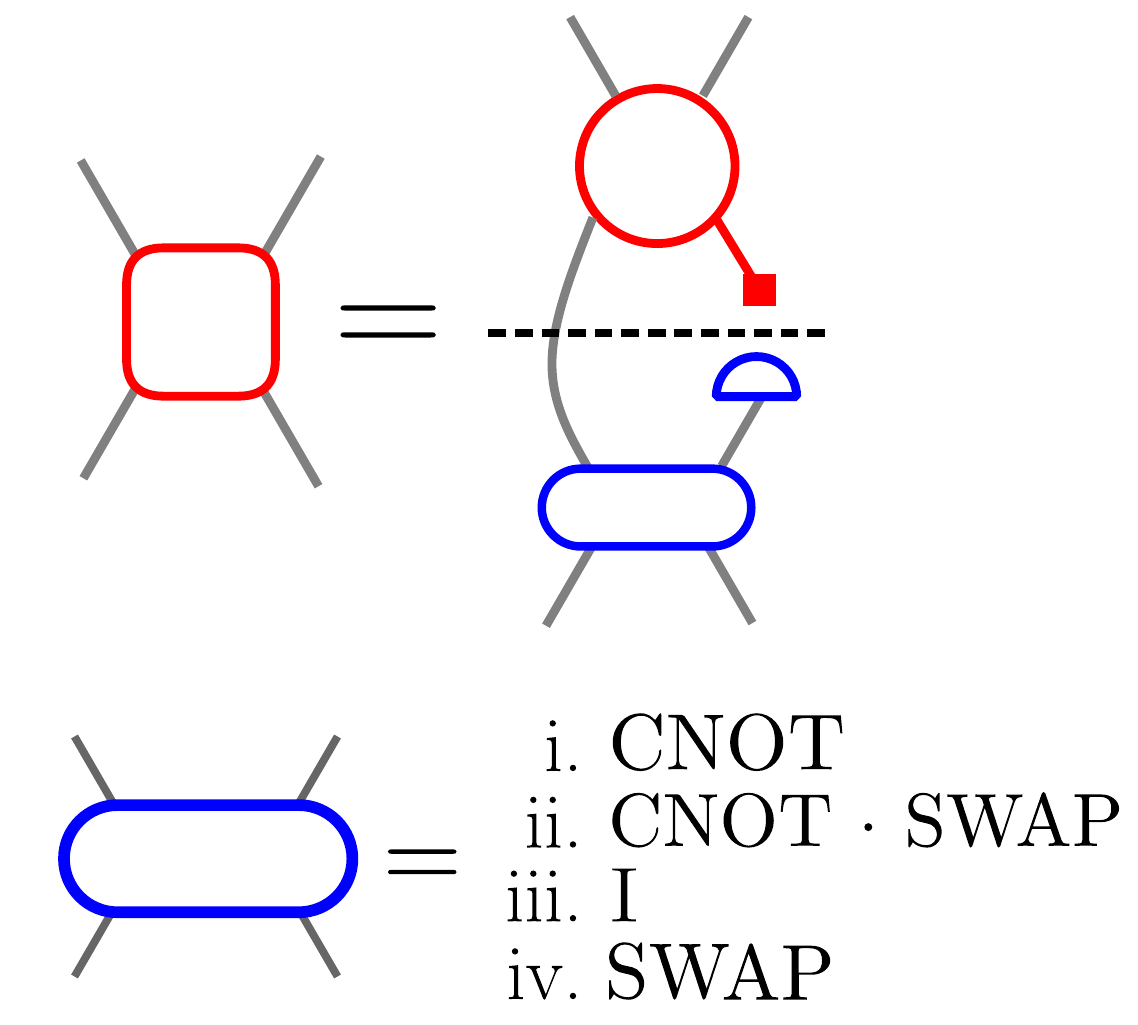}\label{fig:rewired-sketch}}
\hfill
\subfloat[]{
\includegraphics[width=0.55\linewidth]{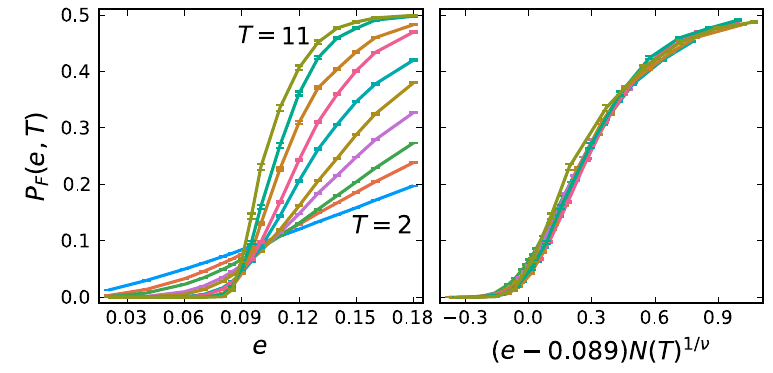}\label{fig:rewired-fail}}
\caption{(a) Interaction between blocks in an even layer of checks for state preparation of $\lstate{0}$. Dashed line indicates where a layer of perfect stabilizer measurements is inserted, to assess the reliability of the state on that time slice.  (b) Logical failure probability, averaged over $\lstate{+}$ and $\lstate{0}$, using the dynamical rewiring protocol. Right panel shows a scaling collapse of the form~\autoref{eq:scaling} with $N(T)=2^T$ the number of physical qubits, $p_c=0.089, \nu = 3.5$, depths 4-11. {Error bars are defined as in~\autoref{eq:error-pfail}, with $N_{shots}=1000$ at $T=11$ and $N_{shots}=2^{10-T}\cdot 1000$ for smaller $T$.}}
\end{figure}

It is infeasible to try all possible pairings of systems and ancillas in every layer, so we opt for the most basic form of optimization: keeping the same butterfly network structure, but changing the interaction between pairs. Consider state preparation of $\lstate{0}$. In the default implementation of the checking procedure, the left block $B_L$ acts as the ``system'' and the right block $B_R$ acts as the ancilla. After (1) a transversal CNOT with $B_L$ as control and $B_R$, we (2) measure all qubits in $B_R$ in the $Z$ basis, then (3) reset those qubits to $\ket{+}$ ($\ket{0}$) after odd (even) layers, respectively. The reset qubits are then (4) fed in as inputs to the next level of encoding on $B_L$.

As shown in~\autoref{fig:rewired-sketch}, in the dynamical rewiring protocol, we optimize across four different transversal gates in step (1): (i) CNOT (default), (ii) $\mathrm{CNOT} \cdot \mathrm{SWAP}$ (meaning $B_L$ now acts as the ancilla, and $B_R$ as system), (iii) identity gate (meaning treat $B_L$ as the system and do not check it), and (iv) SWAP (meaning we choose $B_R$ as the system and do not check it). In options (iii) and (iv), the subsequent measurement step can be omitted, since it only tells us about the syndrome on the ancilla block which is then reset.

If the errors are fully heralded, we choose between the four wirings as follows:
\begin{itemize}
\item Given an erasure pattern up to depth $t$, classically simulate preparation of the state up to depth $t$ with wiring i, ii, iii, or iv in the $t$-th layer of checks, up to the dashed line in~\autoref{fig:rewired-sketch}, followed by perfect stabilizer measurements. Simulating wirings i and ii involves sampling erasure errors after the check gates and before the measurements (which have not been observed yet).
\item Choose the wiring that yields the lowest entropy for prepared state. In case of a tie, default to i or ii for reasons discussed below. 
\end{itemize}

Under the same (fully heralded) noise model as~\autoref{fig:heralded}, we can again infer the failure probability just by simulating the stabilizer evolution and recording the entropy of the final state. The data up to depth 11 in~\autoref{fig:rewired-fail} indicate roughly the same threshold as the default implementation: the crossing between system sizes drifts to smaller $e$ as $T$ increases, but a reasonable scaling collapse is obtained with $e_c = 0.089$. {As shown in the right panel, the transition is also broader:  whereas the original protocol has a scaling collapse to $(e-e_c) (T 2^T)^{1/\nu}$ with $\nu \approx 2.85$, the rewired protocol has a scaling collapse to $(e-e_c) (T 2^T)^{1/\nu}$ with $\nu \approx 3.5$.}

\begin{figure}[hbtp]
\includegraphics[width=\linewidth]{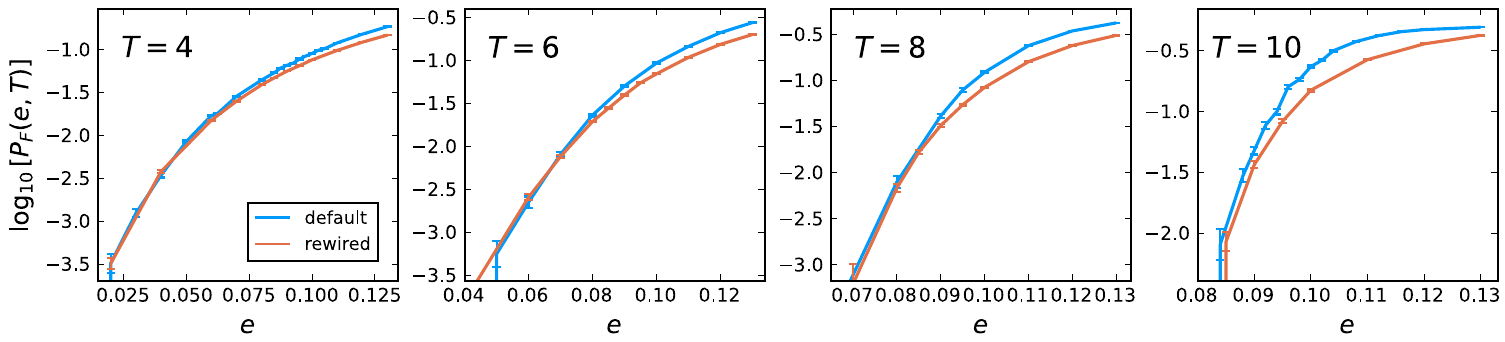}
\caption{Comparison of $\log[P_F(e,T)]$, averaged over $\lstate{0}$ and $\lstate{+}$ (\autoref{eq:ave-log}), in the default implementation (blue) vs. the circuit with dynamical rewiring (orange), under erasure errors.\label{fig:rewired-comp} {Error bars are defined from error propagation of the standard error of the mean for each logical state.}}
\end{figure}

While the above results suggest that the dynamical rewiring protocol is ineffective at large depths, it does lead to noticeable gains at shallow depths and error rates near and above threshold. The logical failure probability in the default and rewired protocols are compared in~\autoref{fig:rewired-comp}. 

{A tentative explanation for the failure of the rewiring scheme to yield asymptotic improvements over the default implementation is the following. For a fixed erasure pattern, consider the final state produced by the default circuit $\rho_1$ (all wirings (i)) vs. the state $\rho_2$ produced by the rewired circuit. For $e<e_c$ almost all pairs at sufficiently large $t$ use wiring (i) or (ii). The choice of (ii) vs. (i) is simply a rearrangement of the erasure locations. While for some patterns this rearrangement may result in $0 = S(\rho_2) < S(\rho_1) = 1$, a rewiring protocol that only uses wirings (i) or (ii) should not have a threshold above that of the default, since for $e>e_c$ \textit{any} rearrangement of a typical erasure pattern will, with high probability, still produce a mixed state.}

{For $e$ near $e_c$, a significant fraction of pairs $B_L, B_R$ are instead assigned a ``no-check'' wiring (iii) or (iv). Consider a protocol in which wiring (iii) or (iv) is \textit{always} chosen. Since there are no midcircuit checks, the threshold of this protocol cannot exceed that of the singletree preparation + distillation, $e_{single}$. However, for both heralded and unheralded errors, the multitree thresholds obtained in this work exceed the singletree thresholds found in Ref.~\cite{Sommers2024tree}.}

{In contrast, a variation of the multitree circuit which prepares logical Bell states, requiring both types of ancilla checks in each layer, has a threshold $e_{c,Bell}$ below $e_{single}$. Thus, an optimal rewiring scheme for that circuit, least in the regime $e_{c,Bell} < e < e_{single}$, is a filtering scheme in which, in each step $t$, no checks are performed at all, and we simply choose the $2^{T-t-1}$ blocks with the lowest entropy to pass on to the next layer. This example demonstrates how adaptive rewiring can be beneficial in some circumstances. A criterion to choose between different wirings based on prior syndromes could also be defined for unheralded error models, with caveat that, since the choice of wiring must be made in real time, the evaluation time of the criterion must be much shorter than the coherence time of the quantum device.}

\section{Decoding using tensor networks}\label{app:tensor-networks}
When the errors are unheralded, or imperfectly heralded, it is no longer the case that each logical class has either zero probability or equal probability. Determining the most likely class is thus a more challenging task, for which we use the tensor enumerator techniques of Ref.~\cite{Cao2024,Cao2024expansion}. In this section, we set up the notation and specialize to the GS code, closely following the presentation in Ref.~\cite{Sommers2024tree}.

\subsection{Notation}
The basic idea behind the decoders for unheralded noise used in this work is that the probability of a logical class  can be expressed as the output of a tensor network. To do so, let us introduce the basis vectors $e_I, e_X, e_Z, e_Y$.\footnote{In subsequent analysis, since the decoding task is ``essentially classical'', we will only need two basis vectors (cf.~\autoref{eq:classical-basis}).} Each physical qubit is associated with the vector
\begin{equation}
\begin{tikzpicture}
	\begin{pgfonlayer}{nodelayer}
		\node [style=none] (0) at (0, -1) {};
		\node [style=physical] (1) at (0, 0.75) {};
	\end{pgfonlayer}
	\begin{pgfonlayer}{edgelayer}
		\draw [style=gray] (1) to (0.center);
	\end{pgfonlayer}
\end{tikzpicture}
 = e_I.
\end{equation}

Each two-qubit Clifford gate $U$ is associated with a four-legged tensor (a \textit{tensor enumerator}~\cite{Cao2024}),
\begin{equation}\label{eq:gate-enum}
\begin{tikzpicture}
	\begin{pgfonlayer}{nodelayer}
		\node [style=U] (0) at (0, 0) {};
		\node [style=none] (1) at (-1, 1) {};
		\node [style=none] (2) at (1, 1) {};
		\node [style=none] (3) at (-1, -1) {};
		\node [style=none] (4) at (1, -1) {};
		\node [style=none] (5) at (-1.2, 1.2) {$\scriptstyle{\alpha}$};
		\node [style=none] (6) at (1.2, 1.2) {$\scriptstyle{\beta}$};
		\node [style=none] (7) at (1.2, -1.2) {$\scriptstyle{j}$};
		\node [style=none] (8) at (-1.2, -1.2) {$\scriptstyle{i}$};
	\end{pgfonlayer}
	\begin{pgfonlayer}{edgelayer}
		\draw [style=gray] (1.center) to (4.center);
		\draw [style=gray] (3.center) to (2.center);
	\end{pgfonlayer}
\end{tikzpicture} = R^{\alpha\beta}_{i j}
\end{equation}
with $R^{\alpha \beta}_{ij} = 1$ if $P_i P_j$ evolves to $\pm P_{\alpha}P_\beta$ under the gate. This object is an unsigned version of the \textit{circuit tensor} for Clifford gates, defined in Ref.~\cite{Kukliansky2024}.

Each stabilizer input --- i.e., feeding in a fresh qubit in the +1 eigenstate of $P_\alpha$ -- is associated with the vector $e_I + e_{P_\alpha}$, which we will denote by
\begin{equation}\label{eq:stabilizer}
\begin{tikzpicture}
	\begin{pgfonlayer}{nodelayer}
		\node [style=stabilizer] (0) at (0, -0.5) {$\alpha$};
		\node [style=none] (1) at (0, 1) {};
	\end{pgfonlayer}
	\begin{pgfonlayer}{edgelayer}
		\draw [style=gray] (1.center) to (0);
	\end{pgfonlayer}
\end{tikzpicture}
 = e_I + e_{P_\alpha},
\end{equation}
the circuit tensor for state preparation~\cite{Kukliansky2024}.
 
Finally, we'll need a way to ``insert errors'' into the tensor network. For convenience, let's use $\pauliplus$ to denote
\begin{equation}\label{eq:pauliplus}
i \pauliplus j = k\quad \mathrm{if} \quad P_i P_j \sim P_k
\end{equation}
where $\sim$ means ``equal up to a phase.''\comment{As in Ref.~\cite{Sommers2024tree}, we adopt a nonstandard convention $P_0 = I, P_1 = X, P_2 = Z, P_3 = Y.$ ordering of $Y$ and $Z$ is nonstandard, chosen for consistency with the binary symplectic representation of stabilizer circuits~\cite{Gottesman2024}, where $X = \begin{pmatrix} 1 & 0 \end{pmatrix}, Z = \begin{pmatrix} 0  & 1 \end{pmatrix}, Y = \begin{pmatrix} 1 & 1\end{pmatrix}$.}

Then, define
\begin{equation}\label{eq:open-error}
\tilde{Q}^{\alpha\beta}_{ij} = \begin{tikzpicture}
	\begin{pgfonlayer}{nodelayer}
		\node [style=none] (0) at (0, 1.25) {};
		\node [style=none] (1) at (0, -1.25) {};
		\node [style=error] (2) at (0, 0) {};
		\node [style=none] (4) at (1.75, 0) {};
		\node [style=none] (5) at (0, 1.575) {$i$};
		\node [style=none] (7) at (0, -1.575) {$j$};
		\node [style=none] (8) at (2, 0) {$\beta$};
		\node [style=none] (9) at (0, 0) {$\alpha$};
	\end{pgfonlayer}
	\begin{pgfonlayer}{edgelayer}
		\draw [style=gray] (0.center) to (1.center);
		\draw (9.center) to (4.center);
	\end{pgfonlayer}
\end{tikzpicture} = \delta_{\alpha \pauliplus \beta, i\pauliplus j} p_\beta
\end{equation}
where $p_\beta$ is the probability of applying $P_\beta$ at that location. Finally, as shorthand, define
\begin{equation}
Q^{\alpha} = \begin{tikzpicture}
	\begin{pgfonlayer}{nodelayer}
		\node [style=none] (0) at (0, 1) {};
		\node [style=none] (1) at (0, -1) {};
		\node [style=error] (2) at (0, 0) {};
		\node [style=none] (3) at (0, 0) {$\alpha$};
	\end{pgfonlayer}
	\begin{pgfonlayer}{edgelayer}
		\draw [style=gray] (0.center) to (1.center);
	\end{pgfonlayer}
\end{tikzpicture}
 = \sum_\beta \tilde{Q}^{\alpha\beta} \Rightarrow Q^{\alpha}_{ij} = p_{\alpha \pauliplus i \pauliplus j}.
\end{equation}

\subsection{Decoding errors in the bulk of the singletree circuit}
As a warm-up for decoding in multitrees, let us review the decoding procedure for the singletree circuit~\cite{Sommers2024tree}. Assign each possible error location (spacetime qubit) a label, $j$, and suppose the probability of applying the Pauli $P_{\alpha}$ at this location is $p_\alpha(j)$. Given a syndrome $s$ measured in the last layer, consider the reference fault $F_s$, which is a Pauli operator defined on the spacetime qubits. If the Pauli applied at location $j$ is $P_{\alpha_j}$, then we insert the tensor $Q^{\alpha_j}$, evaluated at $\vec{p} = \vec{p}(j)$. In a model with errors on each link of a depth $t$ tree,

\begin{equation}\label{eq:coset-enumerator-bulk}
    \vec{A}^{(L, F_s)}_t = \scalebox{1.1}{\begin{tikzpicture}
	\begin{pgfonlayer}{nodelayer}
		\node [style=U] (0) at (-3.5, 0) {};
		\node [style=U] (1) at (3.5, 0) {};
		\node [style=U] (2) at (0, -2.5) {};
		\node [style=U] (3) at (-5.5, 2.5) {};
		\node [style=U] (4) at (-1.5, 2.5) {};
		\node [style=U] (5) at (1.5, 2.5) {};
		\node [style=U] (6) at (5.5, 2.5) {};
		\node [style=error] (9) at (-6, 3.5) {};
		\node [style=error] (10) at (-5, 3.5) {};
		\node [style=error] (11) at (-2, 3.5) {};
		\node [style=error] (12) at (-1, 3.5) {};
		\node [style=error] (13) at (1, 3.5) {};
		\node [style=error] (14) at (2, 3.5) {};
		\node [style=error] (15) at (5, 3.5) {};
		\node [style=error] (16) at (6, 3.5) {};
		\node [style=physical] (24) at (-6.5, 4.5) {};
		\node [style=physical] (25) at (-4.5, 4.5) {};
		\node [style=physical] (26) at (-2.5, 4.5) {};
		\node [style=physical] (27) at (-0.5, 4.5) {};
		\node [style=physical] (28) at (0.5, 4.5) {};
		\node [style=physical] (29) at (2.5, 4.5) {};
		\node [style=physical] (30) at (4.5, 4.5) {};
		\node [style=physical] (31) at (6.5, 4.5) {};
		\node [style=none] (36) at (0, -4.5) {};
		\node [style=none] (39) at (-6, 3.5) {$\scriptstyle{\alpha_1}$};
		\node [style=error] (47) at (-4.5, 1.25) {};
		\node [style=error] (48) at (-2.5, 1.25) {};
		\node [style=error] (49) at (2.5, 1.25) {};
		\node [style=error] (50) at (4.5, 1.25) {};
		\node [style=error] (51) at (-1.75, -1.25) {};
		\node [style=error] (52) at (1.75, -1.25) {};
		\node [style=error] (53) at (0, -3.75) {};
		\node [style=none] (54) at (-5, 3.5) {$\scriptstyle{\alpha_2}$};
		\node [style=none] (55) at (-2, 3.5) {$\scriptstyle{\alpha_3}$};
		\node [style=none] (56) at (-1, 3.5) {$\scriptstyle{\alpha_4}$};
		\node [style=none] (57) at (1, 3.5) {$\scriptstyle{\alpha_5}$};
		\node [style=none] (58) at (2, 3.5) {$\scriptstyle{\alpha_6}$};
		\node [style=none] (59) at (5, 3.5) {$\scriptstyle{\alpha_7}$};
		\node [style=none] (60) at (6, 3.5) {$\scriptstyle{\alpha_8}$};
		\node [style=none] (61) at (-4.5, 1.25) {$\scriptstyle{\alpha_9}$};
		\node [style=none] (62) at (-2.5, 1.25) {$\scriptstyle{\alpha_{\scalebox{0.4}{10}}}$};
		\node [style=none] (63) at (2.5, 1.25) {$\scriptstyle{\alpha_{\scalebox{0.4}{11}}}$};
		\node [style=none] (64) at (4.5, 1.25) {$\scriptstyle{\alpha_{\scalebox{0.4}{12}}}$};
		\node [style=none] (65) at (-1.75, -1.25) {$\scriptstyle{\alpha_{\scalebox{0.4}{13}}}$};
		\node [style=none] (66) at (1.75, -1.25) {$\scriptstyle{\alpha_{\scalebox{0.4}{14}}}$};
		\node [style=none] (67) at (0, -3.75) {$\scriptstyle{\alpha_{\scalebox{0.4}{15}}}$};
	\end{pgfonlayer}
	\begin{pgfonlayer}{edgelayer}
		\draw [style=gray] (0) to (2);
		\draw [style=gray] (2) to (1);
		\draw [style=gray] (5) to (1);
		\draw [style=gray] (1) to (6);
		\draw [style=gray] (3) to (0);
		\draw [style=gray] (0) to (4);
		\draw [style=gray] (24) to (3);
		\draw [style=gray] (25) to (3);
		\draw [style=gray] (26) to (4);
		\draw [style=gray] (4) to (27);
		\draw [style=gray] (5) to (28);
		\draw [style=gray] (29) to (5);
		\draw [style=gray] (30) to (6);
		\draw [style=gray] (6) to (31);
		\draw [style=thick] (2) to (36.center);
	\end{pgfonlayer}
\end{tikzpicture}},
\end{equation}
where, since the model of choice has no errors on the stabilizer inputs, we have suppressed those legs. $\mathbf{A}^{(L,E_s)}$ is a \textit{vector coset enumerator}~\cite{Cao2024expansion}: it is, in general, a four-component vector with a logical leg ($L$) left open, whose $e_P$ component is the total probability of a fault that propagates to an error in the logical $\overline{P}$ coset relative to $F_s$.  Then $L^* = \overline{P}_{\alpha^*}$ where $\alpha^* = \mathrm{argmax}(\coset{F_\vec{s}})$.

Summing over the logical leg yields the total probability of the syndrome $s$, $B_t^{(s)}= |\coset{F_s}|$. The logical failure probability for decoding this syndrome is 
\begin{equation}
    P_F(s) = 1 - \frac{ (\coset{F_s})_{\alpha^*}}{|B_t^{(s)}|}.
\end{equation}

We can also calculate probabilities of errors in one or more locations $\{j\}$, conditioned on the observed syndrome, by replacing $Q^{\alpha_j}$ with $\tilde{Q}$ on site $j$. The conditional probability depends only on the syndrome and not on the particular choice of reference fault $F_s$, since the logical leg is summed over.

When errors are partially or fully heralded, our decoder has access not just to the syndrome of violated stabilizers, but also a classical record of corrupted sites. Suppose our error model is fully heralded erasures (\autoref{eq:erasures}), and consider the reference fault $F_s$, which produces the observed syndrome and is supported on the erased qubits.\footnote{In practice, $F_s$ need not be supported on the erased qubits: at least one of the components of $\coset{F_s}$ is guaranteed to be nonzero, since there is \textit{some} error with the correct syndrome supported on the erased sites. Taking $F_s$ to be supported on the erased region guarantees that the $e_I$ component is nonzero.} Then if $j$ is heralded, $Q^{\alpha_j}_{nm} = 1/4$ regardless of which Pauli is applied at location $j$: erasing a qubit means all Paulis are equally likely. Meanwhile, the remaining locations are guaranteed to have suffered no errors, so $Q^{\alpha_j}_{nm} = \delta_{\alpha_j + n + m}$. The output of~\autoref{eq:coset-enumerator-bulk} then takes the form
\begin{equation}\label{eq:coset-heralded}
    \coset{F_s} = B_t^{(s)} \begin{cases}
        e_I & \mathrm{perfect \, decoding \, possible,} \\
        (e_I + e_{\alpha \neq I})/2 & \mathrm{logical \, } P_\alpha \mathrm{\, error \, undetectable,} \\
        (e_I + e_X + e_Z + e_Y)/4 & \mathrm{logical \, state \, fully \, mixed.}
    \end{cases}
\end{equation}

Partial heralding (which is the model relevant to the experiments in~\autoref{app:experiment} of this Supplement) is treated similarly, except that there is some residual error probability on the unheralded sites.
\subsection{Interlude: Logical $T$ States}\label{app:T}
The above discussion shows that arbitrary Pauli noise can be efficiently decoded if the circuit is a singletree. Before proceeding to the multitree state preparation circuit for $\lstate{0}$ and $\lstate{+}$, let us briefly comment upon how the logical fidelity for singletree preparation of $\lstate{T}$ is extracted from the tensor network calculation.

For a given syndrome $\vec{s}$,
\begin{equation}\label{eq:error-s}
\mathbbm{P}_{P_\beta}(s) = \mathrm{prob(correction \, operator \, differs \, from \, actual \, fault \, by \, a \, logical 
\,}\overline{P}_\beta) = \frac{(\coset{F_s})_{\alpha^* \oplus \beta}}{B_t^{(s)}}.
\end{equation}

Averaging over syndromes then yields the average probability of a logical $P_\beta$ error.
\begin{equation}
\mathbbm{P}_{P_\beta} = \sum_{s} (\coset{F_s})_{\alpha^* \oplus \beta}.
\end{equation}

In Ref.~\cite{Sommers2024tree}, a subset of us found that the singletree encoding circuit of a GS code admits a ``coding phase'' at low bulk error rate, in which $\mathbbm{P}_I$ converges to a nontrivial value (above 1/4). For independent bit and phase flips, the critical noise rate was estimated to be $r_c = 0.0066 \pm 0.0004$, while under heralded bit and phase flips, the threshold $e_c = 0.05505...$ was determined analytically. The analytic tractability of the heralded phase diagram again stems from the fact that when physical errors are heralded, the logical class probabilities (and thus the logical error rates for a given syndrome,~\autoref{eq:error-s}) can only take a few discrete values $(0, 1/4, 1/2, 1)$ (cf.~\autoref{eq:coset-heralded}). By recursively tracking the probability across syndromes of each value, we arrive at~\autoref{fig:T-fidelity} for the logical $T$ state fidelity.
\comment{a different basis This fidelity is equal to that of a different $T$ state, fixed point logical fidelity for the \GSC{e} family  We made this choice because for the \GSC{o} family, the fixed which lies in the XY plane of the logical Bloch sphere:
\begin{equation}
\lstate{T}_z \bra{\overline{T}}_z = \frac{1}{2} \left[\overline{I} + \frac{1}{\sqrt{2}}(\overline{X} + \overline{Y})\right].
\end{equation}}

{In the main text, we presented results for the standard logical $T$ state, $\lstate{T}_z \equiv e^{-i\pi \overline{Z}/8} \lstate{+}$. There, we showed the fidelity at even depths using the \GSC{e} family, which places the encoding isometries in opposite order from the rest of the text. We made this choice due to an odd-even effect in the fixed-point logical fidelity. Namely, for integer $\tau$,
\begin{equation}
\lim_{\tau\rightarrow\infty} F_\mathrm{e}(\overline{T}_z, t=2\tau) = \lim_{\tau\rightarrow\infty} F_\mathrm{o}(\overline{T}_z,t=2\tau+1) > \lim_{\tau\rightarrow\infty} F_\mathrm{o}(\overline{T}_z, t=2\tau) = \lim_{\tau\rightarrow\infty}F_\mathrm{e}(\overline{T}_z, t=2\tau+1) .
\end{equation}
That is, a higher asymptotic fidelity is achieved using the \GSC{e} (\GSC{o}) encoding order for even (odd) $t$.}

{Exchanging the order of the encoding isometries is equivalent to changing the basis of logical $T$ state. Define
\begin{equation}
    \lstate{T}_x = \overline{H} \lstate{T}_z = e^{-i \pi \overline{X}/8} \lstate{0},
\end{equation}
which can be used to inject a $T$ gate in the logical $X$ basis. Pushing the Hadamard through the tree, $F_\mathrm{o}(\overline{T}_z)=F_\mathrm{e}(\overline{T}_x)$ and vice versa.}

\subsection{Classical decoding}
In the ensuing analysis, we decode bit and phase flips independently. From the decoder's perspective, this will mean we can act as if there is only one type of error in the system, i.e. only pure bit flips.

If only pure bit flips occur, we can suppress the components $e_Z$ and $e_Y$, reducing the dimension of each leg from 4 to 2. For example, the tensor enumerator associated the CNOT gate is
\begin{equation}
R_{\mathrm{CNOT}} = \begin{tikzpicture}
	\begin{pgfonlayer}{nodelayer}
		\node [style=CNOT] (9) at (0, 0) {};
		\node [style=none] (10) at (-1.25, 1.25) {};
		\node [style=none] (11) at (1.25, 1.25) {};
		\node [style=none] (12) at (-1.25, -1.25) {};
		\node [style=none] (13) at (1.25, -1.25) {};
	\end{pgfonlayer}
	\begin{pgfonlayer}{edgelayer}
		\draw [style=gray] (10.center) to (13.center);
		\draw [style=gray] (12.center) to (11.center);
	\end{pgfonlayer}
\end{tikzpicture} = e_{II}^{II} + e_{XI}^{XX} + e_{IX}^{IX} + e_{XX}^{XI}
\end{equation}
while tensor enumerator for the NOTC gate is the mirror image:
\begin{equation}
\left(R_{\mathrm{NOTC}}\right)^{\beta \alpha}_{ij} = \begin{tikzpicture}
	\begin{pgfonlayer}{nodelayer}
		\node [style=NOTC] (0) at (-2.25, 0) {};
		\node [style=none] (1) at (-3.25, 1) {};
		\node [style=none] (2) at (-1.25, 1) {};
		\node [style=none] (3) at (-3.25, -1) {};
		\node [style=none] (4) at (-1.25, -1) {};
		\node [style=none] (5) at (-3.45, 1.2) {$\scriptstyle{\beta}$};
		\node [style=none] (6) at (-1.05, 1.2) {$\scriptstyle{\alpha}$};
		\node [style=none] (7) at (-1.05, -1.2) {$\scriptstyle{i}$};
		\node [style=none] (8) at (-3.45, -1.2) {$\scriptstyle{j}$};
		\node [style=CNOT] (9) at (2.25, 0) {};
		\node [style=none] (10) at (1.25, 1) {};
		\node [style=none] (11) at (3.25, 1) {};
		\node [style=none] (12) at (1.25, -1) {};
		\node [style=none] (13) at (3.25, -1) {};
		\node [style=none] (14) at (1.05, 1.2) {$\scriptstyle{\alpha}$};
		\node [style=none] (15) at (3.45, 1.2) {$\scriptstyle{\beta}$};
		\node [style=none] (16) at (3.45, -1.2) {$\scriptstyle{j}$};
		\node [style=none] (17) at (1.05, -1.2) {$\scriptstyle{i}$};
		\node [style=none] (18) at (0, 0) {$=$};
	\end{pgfonlayer}
	\begin{pgfonlayer}{edgelayer}
		\draw [style=gray] (1.center) to (4.center);
		\draw [style=gray] (3.center) to (2.center);
		\draw [style=gray] (10.center) to (13.center);
		\draw [style=gray] (12.center) to (11.center);
	\end{pgfonlayer}
\end{tikzpicture}.
\end{equation}

The two-component basis vectors are
\begin{equation}\label{eq:classical-basis}
e_I = \begin{pmatrix} 1 \\ 0 \end{pmatrix}, \quad e_X = \begin{pmatrix} 0 \\ 1 \end{pmatrix}.
\end{equation}

Since $Z$ errors have zero probability, $Z$ stabilizer inputs evaluate to 
\begin{equation}
\begin{tikzpicture}
	\begin{pgfonlayer}{nodelayer}
		\node [style=Z-input] (0) at (0, -1) {};
		\node [style=none] (1) at (0, 1) {};
		\node [style=none] (2) at (1, 0) {$=$};
		\node [style=stabilizer] (3) at (2, -1) {};
		\node [style=none] (4) at (2, 1) {};
		\node [style=none] (5) at (2, -1) {$\scriptstyle{Z}$};
	\end{pgfonlayer}
	\begin{pgfonlayer}{edgelayer}
		\draw [style=gray] (1.center) to (0);
		\draw [style=gray] (4.center) to (5.center);
	\end{pgfonlayer}
\end{tikzpicture} = e_I
\end{equation}
while
\begin{equation}\label{eq:x-input}
\begin{tikzpicture}
	\begin{pgfonlayer}{nodelayer}
		\node [style=X-input] (0) at (0, -1) {};
		\node [style=none] (1) at (0, 1) {};
		\node [style=none] (2) at (1, 0) {$=$};
		\node [style=stabilizer] (3) at (2, -1) {};
		\node [style=none] (4) at (2, 1) {};
		\node [style=none] (5) at (2, -1) {$\scriptstyle{X}$};
	\end{pgfonlayer}
	\begin{pgfonlayer}{edgelayer}
		\draw [style=gray] (1.center) to (0);
		\draw [style=gray] (4.center) to (5.center);
	\end{pgfonlayer}
\end{tikzpicture} = e_I + e_X = \begin{pmatrix} 1 \\ 1 \end{pmatrix}.
\end{equation}
We can alternatively interpret ~\autoref{eq:x-input} as defining an $X$ stabilizer input or, living in a classical world where there are no $X$-type stabilizers, as a gauge degree of freedom.

With $P_0=I, P_1 = X$, the operation $\pauliplus$ (\autoref{eq:pauliplus}) is now addition modulo 2. If bit flip errors occur with probability $r$,~\autoref{eq:open-error} simplifies to
\begin{equation}
\tilde{Q}^{0\beta}_{ij} = (1-r) \delta_{\beta 0} \mathbbm{1}_{ij} + r \delta_{\beta 1} X_{ij}, \qquad \tilde{Q}^{1\beta}_{ij} = (1-r) \delta_{\beta 0} X_{ij} + r \delta_{\beta 1} \mathbbm{1}_{ij}.
\end{equation}

\section{Probability passing for state preparation}\label{app:ppd}
Now we specialize to the multitree state preparation circuit (\autoref{fig:model}b), using the tensor network formalism introduced in the previous section.


\subsection{Bayesian updates}

\begin{figure}

\scalebox{0.8}{\begin{tikzpicture}
	\begin{pgfonlayer}{nodelayer}
		\node [style=cnot] (269) at (-22, 0) {};
		\node [style=cnot] (270) at (-18.25, 0) {};
		\node [style=cnot] (271) at (-14.25, 0) {};
		\node [style=cnot] (272) at (-10.25, 0) {};
		\node [style=error] (273) at (-20.95, -1.6) {};
		\node [style=error] (274) at (-9.45, -1.6) {};
		\node [style=physical] (275) at (-20.25, -2.75) {};
		\node [style=physical] (276) at (-16.75, -2.75) {};
		\node [style=physical] (277) at (-12.25, -2.75) {};
		\node [style=physical] (278) at (-8.75, -2.75) {};
		\node [style=CNOT] (279) at (-16.25, -8.75) {};
		\node [style=physical] (280) at (-13.25, -11) {};
		\node [style=physical] (281) at (-19.25, -11) {};
		\node [style=error] (282) at (-17.75, -10) {};
		\node [style=error] (283) at (-14.75, -10) {};
		\node [style=error] (284) at (-19.05, -7.25) {};
		\node [style=error] (285) at (-13.45, -7.25) {};
		\node [style=NOTC] (286) at (-20.25, -4.5) {};
		\node [style=error] (287) at (-19.85, -6) {};
		\node [style=error] (288) at (-12.65, -6) {};
		\node [style=physical] (289) at (-18.25, -6) {};
		\node [style=physical] (290) at (-11.25, -6) {};
		\node [style=error] (291) at (-17.5, -1.5) {};
		\node [style=NOTC] (292) at (-12.25, -4.5) {};
		\node [style=error] (293) at (-13.25, -1.5) {};
		\node [style=error] (294) at (-21.5, -2.75) {};
		\node [style=error] (295) at (-13.6, -2.75) {};
		\node [style=error] (296) at (-18.75, -2.75) {};
		\node [style=error] (297) at (-10.75, -2.75) {};
		\node [style=CNOT] (298) at (-16.25, 8) {};
		\node [style=Z-input] (299) at (-13.25, 10.25) {};
		\node [style=Z-input] (300) at (-19.25, 10.25) {};
		\node [style=error] (301) at (-17.75, 9.25) {};
		\node [style=error] (302) at (-14.75, 9.25) {};
		\node [style=error] (303) at (-18.8, 6.55) {};
		\node [style=error] (304) at (-13.7, 6.55) {};
		\node [style=NOTC] (305) at (-20.25, 3.5) {};
		\node [style=error] (306) at (-19.85, 5) {};
		\node [style=error] (307) at (-12.65, 5) {};
		\node [style=X-input] (308) at (-18.25, 5) {};
		\node [style=X-input] (309) at (-11.25, 5) {};
		\node [style=NOTC] (310) at (-12.25, 3.5) {};
		\node [style=error] (311) at (-21.25, 1.5) {};
		\node [style=error] (312) at (-13.25, 1.5) {};
		\node [style=error] (313) at (-19.25, 1.5) {};
		\node [style=error] (314) at (-11.25, 1.5) {};
		\node [style=none] (255) at (-20.95, -1.6) {$\scriptstyle{m}$};
		\node [style=none] (256) at (-17.5, -1.5) {$\scriptstyle{m}$};
		\node [style=none] (257) at (-13.25, -1.5) {$\scriptstyle{m}$};
		\node [style=none] (258) at (-9.45, -1.6) {$\scriptstyle{m}$};
		\node [style=none] (251) at (-21.5, -2.75) {$e$};
		\node [style=none] (252) at (-18.75, -2.75) {$e$};
		\node [style=none] (253) at (-13.6, -2.75) {$e$};
		\node [style=none] (254) at (-10.75, -2.75) {$e$};
		\node [style=none] (259) at (-21.25, 1.5) {$e$};
		\node [style=none] (260) at (-19.25, 1.5) {$e$};
		\node [style=none] (261) at (-13.25, 1.5) {$e$};
		\node [style=none] (262) at (-11.25, 1.5) {$e$};
		\node [style=none] (263) at (-19.85, 5) {$c$};
		\node [style=none] (264) at (-12.65, 5) {$c$};
		\node [style=none] (265) at (-18.8, 6.55) {$e$};
		\node [style=none] (266) at (-13.7, 6.55) {$e$};
		\node [style=none] (267) at (-17.75, 9.25) {$i$};
		\node [style=none] (268) at (-14.75, 9.25) {$i$};
		\node [style=none] (245) at (-17.75, -10) {$i$};
		\node [style=none] (246) at (-14.75, -10) {$i$};
		\node [style=none] (247) at (-19.05, -7.25) {$e$};
		\node [style=none] (248) at (-13.45, -7.25) {$e$};
		\node [style=none] (249) at (-12.65, -6) {$c$};
		\node [style=none] (250) at (-19.85, -6) {$c$};
	\end{pgfonlayer}
	\begin{pgfonlayer}{edgelayer}
		\draw [style=gray] (275) to (269);
		\draw [style=gray] (276) to (270);
		\draw [style=gray] (277) to (271);
		\draw [style=gray] (278) to (272);
		\draw [style=black] (279) to (281);
		\draw [style=black] (279) to (280);
		\draw [style=black, bend right] (286) to (279);
		\draw [style=black] (286) to (289);
		\draw [style=black, bend right] (279) to (292);
		\draw [style=black] (292) to (290);
		\draw [style=black, bend right, looseness=0.50] (269) to (286);
		\draw [style=black, bend left, looseness=0.50] (270) to (286);
		\draw [style=black, bend right, looseness=0.50] (271) to (292);
		\draw [style=black, bend left, looseness=0.50] (272) to (292);
		\draw [style=gray] (298) to (300);
		\draw [style=gray] (298) to (299);
		\draw [style=gray, bend left] (305) to (298);
		\draw [style=gray] (305) to (308);
		\draw [style=gray, bend left] (298) to (310);
		\draw [style=gray] (310) to (309);
		\draw [style=gray] (305) to (269);
		\draw [style=gray] (305) to (270);
		\draw [style=gray] (310) to (271);
		\draw [style=gray] (310) to (272);
	\end{pgfonlayer}
\end{tikzpicture}}
\caption{\label{fig:double-tree-syndrome}Multitree used to calculate conditional error probabilities in preparation of $\lstate{0}$. $i$, $e$, $c$, $m$ label error locations associated with stabilizer inputs, encoding gates, check gates, and measurements, respectively. The upright black tree contains the error locations on $S$, while the upside-down gray tree contains the error locations on $A$.}
\end{figure}

Recall that when we couple a depth-$t$ ancilla block $A$ to a system block $S$ via a transversal CNOT and measure all of the ancilla qubits in the $Z$ basis, we are learning the $Z$ syndrome of a code on $S\otimes A$, with the stabilizer group given in~\autoref{eq:block-s}. Consider a noise model with independent bit flip errors at each labeled $Q$-tensor location in~\autoref{fig:double-tree-syndrome}: after each state preparation (input error, $i$, at rate $r_i$\footnote{We can ignore errors on even-layer inputs since the $X$ stabilizers fed into those nodes are unaffected by bit flip errors.}) and two-qubit gate (including the ``encoding gates'' within each block, $e$, and the ``check gates'', $c$, between two blocks, with the same rate $r_g$) and before each measurement ($m$). Since each measurement error is preceded by a check gate error on the ancilla, we combine the errors into one by contracting a physical leg of their respective tensors, i.e., two consecutive error locations with probabilities $\vec{p} = (1-p,p)$ and $\vec{p}' = (1-p', p')$ combine into one location with probability vector $\vec{q}$\footnote{Note that if \textit{either} location is a heralded bit flip, then $\vec{q} = (1/2, 1/2)$.}:
\begin{equation}\label{eq:join}
    \sum_j Q^{\alpha}_{ij}(\vec{p}) Q^{\beta}_{jk}(\vec{p'}) \rightarrow \tilde{Q}^{\alpha\pauliplus\beta}_{ik}(\vec{q}) \qquad \mathrm{where} \quad \vec{q}_i = \sum_j \vec{p}_j \vec{p}'_{j\pauliplus i}.
\end{equation}

In subsequent analysis, a location labeled $m$ will therefore be understood to have bit flip probability $p_g(1-p_m)+p_m(1-p_g)$. Consecutive $e$ and $c$ tensors are \textit{not} absorbed into one, because an encoding gate error in layer $t$ of the system influences the syndrome $s_t$, while a check gate error in layer $t$ on the system is not detected until $s_{t+1}$.

Then, given a reference fault $F = (\vec{i}, \vec{e}, \vec{c}, \vec{m})$, the double tree shown in~\autoref{fig:double-tree-syndrome} calculates the total probability of a fault that produces the same syndrome as $F$. The upside-down tree (gray lines) is associated with the state preparation on $A$, while the upright tree (black lines) is associated with the state preparation on $S$. The two trees are joined by a layer of tensors (``check nodes'')
\begin{equation}
\scalebox{0.8}{\begin{tikzpicture}
	\begin{pgfonlayer}{nodelayer}
		\node [style=cnot] (0) at (0, 0) {};
		\node [style=none] (1) at (0, -2) {};
		\node [style=none] (2) at (-2, 2) {};
		\node [style=none] (3) at (2, 2) {};
		\node [style=CNOT] (4) at (5, 0) {};
		\node [style=X-input] (5) at (3, -2) {};
		\node [style=none] (6) at (3, 2) {};
		\node [style=none] (7) at (7, 2) {};
		\node [style=none] (8) at (7, -2) {};
		\node [style=none] (9) at (2.5, 0) {$=$};
	\end{pgfonlayer}
	\begin{pgfonlayer}{edgelayer}
		\draw [style=gray] (2.center) to (0);
		\draw [style=gray] (0) to (3.center);
		\draw [style=gray] (0) to (1.center);
		\draw [style=gray] (6.center) to (8.center);
		\draw [style=gray] (5) to (7.center);
	\end{pgfonlayer}
\end{tikzpicture}} = e_I^{II} + e_I^{XX} + e_X^{XI} + e_X^{IX}.
\end{equation}
Now suppose we want to find the full probability distribution of faults on each error location in the system subtree, conditioned on the observed syndrome, is obtained by leaving an open leg on each $Q^{\alpha}$-tensor in the system tree, i.e. placing the tensor $\tilde{F}$ there (\autoref{eq:open-error}). Then, the $(\beta_1,\beta_2,...,\beta_n)$ entry in the resulting $2^n$-element tensor is the conditional probability of the fault $P_{\beta_1} P_{\beta_2} \cdots P_{\beta_n}$ on the $n$ possible error locations. 

\subsection{Optimal vs. approximate decoders}
Where various decoders differ is in the computation of $I(F_{\vec{s}}), X(F_{\vec{s}})$. In the optimal decoder, $I(F_{\vec{s}})$ and $X(F_{\vec{s}})$ can be expressed in terms of conditional probabilities:
\begin{equation}\label{eq:optimal}
L(F_{\vec{s}}) = \mathrm{prob}(\vec{\tilde{s}}) \sum_{F: \eff(F) \in \finalS P_{s_T} \overline{L}} \mathbb{P}(F|\vec{\tilde{s}})
\end{equation}

Explicitly, in each round of syndrome measurements coupling a pair of states $S$ and $A$, we update the probability distribution on the $S$ subtree according to~\autoref{fig:double-tree-syndrome}. In round 1, the prior probabilities $\mathbb{P}(F)$ are uncorrelated, but as we update $\mathbb{P}(F)\rightarrow \mathbb{P}(F|s_1) \rightarrow \mathbb{P}(F|\{s_1,s_2\})...$, in subsequent rounds, the distribution becomes increasingly correlated.\footnote{So, contrary to the depiction in~\autoref{fig:double-tree-syndrome}, the conditional probabilities entering the second round of updates would not be independent.} This belief propagation/probability passing is implemented by the light blue lines between ``stacks'' of multitrees in~\autoref{fig:optimal}. Then, following the final round of stabilizer measurements, the logical leg on the root is left open to evaluate the weights of the two logical classes (with respect to the final stabilizer group $\finalS$) according to this updated error distribution. In particular, $\mathbbm{P}(F|\tilde{\vec{s}}) = 0$ if fault $F$ is inconsistent with the observed syndrome in any layer.

\begin{figure}[t]
\centering
\scalebox{0.45}{\input{fault-tolerant-network-color2.tikz}}
\caption{Tensor network for the optimal decoder, depth 4. Light blue lines propagate information from one level to the next, preserving all the correlations between error probabilities imposed by the syndrome $\tilde{\vec{s}}$. In the approximate decoder, the red shaded parts of the tensor network are omitted, and the error probabilities inserted on each first-order ancilla block (upside-down unshaded trees) are the updated marginals (cf.~\autoref{fig:network-b}). \label{fig:optimal}}
\end{figure}

The normalization of~\autoref{eq:optimal} is such that
\begin{align}
\mathrm{prob}(\vec{s}) = I(F_{\vec{s}}) + X(F_{\vec{s}}), \qquad \sum_{\vec{s}} I(F_{\vec{s}}) + X(F_{\vec{s}}) = 1.
\end{align}

From the spacetime code perspective, the optimal decoder only sums over faults consistent with the complete spacetime syndrome. Rewriting $\sum \mathbb{P}(F,\tilde{\vec{s}}) = \sum \mathbb{P}(F) \delta[\tilde{\sigma}(F) = \tilde{\vec{s}}]$ yields:
\begin{equation}
L(F_{\vec{s}}) = \sum_{F: \, \eff(F) \in \finalS P_{s_T} \overline{L}} \mathbb{P}(F) \delta[\tilde{\sigma}(F) = \tilde{\vec{s}}],
\end{equation}
which, comparing to~\autoref{eq:recovery-ft}, implies that the optimal recovery probability averaged over all syndromes is
\begin{equation}
P_R^{opt} = \sum_{\vec{s}} \max(I(F_{\vec{s}}), X(F_{\vec{s}})) = \sum_{\vec{s}} \mathrm{prob}(\vec{s}) \frac{\max(I(F_{\vec{s}}), X(F_{\vec{s}}))}{I(F_{\vec{s}}) + X(F_{\vec{s}})} = \sum_F \mathbb{P}(F) \frac{\max(I(F), X(F))}{I(F) + X(F)}.
\end{equation}

If we had a tractable way to evaluate~\autoref{fig:optimal}, we could get an unbiased estimator for the recovery probability, without explicit computation of the ``reference fault'' $F_{\vec{s}}$: sample $F$, calculate $I(F)$ and $X(F)$, and return $\max(I(F), X(F))/(I(F) + X(F))$ for each sample. Unfortunately, the tensor network is not efficiently contractible because of the entanglement in the stacked multitree structure. Each pair of trees joined by check nodes can be folded over, and nodes joined by blue lines stacked on top of each other, to bring the tensor network back into a tree form, but with a bond dimension of $2^{2^{T-t+1}-1}$ at depth $t$ from the root, which is exponential in the system size.

To approximate this probability passing step in a more efficient way, we only keep the correlations to ``first order.'' 
Concretely, let $T$ denote the depth of the final state. Label the ancilla blocks by their maximal depth before measurement, i.e. there are $2^{T-t-1}$ blocks of depth $t$. The system directly interacts with one depth $1$ block $A^{(1)}_1$, one depth $2$ block $A^{(1)}_2$,..., and one depth $t-1$ block $A^{(1)}_{t-1}$. It also has a ``second-order'' interaction with $t-2$ depth $1$ blocks $A^{(2,1)}_1$,..., $A^{(2,t-2)}_1$; $t-3$ depth $2$ blocks $A^{(2,1)}_2,...,A^{(2,t-3)}_2$; and so on. 

\begin{figure}[hbtp]
\centering
\subfloat[]{
\includegraphics[height=0.35\textheight]{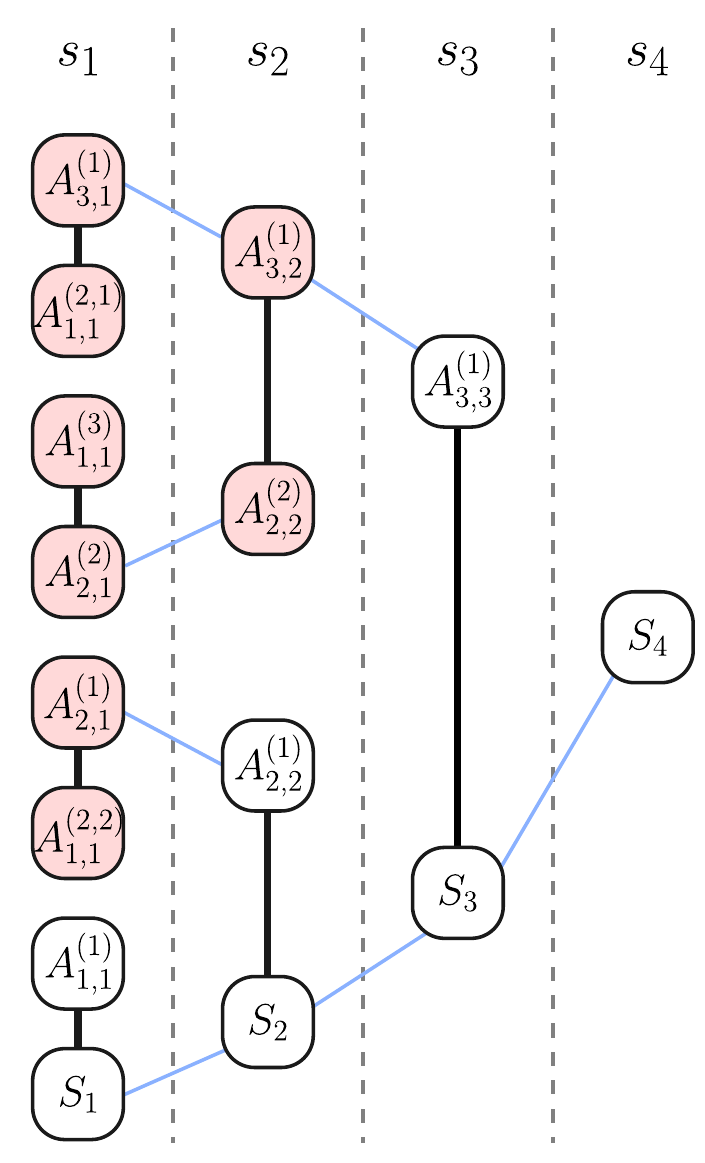}\label{fig:network-a}}
\hspace{15pt}
\subfloat[]{
\includegraphics[height=0.35\textheight]{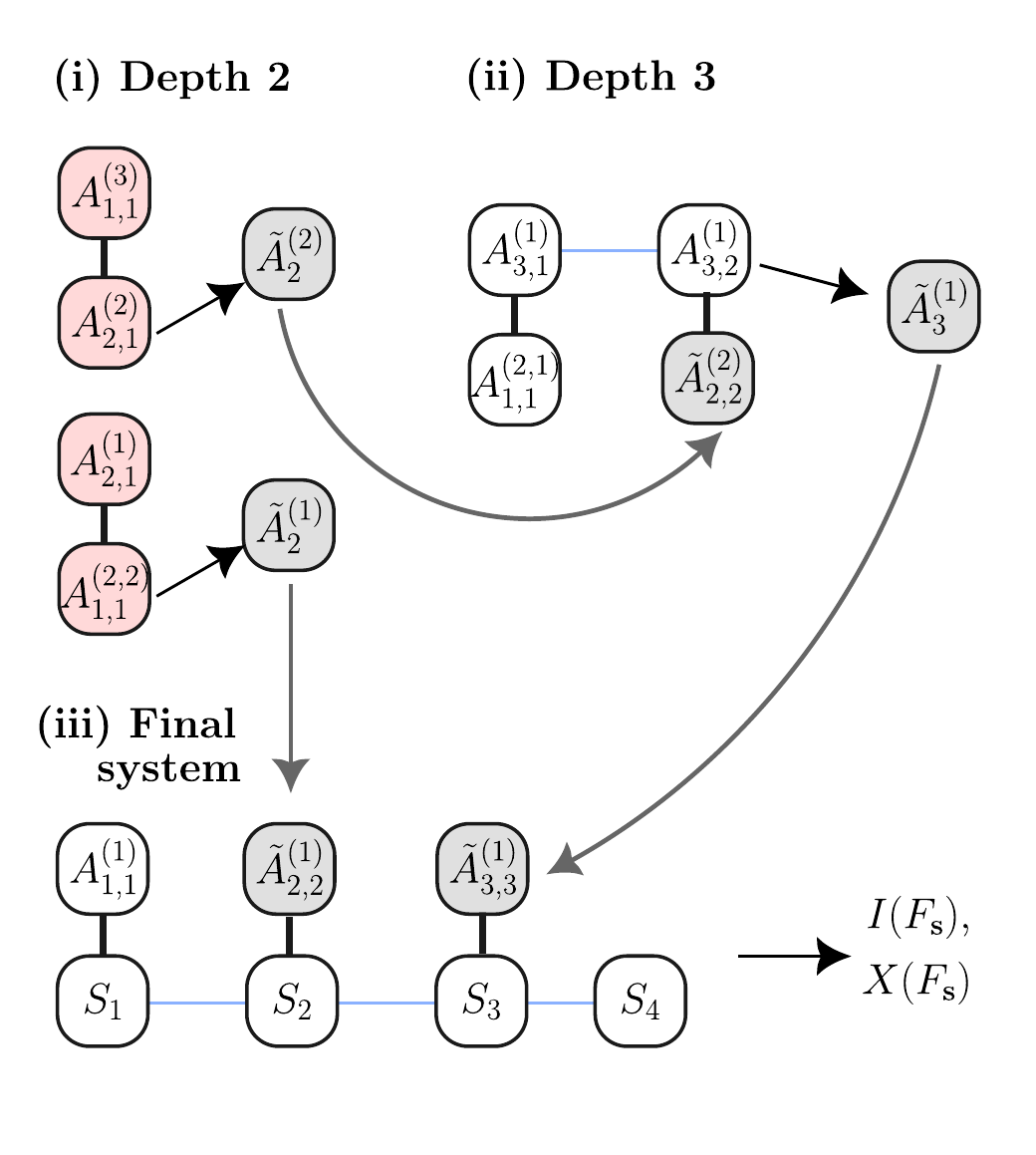}\label{fig:network-b}}
\caption{(a) Schematic depiction of the optimal decoder for depth 4 state preparation. Thick black vertical lines indicate pairs of blocks involved in the same check, and light  blue lines indicate probability passing within a block. Light red boxes correspond to the shaded regions of~\autoref{fig:optimal}, and are left out of the final tensor network contraction in the first-order stacked probability passing decoder. The first-order decoder is shown step by step in (b). Gray boxes labeled $\tilde{A}_t$ are the outputs of marginal updates on block $A$ up to $t-1$. These are then fed into the next steps (gray arrows), with the final step yielding an estimate of the logical class probabilities. To avoid clutter, if there is only one block of depth $t$ at order $n$, we omit the label $m=1$ in $A^{(n,m)}_t$.}
\end{figure}

Let $A^{(n,m)}_{t,t'}$ denote the subtree in~\autoref{fig:optimal} containing the error locations on the block $A^{(n,m)}_t$ that can influence the syndrome $s_{t'}$. For example, if $t'<t$ (so this block is acting as the ``system'' in this measurement round), then $s_t'$ is sensitive to input errors entering the first $t'$ layers of encoding gates on this block; the gate errors after the first $t'$ layers of encoding gates; and the gate errors after the first $t'-1$ layers of ``check gates''. If $t'=t$, then $s_t'$ is also sensitive to the gate errors after layer $t'$ of check gates, and the measurement errors which immediately follow them (which can be absorbed into the preceding gate errors with $p\rightarrow p_g + p_m - 2 p_g p_m$). We similarly let $S_t$ denote the subtree of the ultimate system containing the error locations that can influence syndrome $s_t$.~\autoref{fig:optimal} can then be condensed into the schematic form shown in~\autoref{fig:network-a}.

Our stacked probability passing decoder keeps only the direct first-order connections as follows:
\begin{enumerate}
\item \label{step0} Initialize the marginal probability at each error location on each block according to the specified error model and, if applicable, the heralding information.
    \item \label{step1}For $t=2,...,T-1$, we calculate the marginal probabilities on each $t$ block, using the part of the syndromes $s_1,...,s_{t-1}$ which directly involve that block and its first-order connections. These updates are sketched in~\autoref{fig:network-b}(i)-(ii).
    \item \label{step2}In the final step, sketched in~\autoref{fig:network-b}(iii) for $T=4$, we estimate $I(F_\vec{s}), X(F_\vec{s})$ by contracting a stacked tree tensor network, which only involves the first-order ancillas (now with updated marginal probabilities), and leaving the logical leg open. At distance $t$ from the root, the stack has a depth of $2(T-t)+1$: one layer of the stack for each $S_{t'}$ and $\tilde{A}^{(1)}_{t',t'}$ with $t'\geq T$. Thus, the maximal bond dimension is $2^{2T+1}$, which is only polynomial in the system size $N=2^T$.
\end{enumerate}
Let us briefly comment upon the ``unstacked'' version of the probability passing decoder. In that version, no correlations are kept at all, as the Bayesian updates are performed using only the most recent round of checks. Namely, in step $t$, we just update the marginals on each block of depth $>t$, only using the part of $s_t$ involving that block and its corresponding depth-$t$ ancilla. The final decoding step uses an updated probability distribution on the system ($\mathbbm{P}(F) \rightarrow \mathbbm{P}'(F)$) but, unlike the optimal decoder, general assigns nonzero probability to errors inconsistent with $\tilde{\vec{s}}$.

The unstacked decoder achieves roughly the same threshold as the stacked decoder, but inferior subthreshold performance. To explain why stacking -- i.e., simultaneously using information from multiple rounds of checks involving a block -- boosts the subthreshold performance, note that it appears in~\autoref{fig:network-b} in two different ways: In ~\autoref{step1} we are using a stacked tree to calculate marginal probabilities (leaving an error leg open) while ~\autoref{step2} uses the stacked tree to calculate logical class probabilities. Empirically, we find that the utility of stacking in the first step is negligible. This is because, while stacking (e.g. the blue line in~\autoref{fig:network-b}(ii)) retains some correlations in the conditional probability distribution of errors on that block, these correlations are less important if we are only trying to update the marginal probabilities anyway. In contrast, the stacking in the final step is essential to the subthreshold performance, as it ensures that $I(F_\vec{s}), X(F_\vec{s})$ only sum over errors consistent with the spacetime syndrome directly involving the final system block. 

\subsection{Theoretical results}

\begin{figure}[hbtp]
\centering
\subfloat[]{\includegraphics[width=0.65\linewidth]{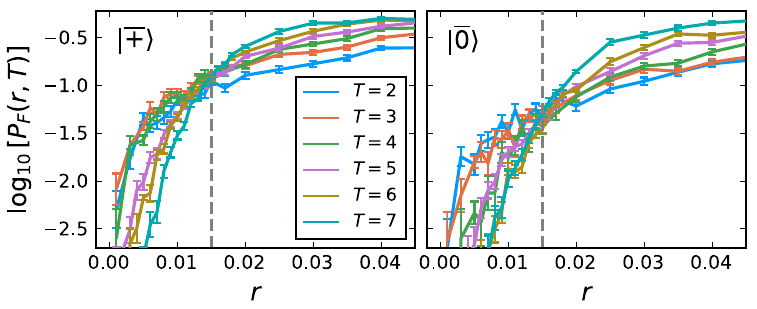}\label{fig:ppd-sep}}\hfill
\subfloat[]{\includegraphics[width=0.34\linewidth]{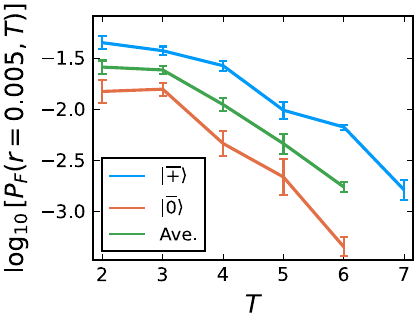}\label{fig:ppd-subthreshold}}
\caption{Logarithm of the failure probability of state preparation in a noise model with independent bit and phase flips. (a) Average, $\lstate{+}$, and $\lstate{0}$ as a function of $p$ for different $T$. The dashed gray line in each panel is the estimated threshold $r_c \approx 0.015$. (b) Average, $\lstate{+}$, and $\lstate{0}$ as a function of $T$, well below threshold at $r=0.005$. Between 1000 and $50\, 000$ shots were gathered at each $T$, {and error bars are determined from error propagation of the one sigma confidence interval on $N_{shots}$ Bernoulli trials.}}
\end{figure}

{\autoref{fig:unheralded} of the main text shows the logarithm of the logical failure probability under unheralded independent bit and phase flips, averaged over $\lstate{+}$ and $\lstate{0}$, exhibiting a threshold of $r_c \approx 0.015$. The failure probability is plotted separately  for the two logical states in~\autoref{fig:ppd-sep}. Since the approximate decoder no longer determines $I(F_\vec{s}), X(F_\vec{s}))$ exactly, we cannot directly compute the logical failure probability for a given sampled syndrome. Instead, for each sampled $F$, we must first find $F_\vec{s}$, run the decoder, and record success or failure based on whether $L^* F_s$ belongs to the same coset at $F$. Each sample is therefore a Bernoulli trial, with the mean failure probability given by~\autoref{eq:pfail-bern} and the one sigma confidence interval $\sqrt{P_F(1-P_F)/N_{shots}}$.}

To examine the subthreshold performance more closely, we fix $r=0.005 < r_c$ and gather up to $50\, 000$ shots for the largest $T$. The trends in~\autoref{fig:ppd-subthreshold} suggest that the logical failure rate decreases exponentially with $T$, unlike for the optimal decoder with erasure errors, where the decay is doubly exponential (\autoref{fig:subthreshold-state-prep}).

\begin{figure}[hbtp]
\includegraphics[width=\linewidth]{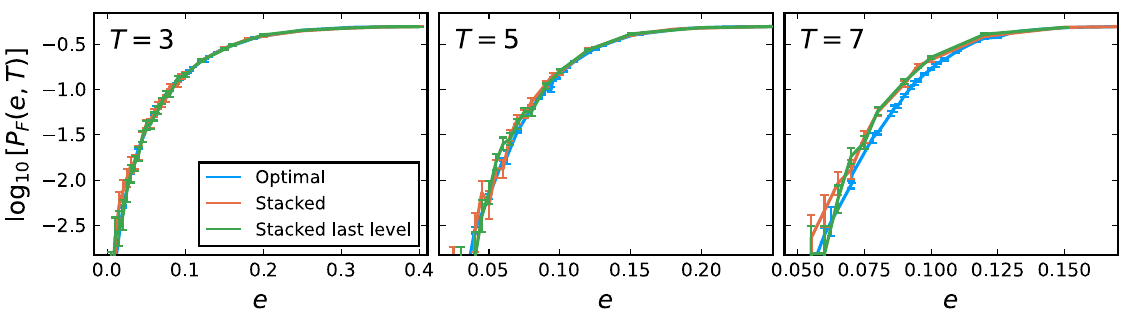}
\caption{Comparison of decoders for fully heralded errors in the state preparation circuit of $\lstate{+}$. Blue curve is the optimal spacetime decoder, orange is the first-order stacked decoder, and green is an approximate decoder which uses stacking only in ~\autoref{step2}. Error bars are the standard error of the mean, across at least $650$ samples.~\label{fig:optimal-vs-ppd}}
\end{figure}

The decoder could be iteratively improved by keeping higher-order connections. We note, however, that when the errors are fully heralded -- enabling an efficient optimal decoder -- the optimal decoder only becomes noticeably better than the first-order stacking decoder at small $e$ and $T=7$. A comparison is shown in~\autoref{fig:optimal-vs-ppd}. 

It is worth noting that when errors are fully heralded, even when we use the approximate decoder, the inferred  logical class probabilities $I_{approx}(F_\vec{s}), X_{approx}(F_\vec{s})$ still enjoy the property that either $I_{approx}(F_\vec{s}) = X_{approx}(F_\vec{s})$, or only one is nonzero. {In the latter case, the approximate decoder is guaranteed to succeed.} However, there can be situations where the approximate decoder finds $I_{approx}(F_\vec{s}) = X_{approx}(F_\vec{s})$ (and thus guesses correctly with only probability 1/2), but the optimal decoder would have found only one class, say $I(F_\vec{s})$, to have nonzero probability ($r_M = S(\rho)=0$, cf.~\autoref{eq:rM}). This can occur if all the faults in the class $\finalS P_{s_T} \overline{X}$ are inconsistent with the erasure pattern or syndrome information in earlier layers, but are consistent with the partial information kept in the stacked probability passing algorithm. 
\section{Results on quantum hardware}\label{app:experiment}
In this section, we provide additional detail on the state preparation demonstrations performed {via cloud computing} on Quantinuum's 32-qubit processor, which led to the results in~\autoref{fig:qtuum}. 
\begin{figure}[htp]
\subfloat[]{
\includegraphics[width=0.65\linewidth]{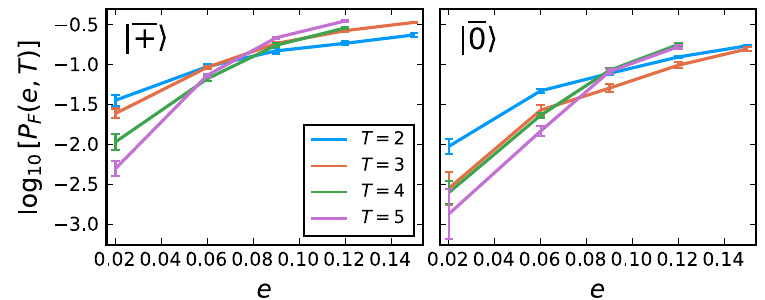}\label{fig:qtuum-sep}}
\subfloat[]{
\includegraphics[width=0.35\linewidth]{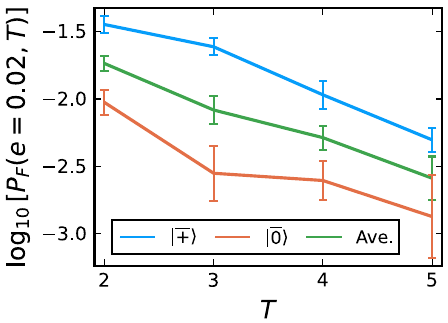}\label{fig:qtuum-sub}}
\caption{Results from the heralded bit/phase flips state preparation circuit on Quantinuum System H2. (a) Log of the logical failure probability for $\lstate{+}$ and $\lstate{0}$ as a function of $e_g=e_m=e$, for circuit depths ranging from $T=2$ to $T=5$ (cf.~\autoref{fig:qtuum} in the main text). (b) Log of the logical failure probability for $\lstate{+}$ (blue), $\lstate{0}$ (orange), and their average (green) well below threshold, as a function of circuit depth at added error rate $e = 0.02$. {Error bars are defined from \autoref{eq:SE-resample}, propagated to the logarithm.}\label{fig:experiment}}
\end{figure}
\subsection{Heralded bit/phase flips}
The first experiment features heralded bit flips (for $\lstate{0}$) and phase flips (for $\lstate{+}$) on each qubit following a two-qubit gate, as well as heralded measurement errors. The gate errors which do not immediately precede a measurement are added during the run of the quantum circuit, by prepending a quantum circuit that generates two random bits for each possible error location. The first random bit is $1$ with probability $e$, and sets a classical flag register to $\ket{1}$ --- information passed on to the decoder. The second bit is $1$ or $0$ with equal probability, and serves as a control to an $X$ gate at the heralded location. The decoder is not privy to the second bit.

We ran between $400$ and $1000$ shots at each error rate and circuit depths, using more shots at smaller $p$, where each shot has an independently sampled pattern of gate errors. The measurement errors and gate errors immediately preceding them are added in classical post-processing: Given the classical bit strings of measurement outcomes at depths $t=1,2,...,T-1$, each bit suffers a heralded error with probability $2e-e^2$ (since the probability that the qubit survives two possible erasures is $(1-e)^2$), and the syndrome $s_t$ is computed from the corrupted bit string. The final layer of measurements, at depth $T$, is a destructive measurement of the system. A final layer of encoding gate errors is also added in post-processing (heralded bit flips with probability $e$). This destructive measurement tells us both the final syndrome $s_T$ (given to the decoder) and the final logical syndrome (not given to the decoder). For each quantum circuit shot, we generate $M=500$-1000 samples of error patterns in post-processing. Although these samples are not independent (as they share the same pattern of gate errors on the hardware), the resampling step is a way to get better statistics out of a fairly small number of hardware runs. {With $y_{ij} = 0, 1$ denoting success or failure, respectively, in the $j$th resample of shot $i$, the average failure probability for state preparation of $\lstate{\psi}$ is
\begin{equation}
    P_F(\lstate{\psi}) = \frac{1}{M N_{shots}}\sum_{i=1}^{N_{shots}}\sum_{j=1}^{M} y_{ij} \equiv \frac{1}{N_{shots}} \sum_{i=1}^{N_{shots}} \overline{y}_i.
\end{equation}
The reported uncertainty is the standard error across the $\overline{y}_i$, i.e.
\begin{equation}\label{eq:SE-resample}
\mathrm{SE}(\overline{y}_i) = \sqrt{\frac{(\overline{y}_i - \overline{y})^2}{N_{shots} (N_{shots}-1)}}
\end{equation}}

For each sample, in~\autoref{step0} of the stacked probability decoder, the probability vector on the unheralded locations is initialized to the native noise rate:
\begin{equation}\label{eq:gate-mem}
    \vec{p} = \begin{pmatrix} 1 - r_g - r_m + 2 r_g r_m, r_g + r_m - 2 r_g r_m
    \end{pmatrix}
\end{equation}
on sites immediately preceding measurements (cf.~\autoref{eq:join}), and
\begin{equation}\label{eq:gate}
    \vec{p} = \begin{pmatrix} 1-r_g, & r_g
    \end{pmatrix}
\end{equation}
after gates in the bulk of the circuit, where $r_g = 0.0005, r_m=0.0025$.
Meanwhile, on heralded locations, $\vec{p} = \begin{pmatrix} 1/2, & 1/2 \end{pmatrix}$. 

The stacked probability passing decoder succeeds if its guess of the logical class matches the measured logical syndrome. As in the classical simulations, $\lstate{0}$ exhibits a significantly lower failure rate than $\lstate{+}$. The logical failure probabilities for the two states are shown separately in~\autoref{fig:qtuum-sep}. At $e=0.02$, the logical failure probability decreases with circuit depth (\autoref{fig:qtuum-sub}). 

\subsection{Heralded coherent errors}
\begin{figure}[hbtp]
\includegraphics[width=0.65\linewidth]{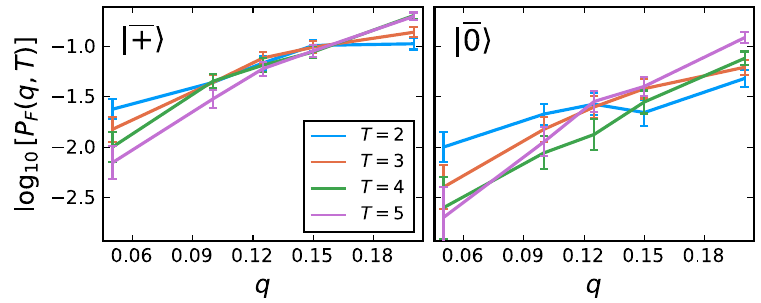}
\caption{Results of the state preparation {demonstration on Quantinuum System H2} with heralded coherent errors introduced at rate $p$. The two panels show the log of the logical failure probability for $\lstate{+}$ (left) and $\lstate{0}$ (right) as a function of added error rate $e$, for circuit depths ranging from $T=2$ to $T=5$ (cf.~\autoref{fig:qtuum} in the main text).\label{fig:coherent-sep} {Error bars denote the one sigma confidence interval over 500-1000 Bernoulli trials.}}
\end{figure}

In the second experiment, the added noise is heralded coherent errors. Unlike with the incoherent noise channel, we now need to generate only one random bit per error location: if a site is flagged, then a $\pi/4$ rotation is deterministically applied. These rotations can only be applied on the quantum level, so we only add gate errors and do not introduce further errors in classical post-processing. The initialization on unheralded sites is the same as in the first experiment [\autoref{eq:gate-mem} and \autoref{eq:gate}], while on heralded sites, using the Pauli twirl approximation [\autoref{eq:twirl}],
\begin{equation}
\vec{p} = \begin{pmatrix} \cos^2(\pi/8), & \sin^2(\pi/8) \end{pmatrix}.
\end{equation}

Between 500 and 1000 shots are taken at each error rate. Logical failure probabilities for $\lstate{+}$ and $\lstate{0}$ are plotted separately in~\autoref{fig:coherent-sep}. While $\lstate{+}$ exhibits a clean crossing and subthreshold phase, the data for $\lstate{0}$ are less clear.

\section{Quantum memory with erasure errors}\label{app:memory}
In~\autoref{sect:memory} of the main text, we presented results on the ``memory threshold'' under Steane syndrome measurements subject to heralded errors. In this section, we discuss several variations on the syndrome extraction circuit and present scaling collapses for both $n(T)=d(T)$ and $n(T)=1$ rounds of checks.

\subsection{Review of Steane Error Correction}
Given a logical state encoded in a CSS code, its syndrome can be measured fault-tolerantly via the logical circuit shown in~\autoref{fig:steane-circuit} in the main text. This circuit, followed by applying the deduced correction operator, is the Steane error correction (EC) gadget, introduced in Ref.~\cite{Steane1997}. For thorough introduction to the protocol, the reader is directed to Ref.~\cite{Gottesman2024}.

A single round of Steane EC consumes two ancilla states, encoded in the same code as the system. An ancilla in the state $\lstate{+}$ is the target of a transversal CNOT and is then measured in the $Z$ basis. $X$ propagates from control to target under CNOT, so if the ancilla, transversal gate, and measurement were noiseless, the measurement would tell us the $Z$ syndrome on the encoded system. In practice, all of these steps are noisy, so we actually learn the joint $Z$ syndrome of the system and ancilla. This is the same logic as that behind the check layer of the state preparation circuit (cf.~\autoref{eq:block-s}), with a crucial difference: because the ancilla is encoded in $\lstate{+}$, the measurement reveals nothing about the $Z$ logical on the system. Likewise, an ancilla in the state $\lstate{0}$ is control of a transversal CNOT and is then measured in the $X$ basis. $Z$ propagates from target to control, so the measurement tells us the joint $X$ syndrome of the syndrome and ancilla, again without revealing or collapsing the logical information on the system.

In the transversal CNOT, errors of the opposite type ($Z$ for $\lstate{+}$, $X$ for $\lstate{0}$) back-propagate from the ancilla to the system. However, since the gates are transversal, the spread of errors is controlled, with one error on the ancilla only propagating to a single qubit of the system. Consequently, a single round of syndrome measurement suffices for fault tolerance~\cite{Gottesman2024}. As usual, to assess the performance of an isolated gadget, a round of perfect stabilizer measurements is appended to the circuit. 

In the main text, we nevertheless performed $n(T) = d(T)$ rounds of syndrome measurement [\autoref{fig:memory}c(i)] as well as just one [\autoref{fig:memory}c(ii)]. We did this to apply a more stringent test of quantum memory, demonstrating that logical information can survive to a circuit depth that diverges with the system size. The particular scaling with code distance was chosen in analogy to the surface code, where $d(L)$ layers of syndrome extraction are necessary. Rather than decoding each round separately, we gather the syndrome information from all $n$ rounds and the final perfect measurement. 

Since the initial state of the system is noiseless, whereas each round of syndrome measurement is noisy, the mutual information with the reference after the final layer of perfect stabilizer measurements tends to decrease with the number of noisy rounds. 
Consequently, the threshold of a \textit{single} round of Steane EC, with the same model of heralded errors, is higher than that of $n(T) = d(T)$ rounds. 

\subsection{Optimizations}
In designing the circuit for one round of Steane EC, there are several variations which can have a significant effect on the logical failure probability.

First, in each round of checks, we have the choice of performing checks against the $\lstate{+}$ ancilla first, or the $\lstate{0}$ ancilla first. The order of these checks has minimal effect on the final recovery probability, so in the remaining discussion we perform the $\lstate{+}$ checks first.

We also have a choice of which types of checks are performed in the multitree circuit that prepares each ancilla. Recall that in the state preparation gadget in isolation, the optimal strategy is to always measure in the basis that matches the root logical---$X$ basis for $\lstate{+}$, $Z$ basis for $\lstate{0}$---since the problem of decoding a fixed logical state requires no information about errors of the opposite type. In contrast, when these ancillas are used in a Steane EC gadget to infer the syndrome of an arbitrary logical state,  it is essential to also periodically perform ``opposite-type'' checks. A straightforward way to do this is to alternate between ``opposite-type'' and ``matching-type'' checks in the state preparation circuit. Thus, for each of $\lstate{0}$ and $\lstate{+}$, we can choose between performing $Z$ checks in odd layers or in even layers, giving rise to four choices altogether. We empirically find that for most system sizes, the lowest failure probability is achieved when matching-type checks are performed in the levels of opposite parity to $T$. Then the last layer of checks (at depth $T-1$) is of the matching type. A minimal example ($T=3$) is shown in~\autoref{fig:ancilla-circuit} of the main text.

\begin{figure}[hbtp]
\subfloat[]{
\includegraphics[height=0.18\textheight]{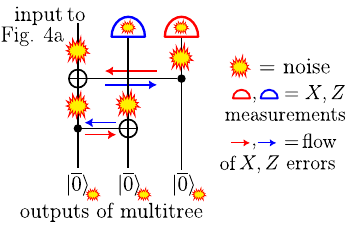}\label{fig:verify}}\hspace{40pt}
\subfloat[]{
\includegraphics[width=0.5\linewidth]{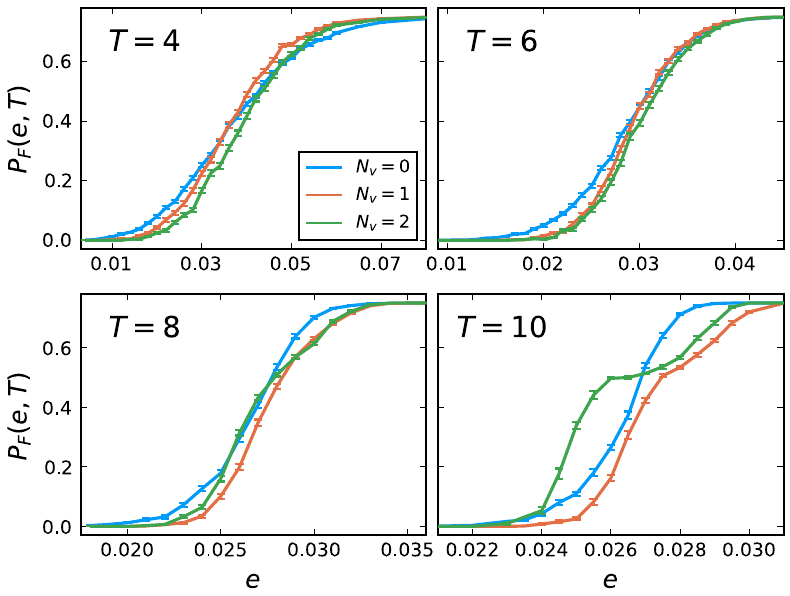}\label{fig:comp}}
\caption{(a) \gs{Optional} ancilla verification circuit for $\lstate{0}$. This circuit uses $N_v=2$ ancillas to check both the $X$ and $Z$ syndrome of the leftmost state, including its root stabilizer. The verified state is then fed into the gadget in~\autoref{fig:steane-circuit}. (b) Logical failure probability using an optimal decoder after $n(T) = 2^{\lfloor T/2 \rfloor}$ rounds of Steane syndrome measurement, with $N_v$ verification ancillas for each $\lstate{0}, \lstate{+}$ used to measure the system. Error bars denote the standard error of the mean across 150-2700 shots.}
\end{figure}

Since the encoded ancillas are now being used as part of a larger gadget, the state preparation circuit no longer has the final round of perfect stabilizer measurements that we used to assess performance of the state preparation protocol. Thus, at the cost of increased overhead, it may be useful to perform extra layers of ancilla verification on the prepared state, before using it to perform syndrome checks. The verification procedure, which itself is subject to noise, is shown in~\autoref{fig:verify} for $\lstate{0}$.~\autoref{fig:comp} compares, for $n(T)=d(T)$, the logical failure probability without verification ($N_{v}=0$), with one extra ancilla for verification (i.e., just checking the $X$ or $Z$ syndrome of the prepared state), and with two extra ancillas (as shown in~\autoref{fig:verify}). 
The optimal choice at one system size can be suboptimal at other system sizes. For example, performing 2-ancilla verification boosts the performance at low $e$ and small $T$, but evidence of an intermediate regime in which only a classical bit is protected emerges for $T\geq 7$. The data presented in the main text were obtained with no ancilla verification.

\subsection{Probability distribution of mutual information}
Recall that the logical failure probability in the quantum memory test is determined from the mutual information between the system and a reference qubit [\autoref{eq:mutual}]. $P_F(e)$ is the probability that the system suffers a logical $X$ \textit{or} $Z$ error (or both). To gain further insight into the mechanisms behind failure, we can examine the probability distribution of the mutual information, i.e.
\begin{equation}\label{eq:pi-mutual}
    \Pi(I(S:R)=k) = [\mathrm{fraction \, of \, runs \, with \, } I(S:R)=k].
\end{equation}

\begin{figure}[hbtp]
\includegraphics[width=\linewidth]{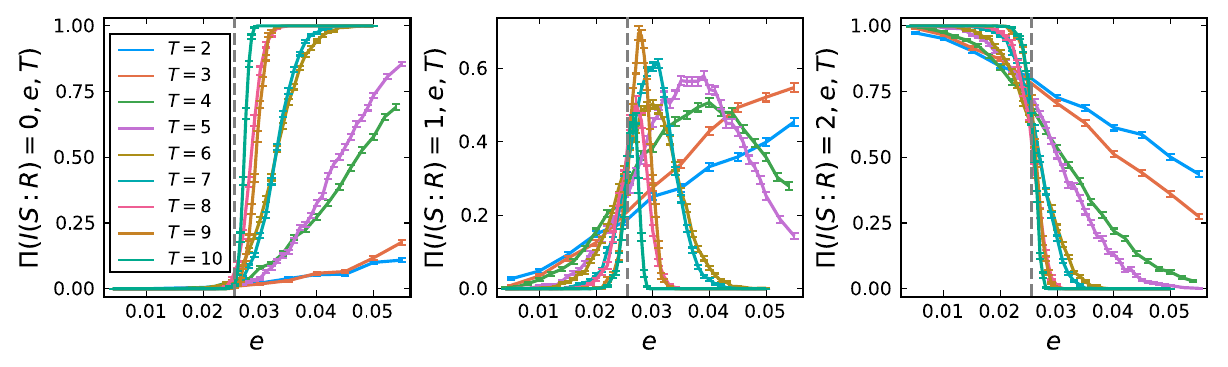}
\caption{Probability of mutual information 0, 1, and 2 respectively between the system and the reference after $2^{\lfloor T/2\rfloor}$ rounds of Steane syndrome measurement + one round of perfect stabilizer measurement, with no ancilla verification. The gray dashed line is at the estimated threshold $e_c = 0.0255$.\label{fig:n0-all} {Error bars are the one sigma confidence interval on $N_{shots}$ Bernoulli trials, with $N_{shots}$ ranging from 500 to 1900.}}
\end{figure}

The quantities $\Pi(I(S:R)=0), \Pi(I(S:R)=1), \Pi(I(S:R)=2)$ are shown for $N_v=0$ in~\autoref{fig:n0-all}. Survival of a classical bit (i.e., a logical $X$ error but no logical $Z$ error, or vice versa) would be signaled by a sustained interval with $I(S:R) = 1$. For $N_v=0$, no such phase is present, and the peak in $\Pi(I(S:R)=1)$ sharpens and shifts to the left as $T$ increases. 

\begin{figure}
\subfloat[]{
\includegraphics[width=\linewidth]{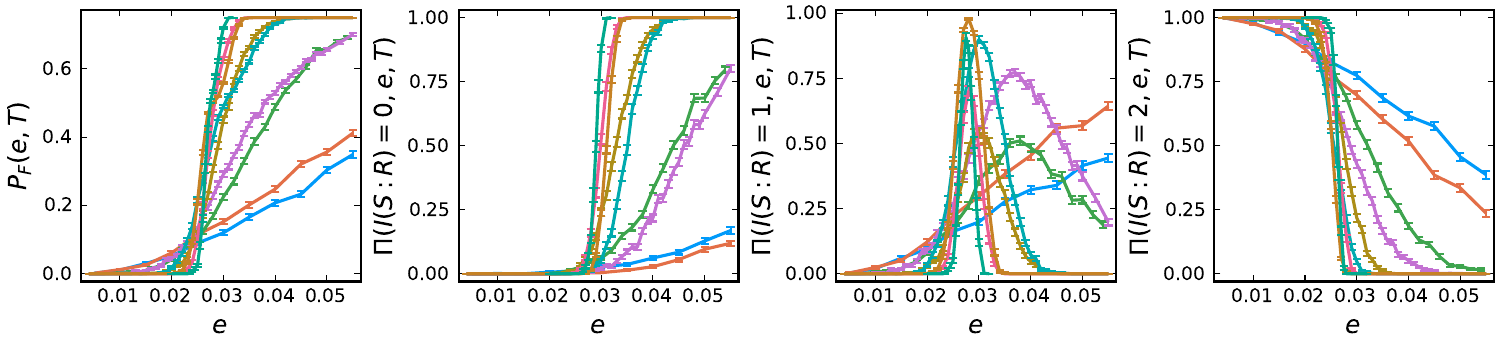}}\\
\subfloat[]{
\includegraphics[width=\linewidth]{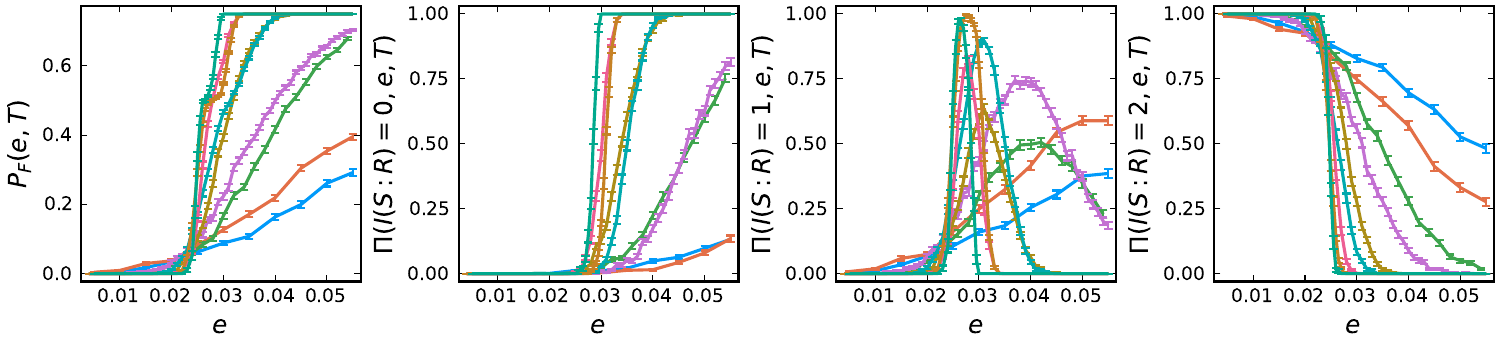}}
\caption{$P_F(e)$ [\autoref{eq:mutual}] and $\Pi(I(S:R)=k)$, $k=0,1,2$ after $2^{\lfloor T/2\rfloor}$ rounds of Steane syndrome measurement + one round of perfect stabilizer measurement, with (a) $N_v=1$ and (b) $N_v=2$. $T=2$ to $T=10$ are shown, with the same colors as~\autoref{fig:n0-all}.\label{fig:n1-n2}. {Uncertainties are over 50-1100 shots.}}
\end{figure}

For comparison,~\autoref{fig:n1-n2} shows $\Pi(I(S:R)=k)$ and $P_F(p)$ for $N_v=1$ and $N_v=2$.

\subsection{Scaling collapses}

\begin{figure}[hbtp]
\subfloat[]{
\centering
\includegraphics[width=0.8\linewidth]{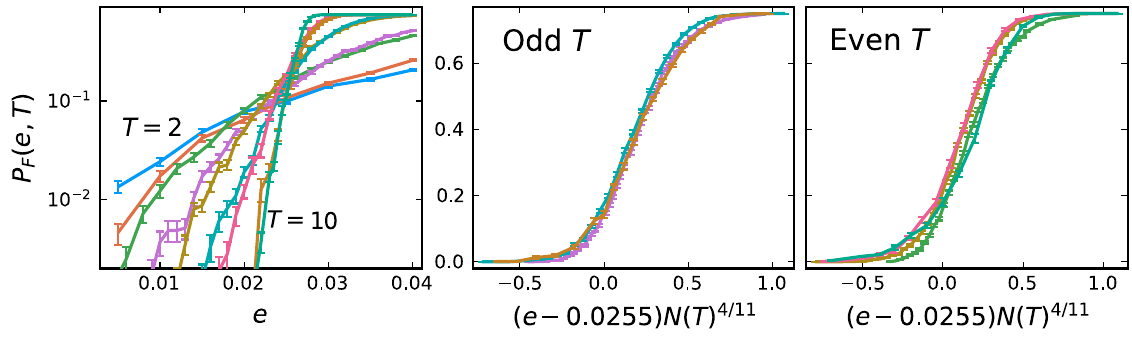}}\\
\subfloat[]{
\centering
\includegraphics[width=0.8\linewidth]{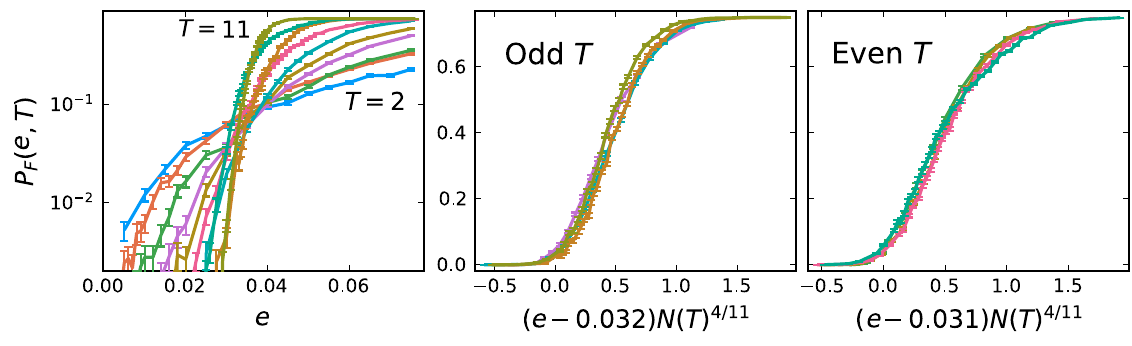}}
\caption{\label{fig:memory-scaling}Logical failure probability of one round of Steane EC followed by perfect stabilizer measurements, with erasure errors. (a) $n(T) = d(T) = 2^{\lfloor T/2 \rfloor}$. (b) $n(T)=1$. In both (a) and (b), the left panel is the unscaled data (on a log scale), center panel is a scaling collapse for odd $T\geq 3$, and right panel is a scaling collapse for even $T\geq 4$. In (b), the odd and even scaling collapses use slightly different $e_c$ (0.032 vs. 0.031), for a slightly better visual collapse. {Error bars denote the standard error of the mean across at least 500 shots.}}
\end{figure}

In the main text, we presented (the logarithm of) the logical failure probability for even $T$. In~\autoref{fig:memory-scaling}, we present the data for both parities of $T$, along with separate scaling collapses for odd and even $T$.

A single round of Steane syndrome extraction has $(3T+2)2^{T+1}$ possible error locations: $(3T-1+1+2) 2^T$ error locations on each of the two ancillas, $2^{T+1}$ additional gate errors incurred during each transversal CNOT, and $2^{T}$ measurement errors on each ancilla. We therefore define the effective system size to be $N(T) = (3T+2) 2^{T+1} n(T)$. Then, for both $n(T)=1$ and $n(T)=2^{\lfloor T/2 \rfloor}$, and both parities of $T$, the best scaling collapse is obtained using $\nu \approx 2.75$. 

An equally good collapse but slightly smaller exponent, $\nu \approx 2.4$, is favored if we instead define the ``effective system size'' without factors of $T$, i.e., $N(T) = 2^T d(T)$.

\section{Probability passing decoder for quantum memory}\label{app:memory-ppd}
If the errors are unheralded, we again must resort to an approximate decoder. In this section, we outline the steps of the decoder, bechmarking it against the optimal decoder for the heralded error model before applying it to a model of unheralded errors. 

\subsection{Decoding method}
We consider a simplified error model in which bit flips and phase flips are independent, such that the associated error probabilities can be updated independently and the final logical classes can be decoded independently. We then implement the following strategy:

\begin{enumerate}
\item Initialize the system to have zero error probabilities $(p_X, p_Z)$ on each qubit.
\item For ancilla state $\lstate{+}, \lstate{0}$:
\begin{enumerate}
\item Compute the marginal error probability at each possible error location, using unstacked probability passing. While the conditional error probabilities could be estimated more accurately using first-order stacking, we find empirically that this has a negligible effect on the performance of the decoder.
\item Given a local noise model on the ancilla and the system, and the measurement outcomes on the ancilla, compute the marginal conditional probability of the corresponding error type on each system qubit. For $\lstate{+}$, measurements are in the $Z$ basis and the marginal bit flip probabilities are updated, and vice versa for $\lstate{0}$.
\item Errors of the opposite type (e.g., phase-flip errors for $Z$ checks) leak via the transversal CNOT/NOTC onto the system. Accordingly, multiply the existing error on $S$ by the error propagated from the ancilla and update the marginal error probabilities.
\item Sample a new error on the system due to the transversal gate, multiply it into the existing error, and update probabilities of both error types.
\end{enumerate}
\item Repeat step 2 $n(T)$ times.
\item Given the syndromes $s_x, s_z$ read out in the final round of perfect measurements, decode $X$ and $Z$ errors separately according to the updated noise model on the system. That is, for $s = s_x, s_z$:
\begin{enumerate}
    \item Find a canonical error $E_s$, supported on the system, with syndrome $s$.
    \item Sum over the probabilities of all faults which propagate to errors in the logical cosets $E_s, \overline{X} E_s$ (for $s_z$) or $E_s, \overline{Z} E_s$ (for $s_x$). Apply an operator in the more likely class $L^*$.
    \item If $\overline{L^*} E_s$ implements a logical error, record failure.
\end{enumerate}
\end{enumerate}

As with state preparation, the optimal decoder would update not just the marginals, but the full probability distribution on the system, such that the logical class probabilities computed in the final step use the complete information from all rounds of syndrome measurements. In the approximate probability passing decoder, only partial information is passed on from one layer to the next. If the number of rounds $n$ does not increase with $T$, we could incorporate stacking into the final decoding step, potentially improving the subthreshold behavior but also increasing the complexity of the decoder. We have chosen not to pursue this avenue here.

\subsection{Heralded errors}
\begin{figure}[hbtp]
\subfloat[]{
\includegraphics[width=\linewidth]{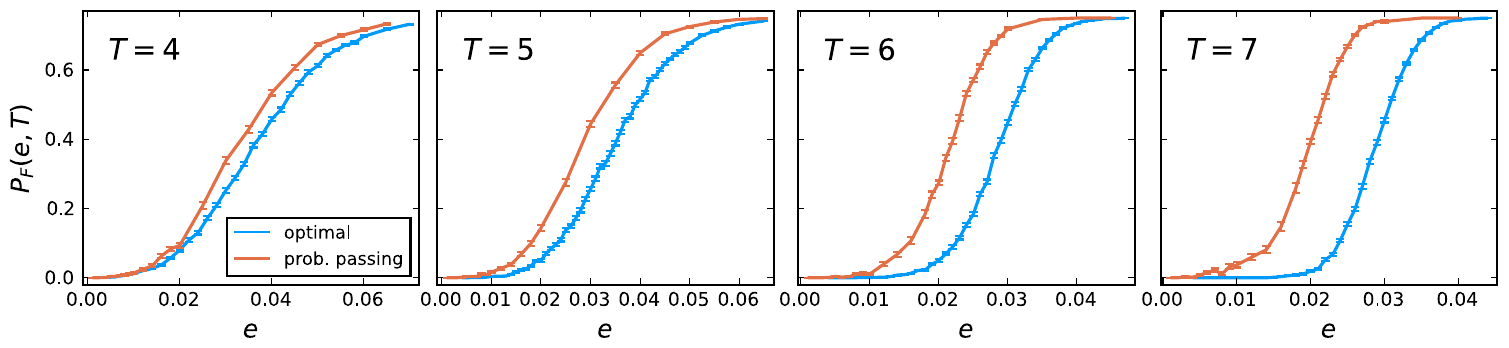}
\label{fig:opt-vs-ppd-d}}\\
\subfloat[]{
\includegraphics[width=\linewidth]{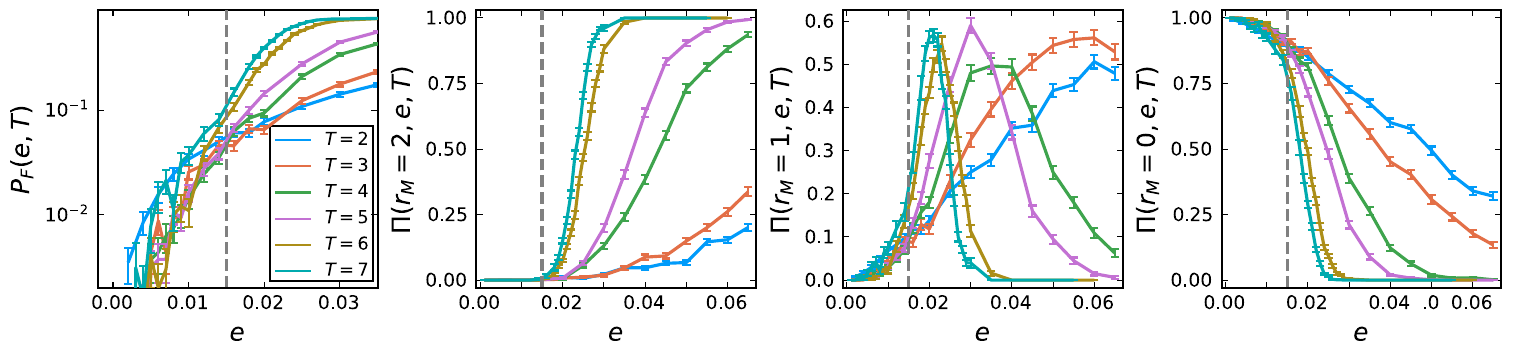}\label{fig:round-d-approx}}
\caption{(a) Comparison of the optimal decoder (with no ancilla verification) to the probability passing decoder, for $n(T) = d(T)$ rounds of Steane EC, heralded errors. (b) Logical failure probability of the probability passing decoder (leftmost panel) as well as the fraction of runs with $r_M = 2, 1, 0$ (cf.~\autoref{fig:n0-all}). The dashed gray line in each panel marks the pseudothreshold $e_c=0.015$. {Error bars denote the standard error of the mean (for $P_F(e)$) and the one sigma confidence interval for $N_{shots}$ Bernoulli trials (for $\Pi(r_M=k)$), with $N_{shots}$ ranging from $750$ to $1800$.} \label{fig:round-d}}
\end{figure}

We first benchmark the probability passing decoder against the optimal decoder for erasure errors. The performance of the two decoders is compared for $n=d(T)$ rounds in~\autoref{fig:opt-vs-ppd-d}. As $T$ increases, the gap in performance widens, and indeed, the estimated transition point $e_c \approx 0.015$ is well below that of the optimal decoder, $e_c \approx 0.0255$ (cf.~\autoref{fig:memory}).

For $e<0.015$, the logical failure rate of the approximate decoder is roughly independent of $T$ for the accessible system sizes, so we call $e_c$ a \textit{pseudothreshold}. This can be seen in~\autoref{fig:round-d-approx}, which shows the logical failure probability on a log scale, along with $\Pi(r_M=k)$ for $k=0,1,2$. Here, borrowing notation from~\autoref{app:erasures} of this Supplement, $r_M$ denotes the number of logical basis elements for which the decoder makes a random guess (\autoref{eq:rM}). In the optimal decoder, $r_M = 2 - I(S:R)$, but in the probability passing decoder, this relationship is weakened to an inequality, $r_M \geq 2 - I(S:R)$. Explicitly, $r_M = r_{M,X} + r_{M,Z}$ where
\begin{align}
r_{M,(X,Z)} = 0 &\Rightarrow \mathrm{Guaranteed \, success \, decoding \, } X, Z \mathrm{\, syndrome} \\
r_{M,(X,Z)} = 1 &\Rightarrow \mathrm{probability \, 1/2 \, of \,} \overline{X}, \overline{Z} \, \mathrm{error}.
\end{align}

\begin{figure}[hbtp]
\subfloat[]{\includegraphics[width=\linewidth]{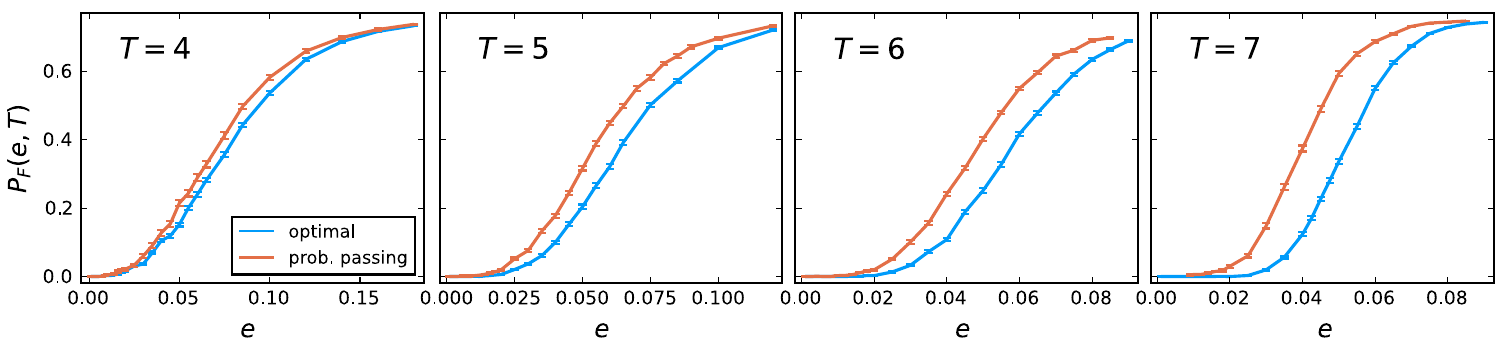}\label{fig:opt-vs-ppd1}}\\
\subfloat[]{
\includegraphics[width=\linewidth]{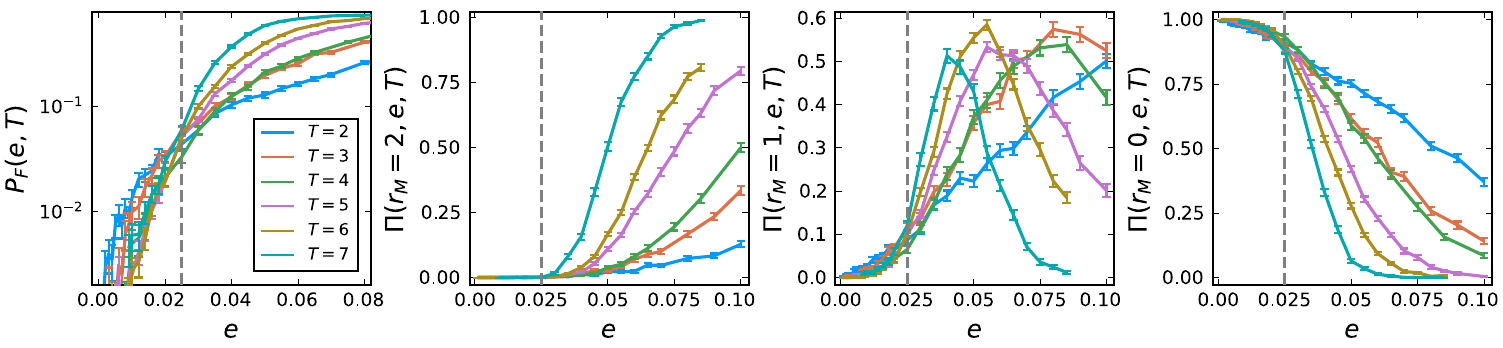}\label{fig:round1-approx}}
\caption{\label{fig:round1} Same as~\autoref{fig:round-d}, but with $n(T)=1$ round of Steane EC. The gray dashed line in (b) marks the pseudothreshold $e_c \approx 0.025$.}
\end{figure}

Qualitatively similar behavior is found when we limit to $n=1$ rounds of Steane EC: switching from the optimal to the probability passing decoder reduces $e_c$ (from $e_c \approx 0.031$, to $e_c \approx 0.025$) and turns it into a pseudothreshold. The results are shown in~\autoref{fig:round1}.

\subsection{Unheralded errors}
\begin{figure}[hbtp]
\centering
\includegraphics[width=\linewidth]{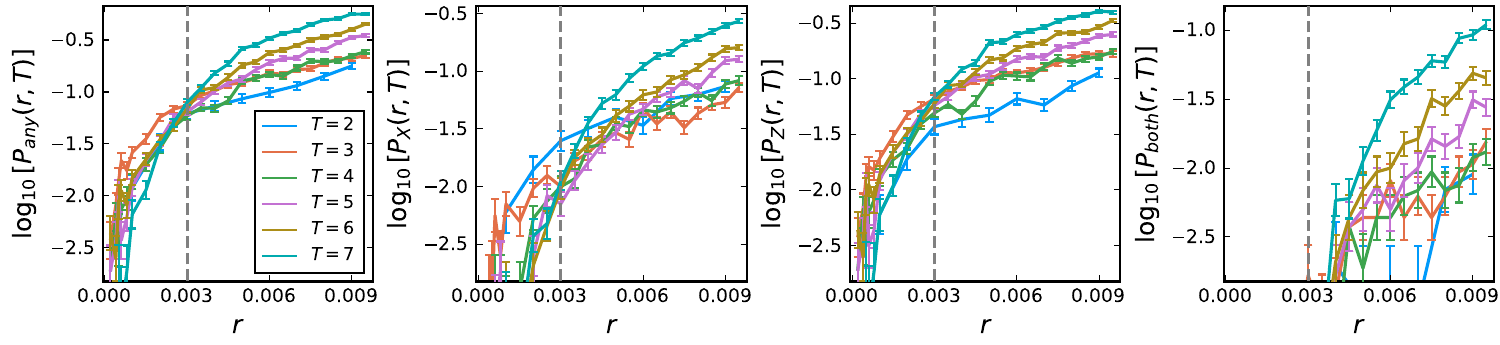}
\caption{Logical failure probability after 1 round of Steane EC, unheralded errors.\label{fig:memory-ppd} The gray dashed line marks the estimated pseudothreshold $r_c \approx 0.003$. {Error bars denote the 
 one sigma confidence interval over at least 1000 Bernoulli trials.}}
\end{figure}

The probabilities of a logical bit flip and logical phase flip after $n=1$ round of checks are shown in~\autoref{fig:memory-ppd}, along with the probability that either logical error occurs ($P_{any}$) and the probability that both occur ($P_{both}$). While we are again only able to sample up to depth $T=7$ due to the complexity of the decoder and the very low logical failure rates, a crossing around $r_c\approx 0.003$ is visible. As with state preparation, where logical $X$ errors (failure in preparation of $\lstate{0}$) are less common than logical $Z$ errors (failure in preparation of $\lstate{+}$), we find here that $P_Z(r,T) > P_X(r,T)$, but both quantities show an approximate crossing at the same $r_c$. For $r<r_c$, $P_{any}$ falls below the sampling precision: i.e., across $\gtrsim 1000$ shots, only 0 or 1 shots suffer both a logical phase flip and a logical bit flip. For the other three quantities below $p_c$, within uncertainties there is no appreciable suppression of the failure probability with increasing $T$ past $T\approx 4$. Thus, as with the approximate decoder for heralded errors, $r_c$ is best considered a pseudothreshold.

{\section{Code switching}\label{app:code-switch}
To code switch between $\GSC{e}(t)$ and $\GSC{o}(t)$, we want to find the same logical representatives for the two codes, to construct a subsystem code whose gauge group is the union of their stabilizer groups and has one logical qubit.}

{Let $P_{\mathrm{o}}(t)$ [$P_{\mathrm{e}}(t)$] denote the time-evolved operator obtained on the leaves of a depth $t$ \GSC{o} (\GSC{e}) tree by feeding the single-qubit Pauli $P$ into the root. $X_{\mathrm{o}}(t)$ is one representative of the logical $X$ operator, where other representatives are obtained through multiplication by stabilizers. In fact, $X(t)$ and $Z(t)$ are both minimum-weight logical representatives: $X_{\mathrm{o}}(t)$ is supported on $2^{\lceil t/2 \rceil}$ qubits, while $Z_{\mathrm{o}}(t)$ is supported on $2^{\lfloor t/2 \rfloor}$ qubits, and vice versa for $X_\mathrm{e}(t), Z_\mathrm{e}(t)$.}

{To obtain the desired subsystem code, it suffices to permute the qubits so that $X_{\mathrm{e}}(t)$ maps onto $X_{\mathrm{o}}(t)$, and $Z_{\mathrm{e}}(t)$ maps onto $Z_{\mathrm{o}}(t)$. Note that owing to the alternating pattern of isometries, the support of $X_{\mathrm{o}}(t)$ is identical to the support of $Z_\mathrm{e}(t)$, and vice versa, for all $t$. If $t$ is even, $X_{\mathrm{o}}(t)$ and $Z_{\mathrm{o}}(t)$ have the same weight, so there exists a permutation mapping the support of $X_{\mathrm{o}}(t)$ to the support of $Z_{\mathrm{o}}(t)$. This permutation consists of disjoint transpositions and thus is its own inverse, so the same permutation maps the support of $Z_{\mathrm{o}}(t)$ to $X_{\mathrm{o}}(t)$.}

{Let us state the permutation explicitly. Label the leaves of a depth $t$ tree as $0,1,...,2^t-1$. Then for even $t=2\tau$, the support of $Z_\mathrm{o}(2\tau)$ is the first $2^\tau$ elements of the Moser-de Bruijn sequence~\cite{oeis}. This is the sequence of distinct sums of powers of 4: $0,1,4,5,16,17,20,21,\dots$ . Denoting the $i$th element of the sequence by $a_i$, $X_\mathrm{o}(2\tau)$ has support on the qubits $\{2a_i\}_{i=1}^{2^\tau}$. Therefore, the necessary permutation at depth $2\tau$ is one which swaps $a_i \leftrightarrow 2a_i$ for $i=2,...,2^\tau$.} 

{Let $\tilde{\mathcal{S}}_{\mathrm{e}}$ denote the stabilizer group of the permuted code $\widetilde{\mathrm{GS}}_\mathrm{e}(2\tau)$, and ${\mathcal{S}}_{\mathrm{o}}$ the stabilizer group of the target code $\GSC{o}(2\tau)$. We can now define the tableau of the subsystem code~\cite{Poulin2005,Albert2023}:
\begin{enumerate}
    \item Gauge group generated by $\langle i,\tilde{\mathcal{S}}_{\mathrm{e}},\mathcal{S}_\mathrm{o}\rangle$, which splits into    \begin{description}
        \item[Stabilizer generators] $g_1,...,g_{s}$, which are elements of both $\widetilde{\mathcal{S}}_\mathrm{e}(2\tau)$ and $\mathcal{S}{o}(2\tau)$. For example, for $\tau=1$, then $g_1=XXXX, g_2=ZZZZ$.
        \item [Nontrivial gauge qubits] $g_{s+1}, h_{s+1},g_{s+2},h_{s+2},...,g_{4^{\tau}-1},h_{4^\tau-1}$, where $g_j \in \mathcal{S}_\mathrm{o}, h_j \in \tilde{\mathcal{S}}_{\mathrm{e}}$, and $(g_i, h_j)$ anticommute if $i=j$ and commute otherwise. For example, for $\tau=1$, $g_3 = XXII, h_3 = ZIZI$.\footnote{{At this point, some readers may recognize the subsystem code for $\tau=1$ as the $[[4,1,1,2]]$ code obtained by gauging out a logical qubit of the $[[4,2,2]]$ code; see Ref.~\cite{Albert2023}.} \gs{This is the smallest realization of a Bacon-Shor code~\cite{Shor1995,Bacon2006}.}}
    \end{description}
    \item ``Bare'' logical qubit $(X_\mathrm{o}(2\tau), Z_\mathrm{o}(2\tau))$, which commutes with the entire gauge group. For example, $X_\mathrm{o}(2) = XIXI, Z_\mathrm{o}(2) = ZZII$.
\end{enumerate}}

{Altogether, the code switching protocol from $\GSC{e}(2\tau)$ to $\GSC{o}(2\tau)$ proceeds as follows:}

{\begin{enumerate}
    \item Apply a layer of SWAP gates on the qubit pairs $(a_i, 2a_i)$ for $i=2,...,2^\tau$. The system is now encoded in the permuted code $\widetilde{\mathrm{GS}}_\mathrm{e}(2\tau)$.
    \item Prepare the ancilla states $\lstate{0},\lstate{+}$ in the $\GSC{o}(2\tau)$ encoding, using the modified multitree state preparation scheme which alternates checks (see \autoref{app:memory} of this Supplement).
    \item Use these ancillas to perform Steane syndrome extraction as in~\autoref{fig:steane-circuit}. If the syndrome measurement were noiseless, this would tell us both the syndrome of the subsystem code (signs of stabilizers $g_1,...,g_s$) and the signs of the new stabilizers $g_{s+1},...,g_{4^\tau-1}$. In practice, the ancillas and measurements have some noise, so the deduced signs on the system may contain errors, but these are local errors that will be handled in the next error correction step. 
    \item For each new stabilizer $g_j$ that got a measurement outcome $-1$, fix up the sign by applying the operator $h_j$, which by definition preserves the logical state.\footnote{{Optionally, if we measure a nonzero syndrome for any of the common stabilizers $g_1,...,g_s$, we could attempt to correct it using this partial syndrome information, but as the intermediate subsystem code has inferior distance, it may be better just to defer correction to the next round of EC.}}
\end{enumerate}}

{What if $t$ is odd? For $t>1$, there still exists a permutation mapping \textit{some} choice of logical $X$ and $Z$ representatives of $\GSC{e}(t)$ to some choice of logical $X$ and $Z$ representatives of $\GSC{o}(t)$. For example, we can use $X_\mathrm{o}(t)$ and, instead of $Z_\mathrm{o}(t)$, find a weight-$2^{\lceil t/2 \rceil}$ logical $Z$ representative of $\GSC{o}(t)$. Then we can apply a sequence of SWAP gates to exchange their supports. We leave it to the curious reader to determine an efficient way to identify a weight-$2^{\lceil t/2 \rceil}$ logical $Z$ representative. Note that for $t=1$, no such representative exists: the only logical $Z$ operators of $\GSC{o}(1)$ are $ZI$ and $IZ$, whereas $X_\mathrm{o}(1) = XX$.}


\begin{thebibliography}{101}%
\makeatletter
\providecommand \@ifxundefined [1]{%
 \@ifx{#1\undefined}
}%
\providecommand \@ifnum [1]{%
 \ifnum #1\expandafter \@firstoftwo
 \else \expandafter \@secondoftwo
 \fi
}%
\providecommand \@ifx [1]{%
 \ifx #1\expandafter \@firstoftwo
 \else \expandafter \@secondoftwo
 \fi
}%
\providecommand \natexlab [1]{#1}%
\providecommand \enquote  [1]{``#1''}%
\providecommand \bibnamefont  [1]{#1}%
\providecommand \bibfnamefont [1]{#1}%
\providecommand \citenamefont [1]{#1}%
\providecommand \href@noop [0]{\@secondoftwo}%
\providecommand \href [0]{\begingroup \@sanitize@url \@href}%
\providecommand \@href[1]{\@@startlink{#1}\@@href}%
\providecommand \@@href[1]{\endgroup#1\@@endlink}%
\providecommand \@sanitize@url [0]{\catcode `\\12\catcode `\$12\catcode `\&12\catcode `\#12\catcode `\^12\catcode `\_12\catcode `\%12\relax}%
\providecommand \@@startlink[1]{}%
\providecommand \@@endlink[0]{}%
\providecommand \url  [0]{\begingroup\@sanitize@url \@url }%
\providecommand \@url [1]{\endgroup\@href {#1}{\urlprefix }}%
\providecommand \urlprefix  [0]{URL }%
\providecommand \Eprint [0]{\href }%
\providecommand \doibase [0]{https://doi.org/}%
\providecommand \selectlanguage [0]{\@gobble}%
\providecommand \bibinfo  [0]{\@secondoftwo}%
\providecommand \bibfield  [0]{\@secondoftwo}%
\providecommand \translation [1]{[#1]}%
\providecommand \BibitemOpen [0]{}%
\providecommand \bibitemStop [0]{}%
\providecommand \bibitemNoStop [0]{.\EOS\space}%
\providecommand \EOS [0]{\spacefactor3000\relax}%
\providecommand \BibitemShut  [1]{\csname bibitem#1\endcsname}%
\let\auto@bib@innerbib\@empty
\bibitem [{\citenamefont {Egan}\ \emph {et~al.}(2021)\citenamefont {Egan}, \citenamefont {Debroy}, \citenamefont {Noel}, \citenamefont {Risinger}, \citenamefont {Zhu}, \citenamefont {Biswas}, \citenamefont {Newman}, \citenamefont {Li}, \citenamefont {Brown}, \citenamefont {Cetina},\ and\ \citenamefont {Monroe}}]{egan2021fault-tolerant-4e9}%
  \BibitemOpen
  \bibfield  {author} {\bibinfo {author} {\bibfnamefont {L.}~\bibnamefont {Egan}}, \bibinfo {author} {\bibfnamefont {D.~M.}\ \bibnamefont {Debroy}}, \bibinfo {author} {\bibfnamefont {C.}~\bibnamefont {Noel}}, \bibinfo {author} {\bibfnamefont {A.}~\bibnamefont {Risinger}}, \bibinfo {author} {\bibfnamefont {D.}~\bibnamefont {Zhu}}, \bibinfo {author} {\bibfnamefont {D.}~\bibnamefont {Biswas}}, \bibinfo {author} {\bibfnamefont {M.}~\bibnamefont {Newman}}, \bibinfo {author} {\bibfnamefont {M.}~\bibnamefont {Li}}, \bibinfo {author} {\bibfnamefont {K.~R.}\ \bibnamefont {Brown}}, \bibinfo {author} {\bibfnamefont {M.}~\bibnamefont {Cetina}},\ and\ \bibinfo {author} {\bibfnamefont {C.}~\bibnamefont {Monroe}},\ }\bibfield  {title} {\bibinfo {title} {Fault-tolerant control of an error-corrected qubit},\ }\href {https://doi.org/10.1038/s41586-021-03928-y} {\bibfield  {journal} {\bibinfo  {journal} {Nature}\ }\textbf {\bibinfo {volume} {598}},\ \bibinfo {pages} {281} (\bibinfo {year} {2021})}\BibitemShut {NoStop}%
\bibitem [{\citenamefont {Postler}\ \emph {et~al.}(2024)\citenamefont {Postler}, \citenamefont {Butt}, \citenamefont {Pogorelov}, \citenamefont {Marciniak}, \citenamefont {Heu{\ss}en}, \citenamefont {Blatt}, \citenamefont {Schindler}, \citenamefont {Rispler}, \citenamefont {M{\"{u}}ller},\ and\ \citenamefont {Monz}}]{Postler2024}%
  \BibitemOpen
  \bibfield  {author} {\bibinfo {author} {\bibfnamefont {L.}~\bibnamefont {Postler}}, \bibinfo {author} {\bibfnamefont {F.}~\bibnamefont {Butt}}, \bibinfo {author} {\bibfnamefont {I.}~\bibnamefont {Pogorelov}}, \bibinfo {author} {\bibfnamefont {C.~D.}\ \bibnamefont {Marciniak}}, \bibinfo {author} {\bibfnamefont {S.}~\bibnamefont {Heu{\ss}en}}, \bibinfo {author} {\bibfnamefont {R.}~\bibnamefont {Blatt}}, \bibinfo {author} {\bibfnamefont {P.}~\bibnamefont {Schindler}}, \bibinfo {author} {\bibfnamefont {M.}~\bibnamefont {Rispler}}, \bibinfo {author} {\bibfnamefont {M.}~\bibnamefont {M{\"{u}}ller}},\ and\ \bibinfo {author} {\bibfnamefont {T.}~\bibnamefont {Monz}},\ }\bibfield  {title} {\bibinfo {title} {{Demonstration of Fault-Tolerant Steane Quantum Error Correction}},\ }\href {https://doi.org/10.1103/PRXQuantum.5.030326} {\bibfield  {journal} {\bibinfo  {journal} {PRX Quantum}\ }\textbf {\bibinfo {volume} {5}},\ \bibinfo {pages} {030326} (\bibinfo {year} {2024})}\BibitemShut {NoStop}%
\bibitem [{\citenamefont {Huang}\ \emph {et~al.}(2023)\citenamefont {Huang}, \citenamefont {Brown},\ and\ \citenamefont {Cetina}}]{Huang2023}%
  \BibitemOpen
  \bibfield  {author} {\bibinfo {author} {\bibfnamefont {S.}~\bibnamefont {Huang}}, \bibinfo {author} {\bibfnamefont {K.~R.}\ \bibnamefont {Brown}},\ and\ \bibinfo {author} {\bibfnamefont {M.}~\bibnamefont {Cetina}},\ }\href {https://arxiv.org/abs/2312.10851v1 http://arxiv.org/abs/2312.10851} {\bibinfo {title} {{Comparing Shor and Steane Error Correction Using the Bacon-Shor Code}}} (\bibinfo {year} {2023}),\ \Eprint {https://arxiv.org/abs/2312.10851} {arXiv:2312.10851} \BibitemShut {NoStop}%
\bibitem [{\citenamefont {Paetznick}\ \emph {et~al.}(2024)\citenamefont {Paetznick}, \citenamefont {da~Silva}, \citenamefont {Ryan-Anderson}, \citenamefont {Bello-Rivas}, \citenamefont {III}, \citenamefont {Chernoguzov}, \citenamefont {Dreiling}, \citenamefont {Foltz}, \citenamefont {Frachon}, \citenamefont {Gaebler}, \citenamefont {Gatterman}, \citenamefont {Grans-Samuelsson}, \citenamefont {Gresh}, \citenamefont {Hayes}, \citenamefont {Hewitt}, \citenamefont {Holliman}, \citenamefont {Horst}, \citenamefont {Johansen}, \citenamefont {Lucchetti}, \citenamefont {Matsuoka}, \citenamefont {Mills}, \citenamefont {Moses}, \citenamefont {Neyenhuis}, \citenamefont {Paz}, \citenamefont {Pino}, \citenamefont {Siegfried}, \citenamefont {Sundaram}, \citenamefont {Tom}, \citenamefont {Wernli}, \citenamefont {Zanner}, \citenamefont {Stutz},\ and\ \citenamefont {Svore}}]{silva2024demonstration-a9a}%
  \BibitemOpen
  \bibfield  {author} {\bibinfo {author} {\bibfnamefont {A.}~\bibnamefont {Paetznick}}, \bibinfo {author} {\bibfnamefont {M.~P.}\ \bibnamefont {da~Silva}}, \bibinfo {author} {\bibfnamefont {C.}~\bibnamefont {Ryan-Anderson}}, \bibinfo {author} {\bibfnamefont {J.~M.}\ \bibnamefont {Bello-Rivas}}, \bibinfo {author} {\bibfnamefont {J.~P.~C.}\ \bibnamefont {III}}, \bibinfo {author} {\bibfnamefont {A.}~\bibnamefont {Chernoguzov}}, \bibinfo {author} {\bibfnamefont {J.~M.}\ \bibnamefont {Dreiling}}, \bibinfo {author} {\bibfnamefont {C.}~\bibnamefont {Foltz}}, \bibinfo {author} {\bibfnamefont {F.}~\bibnamefont {Frachon}}, \bibinfo {author} {\bibfnamefont {J.~P.}\ \bibnamefont {Gaebler}}, \bibinfo {author} {\bibfnamefont {T.~M.}\ \bibnamefont {Gatterman}}, \bibinfo {author} {\bibfnamefont {L.}~\bibnamefont {Grans-Samuelsson}}, \bibinfo {author} {\bibfnamefont {D.}~\bibnamefont {Gresh}}, \bibinfo {author} {\bibfnamefont {D.}~\bibnamefont {Hayes}}, \bibinfo {author} {\bibfnamefont {N.}~\bibnamefont {Hewitt}}, \bibinfo
  {author} {\bibfnamefont {C.}~\bibnamefont {Holliman}}, \bibinfo {author} {\bibfnamefont {C.~V.}\ \bibnamefont {Horst}}, \bibinfo {author} {\bibfnamefont {J.}~\bibnamefont {Johansen}}, \bibinfo {author} {\bibfnamefont {D.}~\bibnamefont {Lucchetti}}, \bibinfo {author} {\bibfnamefont {Y.}~\bibnamefont {Matsuoka}}, \bibinfo {author} {\bibfnamefont {M.}~\bibnamefont {Mills}}, \bibinfo {author} {\bibfnamefont {S.~A.}\ \bibnamefont {Moses}}, \bibinfo {author} {\bibfnamefont {B.}~\bibnamefont {Neyenhuis}}, \bibinfo {author} {\bibfnamefont {A.}~\bibnamefont {Paz}}, \bibinfo {author} {\bibfnamefont {J.}~\bibnamefont {Pino}}, \bibinfo {author} {\bibfnamefont {P.}~\bibnamefont {Siegfried}}, \bibinfo {author} {\bibfnamefont {A.}~\bibnamefont {Sundaram}}, \bibinfo {author} {\bibfnamefont {D.}~\bibnamefont {Tom}}, \bibinfo {author} {\bibfnamefont {S.~J.}\ \bibnamefont {Wernli}}, \bibinfo {author} {\bibfnamefont {M.}~\bibnamefont {Zanner}}, \bibinfo {author} {\bibfnamefont {R.~P.}\ \bibnamefont {Stutz}},\ and\ \bibinfo
  {author} {\bibfnamefont {K.~M.}\ \bibnamefont {Svore}},\ }\href {https://arxiv.org/abs/2404.02280} {\bibinfo {title} {Demonstration of logical qubits and repeated error correction with better-than-physical error rates}} (\bibinfo {year} {2024}),\ \Eprint {https://arxiv.org/abs/2404.02280} {arXiv:2404.02280 [quant-ph]} \BibitemShut {NoStop}%
\bibitem [{\citenamefont {Krinner}\ \emph {et~al.}(2022)\citenamefont {Krinner}, \citenamefont {Lacroix}, \citenamefont {Remm}, \citenamefont {{Di Paolo}}, \citenamefont {Genois}, \citenamefont {Leroux}, \citenamefont {Hellings}, \citenamefont {Lazar}, \citenamefont {Swiadek}, \citenamefont {Herrmann}, \citenamefont {Norris}, \citenamefont {Andersen}, \citenamefont {M{\"{u}}ller}, \citenamefont {Blais}, \citenamefont {Eichler},\ and\ \citenamefont {Wallraff}}]{krinner2021realizing-d73}%
  \BibitemOpen
  \bibfield  {author} {\bibinfo {author} {\bibfnamefont {S.}~\bibnamefont {Krinner}}, \bibinfo {author} {\bibfnamefont {N.}~\bibnamefont {Lacroix}}, \bibinfo {author} {\bibfnamefont {A.}~\bibnamefont {Remm}}, \bibinfo {author} {\bibfnamefont {A.}~\bibnamefont {{Di Paolo}}}, \bibinfo {author} {\bibfnamefont {E.}~\bibnamefont {Genois}}, \bibinfo {author} {\bibfnamefont {C.}~\bibnamefont {Leroux}}, \bibinfo {author} {\bibfnamefont {C.}~\bibnamefont {Hellings}}, \bibinfo {author} {\bibfnamefont {S.}~\bibnamefont {Lazar}}, \bibinfo {author} {\bibfnamefont {F.}~\bibnamefont {Swiadek}}, \bibinfo {author} {\bibfnamefont {J.}~\bibnamefont {Herrmann}}, \bibinfo {author} {\bibfnamefont {G.~J.}\ \bibnamefont {Norris}}, \bibinfo {author} {\bibfnamefont {C.~K.}\ \bibnamefont {Andersen}}, \bibinfo {author} {\bibfnamefont {M.}~\bibnamefont {M{\"{u}}ller}}, \bibinfo {author} {\bibfnamefont {A.}~\bibnamefont {Blais}}, \bibinfo {author} {\bibfnamefont {C.}~\bibnamefont {Eichler}},\ and\ \bibinfo {author} {\bibfnamefont
  {A.}~\bibnamefont {Wallraff}},\ }\bibfield  {title} {\bibinfo {title} {{Realizing repeated quantum error correction in a distance-three surface code}},\ }\href {https://doi.org/10.1038/s41586-022-04566-8} {\bibfield  {journal} {\bibinfo  {journal} {Nature}\ }\textbf {\bibinfo {volume} {605}},\ \bibinfo {pages} {669} (\bibinfo {year} {2022})}\BibitemShut {NoStop}%
\bibitem [{\citenamefont {Acharya}\ \emph {et~al.}(2023)\citenamefont {Acharya}, \citenamefont {Aleiner}, \citenamefont {Allen}, \citenamefont {Andersen}, \citenamefont {Ansmann}, \citenamefont {Arute}, \citenamefont {Arya}, \citenamefont {Asfaw}, \citenamefont {Atalaya}, \citenamefont {Babbush}, \citenamefont {Bacon}, \citenamefont {Bardin}, \citenamefont {Basso}, \citenamefont {Bengtsson}, \citenamefont {Boixo}, \citenamefont {Bortoli}, \citenamefont {Bourassa}, \citenamefont {Bovaird}, \citenamefont {Brill}, \citenamefont {Broughton}, \citenamefont {Buckley}, \citenamefont {Buell}, \citenamefont {Burger}, \citenamefont {Burkett}, \citenamefont {Bushnell}, \citenamefont {Chen}, \citenamefont {Chen}, \citenamefont {Chiaro}, \citenamefont {Cogan}, \citenamefont {Collins}, \citenamefont {Conner}, \citenamefont {Courtney}, \citenamefont {Crook}, \citenamefont {Curtin}, \citenamefont {Debroy}, \citenamefont {Barba}, \citenamefont {Demura}, \citenamefont {Dunsworth}, \citenamefont {Eppens}, \citenamefont
  {Erickson}, \citenamefont {Faoro}, \citenamefont {Farhi}, \citenamefont {Fatemi}, \citenamefont {Burgos}, \citenamefont {Forati}, \citenamefont {Fowler}, \citenamefont {Foxen}, \citenamefont {Giang}, \citenamefont {Gidney}, \citenamefont {Gilboa}, \citenamefont {Giustina}, \citenamefont {Dau}, \citenamefont {Gross}, \citenamefont {Habegger}, \citenamefont {Hamilton}, \citenamefont {Harrigan}, \citenamefont {Harrington}, \citenamefont {Higgott}, \citenamefont {Hilton}, \citenamefont {Hoffmann}, \citenamefont {Hong}, \citenamefont {Huang}, \citenamefont {Huff}, \citenamefont {Huggins}, \citenamefont {Ioffe}, \citenamefont {Isakov}, \citenamefont {Iveland}, \citenamefont {Jeffrey}, \citenamefont {Jiang}, \citenamefont {Jones}, \citenamefont {Juhas}, \citenamefont {Kafri}, \citenamefont {Kechedzhi}, \citenamefont {Kelly}, \citenamefont {Khattar}, \citenamefont {Khezri}, \citenamefont {Kieferová}, \citenamefont {Kim}, \citenamefont {Kitaev}, \citenamefont {Klimov}, \citenamefont {Klots}, \citenamefont
  {Korotkov}, \citenamefont {Kostritsa}, \citenamefont {Kreikebaum}, \citenamefont {Landhuis}, \citenamefont {Laptev}, \citenamefont {Lau}, \citenamefont {Laws}, \citenamefont {Lee}, \citenamefont {Lee}, \citenamefont {Lester}, \citenamefont {Lill}, \citenamefont {Liu}, \citenamefont {Locharla}, \citenamefont {Lucero}, \citenamefont {Malone}, \citenamefont {Marshall}, \citenamefont {Martin}, \citenamefont {{McClean}}, \citenamefont {{McCourt}}, \citenamefont {{McEwen}}, \citenamefont {Megrant}, \citenamefont {Costa}, \citenamefont {Mi}, \citenamefont {Miao}, \citenamefont {Mohseni}, \citenamefont {Montazeri}, \citenamefont {Morvan}, \citenamefont {Mount}, \citenamefont {Mruczkiewicz}, \citenamefont {Naaman}, \citenamefont {Neeley}, \citenamefont {Neill}, \citenamefont {Nersisyan}, \citenamefont {Neven}, \citenamefont {Newman}, \citenamefont {Ng}, \citenamefont {Nguyen}, \citenamefont {Nguyen}, \citenamefont {Niu}, \citenamefont {O’Brien}, \citenamefont {Opremcak}, \citenamefont {Platt}, \citenamefont
  {Petukhov}, \citenamefont {Potter}, \citenamefont {Pryadko}, \citenamefont {Quintana}, \citenamefont {Roushan}, \citenamefont {Rubin}, \citenamefont {Saei}, \citenamefont {Sank}, \citenamefont {Sankaragomathi}, \citenamefont {Satzinger}, \citenamefont {Schurkus}, \citenamefont {Schuster}, \citenamefont {Shearn}, \citenamefont {Shorter}, \citenamefont {Shvarts}, \citenamefont {Skruzny}, \citenamefont {Smelyanskiy}, \citenamefont {Smith}, \citenamefont {Sterling}, \citenamefont {Strain}, \citenamefont {Szalay}, \citenamefont {Torres}, \citenamefont {Vidal}, \citenamefont {Villalonga}, \citenamefont {Heidweiller}, \citenamefont {White}, \citenamefont {Xing}, \citenamefont {Yao}, \citenamefont {Yeh}, \citenamefont {Yoo}, \citenamefont {Young}, \citenamefont {Zalcman}, \citenamefont {Zhang},\ and\ \citenamefont {Zhu}}]{ai2023suppressing-df2}%
  \BibitemOpen
  \bibfield  {author} {\bibinfo {author} {\bibfnamefont {R.}~\bibnamefont {Acharya}}, \bibinfo {author} {\bibfnamefont {I.}~\bibnamefont {Aleiner}}, \bibinfo {author} {\bibfnamefont {R.}~\bibnamefont {Allen}}, \bibinfo {author} {\bibfnamefont {T.~I.}\ \bibnamefont {Andersen}}, \bibinfo {author} {\bibfnamefont {M.}~\bibnamefont {Ansmann}}, \bibinfo {author} {\bibfnamefont {F.}~\bibnamefont {Arute}}, \bibinfo {author} {\bibfnamefont {K.}~\bibnamefont {Arya}}, \bibinfo {author} {\bibfnamefont {A.}~\bibnamefont {Asfaw}}, \bibinfo {author} {\bibfnamefont {J.}~\bibnamefont {Atalaya}}, \bibinfo {author} {\bibfnamefont {R.}~\bibnamefont {Babbush}}, \bibinfo {author} {\bibfnamefont {D.}~\bibnamefont {Bacon}}, \bibinfo {author} {\bibfnamefont {J.~C.}\ \bibnamefont {Bardin}}, \bibinfo {author} {\bibfnamefont {J.}~\bibnamefont {Basso}}, \bibinfo {author} {\bibfnamefont {A.}~\bibnamefont {Bengtsson}}, \bibinfo {author} {\bibfnamefont {S.}~\bibnamefont {Boixo}}, \bibinfo {author} {\bibfnamefont {G.}~\bibnamefont
  {Bortoli}}, \bibinfo {author} {\bibfnamefont {A.}~\bibnamefont {Bourassa}}, \bibinfo {author} {\bibfnamefont {J.}~\bibnamefont {Bovaird}}, \bibinfo {author} {\bibfnamefont {L.}~\bibnamefont {Brill}}, \bibinfo {author} {\bibfnamefont {M.}~\bibnamefont {Broughton}}, \bibinfo {author} {\bibfnamefont {B.~B.}\ \bibnamefont {Buckley}}, \bibinfo {author} {\bibfnamefont {D.~A.}\ \bibnamefont {Buell}}, \bibinfo {author} {\bibfnamefont {T.}~\bibnamefont {Burger}}, \bibinfo {author} {\bibfnamefont {B.}~\bibnamefont {Burkett}}, \bibinfo {author} {\bibfnamefont {N.}~\bibnamefont {Bushnell}}, \bibinfo {author} {\bibfnamefont {Y.}~\bibnamefont {Chen}}, \bibinfo {author} {\bibfnamefont {Z.}~\bibnamefont {Chen}}, \bibinfo {author} {\bibfnamefont {B.}~\bibnamefont {Chiaro}}, \bibinfo {author} {\bibfnamefont {J.}~\bibnamefont {Cogan}}, \bibinfo {author} {\bibfnamefont {R.}~\bibnamefont {Collins}}, \bibinfo {author} {\bibfnamefont {P.}~\bibnamefont {Conner}}, \bibinfo {author} {\bibfnamefont {W.}~\bibnamefont {Courtney}},
  \bibinfo {author} {\bibfnamefont {A.~L.}\ \bibnamefont {Crook}}, \bibinfo {author} {\bibfnamefont {B.}~\bibnamefont {Curtin}}, \bibinfo {author} {\bibfnamefont {D.~M.}\ \bibnamefont {Debroy}}, \bibinfo {author} {\bibfnamefont {A.~D.~T.}\ \bibnamefont {Barba}}, \bibinfo {author} {\bibfnamefont {S.}~\bibnamefont {Demura}}, \bibinfo {author} {\bibfnamefont {A.}~\bibnamefont {Dunsworth}}, \bibinfo {author} {\bibfnamefont {D.}~\bibnamefont {Eppens}}, \bibinfo {author} {\bibfnamefont {C.}~\bibnamefont {Erickson}}, \bibinfo {author} {\bibfnamefont {L.}~\bibnamefont {Faoro}}, \bibinfo {author} {\bibfnamefont {E.}~\bibnamefont {Farhi}}, \bibinfo {author} {\bibfnamefont {R.}~\bibnamefont {Fatemi}}, \bibinfo {author} {\bibfnamefont {L.~F.}\ \bibnamefont {Burgos}}, \bibinfo {author} {\bibfnamefont {E.}~\bibnamefont {Forati}}, \bibinfo {author} {\bibfnamefont {A.~G.}\ \bibnamefont {Fowler}}, \bibinfo {author} {\bibfnamefont {B.}~\bibnamefont {Foxen}}, \bibinfo {author} {\bibfnamefont {W.}~\bibnamefont {Giang}}, \bibinfo
  {author} {\bibfnamefont {C.}~\bibnamefont {Gidney}}, \bibinfo {author} {\bibfnamefont {D.}~\bibnamefont {Gilboa}}, \bibinfo {author} {\bibfnamefont {M.}~\bibnamefont {Giustina}}, \bibinfo {author} {\bibfnamefont {A.~G.}\ \bibnamefont {Dau}}, \bibinfo {author} {\bibfnamefont {J.~A.}\ \bibnamefont {Gross}}, \bibinfo {author} {\bibfnamefont {S.}~\bibnamefont {Habegger}}, \bibinfo {author} {\bibfnamefont {M.~C.}\ \bibnamefont {Hamilton}}, \bibinfo {author} {\bibfnamefont {M.~P.}\ \bibnamefont {Harrigan}}, \bibinfo {author} {\bibfnamefont {S.~D.}\ \bibnamefont {Harrington}}, \bibinfo {author} {\bibfnamefont {O.}~\bibnamefont {Higgott}}, \bibinfo {author} {\bibfnamefont {J.}~\bibnamefont {Hilton}}, \bibinfo {author} {\bibfnamefont {M.}~\bibnamefont {Hoffmann}}, \bibinfo {author} {\bibfnamefont {S.}~\bibnamefont {Hong}}, \bibinfo {author} {\bibfnamefont {T.}~\bibnamefont {Huang}}, \bibinfo {author} {\bibfnamefont {A.}~\bibnamefont {Huff}}, \bibinfo {author} {\bibfnamefont {W.~J.}\ \bibnamefont {Huggins}}, \bibinfo
  {author} {\bibfnamefont {L.~B.}\ \bibnamefont {Ioffe}}, \bibinfo {author} {\bibfnamefont {S.~V.}\ \bibnamefont {Isakov}}, \bibinfo {author} {\bibfnamefont {J.}~\bibnamefont {Iveland}}, \bibinfo {author} {\bibfnamefont {E.}~\bibnamefont {Jeffrey}}, \bibinfo {author} {\bibfnamefont {Z.}~\bibnamefont {Jiang}}, \bibinfo {author} {\bibfnamefont {C.}~\bibnamefont {Jones}}, \bibinfo {author} {\bibfnamefont {P.}~\bibnamefont {Juhas}}, \bibinfo {author} {\bibfnamefont {D.}~\bibnamefont {Kafri}}, \bibinfo {author} {\bibfnamefont {K.}~\bibnamefont {Kechedzhi}}, \bibinfo {author} {\bibfnamefont {J.}~\bibnamefont {Kelly}}, \bibinfo {author} {\bibfnamefont {T.}~\bibnamefont {Khattar}}, \bibinfo {author} {\bibfnamefont {M.}~\bibnamefont {Khezri}}, \bibinfo {author} {\bibfnamefont {M.}~\bibnamefont {Kieferová}}, \bibinfo {author} {\bibfnamefont {S.}~\bibnamefont {Kim}}, \bibinfo {author} {\bibfnamefont {A.}~\bibnamefont {Kitaev}}, \bibinfo {author} {\bibfnamefont {P.~V.}\ \bibnamefont {Klimov}}, \bibinfo {author}
  {\bibfnamefont {A.~R.}\ \bibnamefont {Klots}}, \bibinfo {author} {\bibfnamefont {A.~N.}\ \bibnamefont {Korotkov}}, \bibinfo {author} {\bibfnamefont {F.}~\bibnamefont {Kostritsa}}, \bibinfo {author} {\bibfnamefont {J.~M.}\ \bibnamefont {Kreikebaum}}, \bibinfo {author} {\bibfnamefont {D.}~\bibnamefont {Landhuis}}, \bibinfo {author} {\bibfnamefont {P.}~\bibnamefont {Laptev}}, \bibinfo {author} {\bibfnamefont {K.-M.}\ \bibnamefont {Lau}}, \bibinfo {author} {\bibfnamefont {L.}~\bibnamefont {Laws}}, \bibinfo {author} {\bibfnamefont {J.}~\bibnamefont {Lee}}, \bibinfo {author} {\bibfnamefont {K.}~\bibnamefont {Lee}}, \bibinfo {author} {\bibfnamefont {B.~J.}\ \bibnamefont {Lester}}, \bibinfo {author} {\bibfnamefont {A.}~\bibnamefont {Lill}}, \bibinfo {author} {\bibfnamefont {W.}~\bibnamefont {Liu}}, \bibinfo {author} {\bibfnamefont {A.}~\bibnamefont {Locharla}}, \bibinfo {author} {\bibfnamefont {E.}~\bibnamefont {Lucero}}, \bibinfo {author} {\bibfnamefont {F.~D.}\ \bibnamefont {Malone}}, \bibinfo {author}
  {\bibfnamefont {J.}~\bibnamefont {Marshall}}, \bibinfo {author} {\bibfnamefont {O.}~\bibnamefont {Martin}}, \bibinfo {author} {\bibfnamefont {J.~R.}\ \bibnamefont {{McClean}}}, \bibinfo {author} {\bibfnamefont {T.}~\bibnamefont {{McCourt}}}, \bibinfo {author} {\bibfnamefont {M.}~\bibnamefont {{McEwen}}}, \bibinfo {author} {\bibfnamefont {A.}~\bibnamefont {Megrant}}, \bibinfo {author} {\bibfnamefont {B.~M.}\ \bibnamefont {Costa}}, \bibinfo {author} {\bibfnamefont {X.}~\bibnamefont {Mi}}, \bibinfo {author} {\bibfnamefont {K.~C.}\ \bibnamefont {Miao}}, \bibinfo {author} {\bibfnamefont {M.}~\bibnamefont {Mohseni}}, \bibinfo {author} {\bibfnamefont {S.}~\bibnamefont {Montazeri}}, \bibinfo {author} {\bibfnamefont {A.}~\bibnamefont {Morvan}}, \bibinfo {author} {\bibfnamefont {E.}~\bibnamefont {Mount}}, \bibinfo {author} {\bibfnamefont {W.}~\bibnamefont {Mruczkiewicz}}, \bibinfo {author} {\bibfnamefont {O.}~\bibnamefont {Naaman}}, \bibinfo {author} {\bibfnamefont {M.}~\bibnamefont {Neeley}}, \bibinfo {author}
  {\bibfnamefont {C.}~\bibnamefont {Neill}}, \bibinfo {author} {\bibfnamefont {A.}~\bibnamefont {Nersisyan}}, \bibinfo {author} {\bibfnamefont {H.}~\bibnamefont {Neven}}, \bibinfo {author} {\bibfnamefont {M.}~\bibnamefont {Newman}}, \bibinfo {author} {\bibfnamefont {J.~H.}\ \bibnamefont {Ng}}, \bibinfo {author} {\bibfnamefont {A.}~\bibnamefont {Nguyen}}, \bibinfo {author} {\bibfnamefont {M.}~\bibnamefont {Nguyen}}, \bibinfo {author} {\bibfnamefont {M.~Y.}\ \bibnamefont {Niu}}, \bibinfo {author} {\bibfnamefont {T.~E.}\ \bibnamefont {O’Brien}}, \bibinfo {author} {\bibfnamefont {A.}~\bibnamefont {Opremcak}}, \bibinfo {author} {\bibfnamefont {J.}~\bibnamefont {Platt}}, \bibinfo {author} {\bibfnamefont {A.}~\bibnamefont {Petukhov}}, \bibinfo {author} {\bibfnamefont {R.}~\bibnamefont {Potter}}, \bibinfo {author} {\bibfnamefont {L.~P.}\ \bibnamefont {Pryadko}}, \bibinfo {author} {\bibfnamefont {C.}~\bibnamefont {Quintana}}, \bibinfo {author} {\bibfnamefont {P.}~\bibnamefont {Roushan}}, \bibinfo {author}
  {\bibfnamefont {N.~C.}\ \bibnamefont {Rubin}}, \bibinfo {author} {\bibfnamefont {N.}~\bibnamefont {Saei}}, \bibinfo {author} {\bibfnamefont {D.}~\bibnamefont {Sank}}, \bibinfo {author} {\bibfnamefont {K.}~\bibnamefont {Sankaragomathi}}, \bibinfo {author} {\bibfnamefont {K.~J.}\ \bibnamefont {Satzinger}}, \bibinfo {author} {\bibfnamefont {H.~F.}\ \bibnamefont {Schurkus}}, \bibinfo {author} {\bibfnamefont {C.}~\bibnamefont {Schuster}}, \bibinfo {author} {\bibfnamefont {M.~J.}\ \bibnamefont {Shearn}}, \bibinfo {author} {\bibfnamefont {A.}~\bibnamefont {Shorter}}, \bibinfo {author} {\bibfnamefont {V.}~\bibnamefont {Shvarts}}, \bibinfo {author} {\bibfnamefont {J.}~\bibnamefont {Skruzny}}, \bibinfo {author} {\bibfnamefont {V.}~\bibnamefont {Smelyanskiy}}, \bibinfo {author} {\bibfnamefont {W.~C.}\ \bibnamefont {Smith}}, \bibinfo {author} {\bibfnamefont {G.}~\bibnamefont {Sterling}}, \bibinfo {author} {\bibfnamefont {D.}~\bibnamefont {Strain}}, \bibinfo {author} {\bibfnamefont {M.}~\bibnamefont {Szalay}}, \bibinfo
  {author} {\bibfnamefont {A.}~\bibnamefont {Torres}}, \bibinfo {author} {\bibfnamefont {G.}~\bibnamefont {Vidal}}, \bibinfo {author} {\bibfnamefont {B.}~\bibnamefont {Villalonga}}, \bibinfo {author} {\bibfnamefont {C.~V.}\ \bibnamefont {Heidweiller}}, \bibinfo {author} {\bibfnamefont {T.}~\bibnamefont {White}}, \bibinfo {author} {\bibfnamefont {C.}~\bibnamefont {Xing}}, \bibinfo {author} {\bibfnamefont {Z.~J.}\ \bibnamefont {Yao}}, \bibinfo {author} {\bibfnamefont {P.}~\bibnamefont {Yeh}}, \bibinfo {author} {\bibfnamefont {J.}~\bibnamefont {Yoo}}, \bibinfo {author} {\bibfnamefont {G.}~\bibnamefont {Young}}, \bibinfo {author} {\bibfnamefont {A.}~\bibnamefont {Zalcman}}, \bibinfo {author} {\bibfnamefont {Y.}~\bibnamefont {Zhang}},\ and\ \bibinfo {author} {\bibfnamefont {N.}~\bibnamefont {Zhu}},\ }\bibfield  {title} {\bibinfo {title} {Suppressing quantum errors by scaling a surface code logical qubit},\ }\href {https://doi.org/10.1038/s41586-022-05434-1} {\bibfield  {journal} {\bibinfo  {journal} {Nature}\
  }\textbf {\bibinfo {volume} {614}},\ \bibinfo {pages} {676} (\bibinfo {year} {2023})}\BibitemShut {NoStop}%
\bibitem [{\citenamefont {Bluvstein}\ \emph {et~al.}(2024)\citenamefont {Bluvstein}, \citenamefont {Evered}, \citenamefont {Geim}, \citenamefont {Li}, \citenamefont {Zhou}, \citenamefont {Manovitz}, \citenamefont {Ebadi}, \citenamefont {Cain}, \citenamefont {Kalinowski}, \citenamefont {Hangleiter}, \citenamefont {{Bonilla Ataides}}, \citenamefont {Maskara}, \citenamefont {Cong}, \citenamefont {Gao}, \citenamefont {{Sales Rodriguez}}, \citenamefont {Karolyshyn}, \citenamefont {Semeghini}, \citenamefont {Gullans}, \citenamefont {Greiner}, \citenamefont {Vuleti{\'{c}}},\ and\ \citenamefont {Lukin}}]{bluvstein2023logical-448}%
  \BibitemOpen
  \bibfield  {author} {\bibinfo {author} {\bibfnamefont {D.}~\bibnamefont {Bluvstein}}, \bibinfo {author} {\bibfnamefont {S.~J.}\ \bibnamefont {Evered}}, \bibinfo {author} {\bibfnamefont {A.~A.}\ \bibnamefont {Geim}}, \bibinfo {author} {\bibfnamefont {S.~H.}\ \bibnamefont {Li}}, \bibinfo {author} {\bibfnamefont {H.}~\bibnamefont {Zhou}}, \bibinfo {author} {\bibfnamefont {T.}~\bibnamefont {Manovitz}}, \bibinfo {author} {\bibfnamefont {S.}~\bibnamefont {Ebadi}}, \bibinfo {author} {\bibfnamefont {M.}~\bibnamefont {Cain}}, \bibinfo {author} {\bibfnamefont {M.}~\bibnamefont {Kalinowski}}, \bibinfo {author} {\bibfnamefont {D.}~\bibnamefont {Hangleiter}}, \bibinfo {author} {\bibfnamefont {J.~P.}\ \bibnamefont {{Bonilla Ataides}}}, \bibinfo {author} {\bibfnamefont {N.}~\bibnamefont {Maskara}}, \bibinfo {author} {\bibfnamefont {I.}~\bibnamefont {Cong}}, \bibinfo {author} {\bibfnamefont {X.}~\bibnamefont {Gao}}, \bibinfo {author} {\bibfnamefont {P.}~\bibnamefont {{Sales Rodriguez}}}, \bibinfo {author} {\bibfnamefont
  {T.}~\bibnamefont {Karolyshyn}}, \bibinfo {author} {\bibfnamefont {G.}~\bibnamefont {Semeghini}}, \bibinfo {author} {\bibfnamefont {M.~J.}\ \bibnamefont {Gullans}}, \bibinfo {author} {\bibfnamefont {M.}~\bibnamefont {Greiner}}, \bibinfo {author} {\bibfnamefont {V.}~\bibnamefont {Vuleti{\'{c}}}},\ and\ \bibinfo {author} {\bibfnamefont {M.~D.}\ \bibnamefont {Lukin}},\ }\bibfield  {title} {\bibinfo {title} {{Logical quantum processor based on reconfigurable atom arrays}},\ }\href {https://doi.org/10.1038/s41586-023-06927-3} {\bibfield  {journal} {\bibinfo  {journal} {Nature}\ }\textbf {\bibinfo {volume} {626}},\ \bibinfo {pages} {58} (\bibinfo {year} {2024})}\BibitemShut {NoStop}%
\bibitem [{\citenamefont {Acharya}\ \emph {et~al.}(2025)\citenamefont {Acharya}, \citenamefont {Abanin}, \citenamefont {Aghababaie-Beni}, \citenamefont {Aleiner}, \citenamefont {Andersen}, \citenamefont {Ansmann}, \citenamefont {Arute}, \citenamefont {Arya}, \citenamefont {Asfaw}, \citenamefont {Astrakhantsev}, \citenamefont {Atalaya}, \citenamefont {Babbush}, \citenamefont {Bacon}, \citenamefont {Ballard}, \citenamefont {Bardin}, \citenamefont {Bausch}, \citenamefont {Bengtsson}, \citenamefont {Bilmes}, \citenamefont {Blackwell}, \citenamefont {Boixo}, \citenamefont {Bortoli}, \citenamefont {Bourassa}, \citenamefont {Bovaird}, \citenamefont {Brill}, \citenamefont {Broughton}, \citenamefont {Browne}, \citenamefont {Buchea}, \citenamefont {Buckley}, \citenamefont {Buell}, \citenamefont {Burger}, \citenamefont {Burkett}, \citenamefont {Bushnell}, \citenamefont {Cabrera}, \citenamefont {Campero}, \citenamefont {Chang}, \citenamefont {Chen}, \citenamefont {Chen}, \citenamefont {Chiaro}, \citenamefont {Chik},
  \citenamefont {Chou}, \citenamefont {Claes}, \citenamefont {Cleland}, \citenamefont {Cogan}, \citenamefont {Collins}, \citenamefont {Conner}, \citenamefont {Courtney}, \citenamefont {Crook}, \citenamefont {Curtin}, \citenamefont {Das}, \citenamefont {Davies}, \citenamefont {{De Lorenzo}}, \citenamefont {Debroy}, \citenamefont {Demura}, \citenamefont {Devoret}, \citenamefont {{Di Paolo}}, \citenamefont {Donohoe}, \citenamefont {Drozdov}, \citenamefont {Dunsworth}, \citenamefont {Earle}, \citenamefont {Edlich}, \citenamefont {Eickbusch}, \citenamefont {Elbag}, \citenamefont {Elzouka}, \citenamefont {Erickson}, \citenamefont {Faoro}, \citenamefont {Farhi}, \citenamefont {Ferreira}, \citenamefont {Burgos}, \citenamefont {Forati}, \citenamefont {Fowler}, \citenamefont {Foxen}, \citenamefont {Ganjam}, \citenamefont {Garcia}, \citenamefont {Gasca}, \citenamefont {Genois}, \citenamefont {Giang}, \citenamefont {Gidney}, \citenamefont {Gilboa}, \citenamefont {Gosula}, \citenamefont {Dau}, \citenamefont {Graumann},
  \citenamefont {Greene}, \citenamefont {Gross}, \citenamefont {Habegger}, \citenamefont {Hall}, \citenamefont {Hamilton}, \citenamefont {Hansen}, \citenamefont {Harrigan}, \citenamefont {Harrington}, \citenamefont {Heras}, \citenamefont {Heslin}, \citenamefont {Heu}, \citenamefont {Higgott}, \citenamefont {Hill}, \citenamefont {Hilton}, \citenamefont {Holland}, \citenamefont {Hong}, \citenamefont {Huang}, \citenamefont {Huff}, \citenamefont {Huggins}, \citenamefont {Ioffe}, \citenamefont {Isakov}, \citenamefont {Iveland}, \citenamefont {Jeffrey}, \citenamefont {Jiang}, \citenamefont {Jones}, \citenamefont {Jordan}, \citenamefont {Joshi}, \citenamefont {Juhas}, \citenamefont {Kafri}, \citenamefont {Kang}, \citenamefont {Karamlou}, \citenamefont {Kechedzhi}, \citenamefont {Kelly}, \citenamefont {Khaire}, \citenamefont {Khattar}, \citenamefont {Khezri}, \citenamefont {Kim}, \citenamefont {Klimov}, \citenamefont {Klots}, \citenamefont {Kobrin}, \citenamefont {Kohli}, \citenamefont {Korotkov}, \citenamefont
  {Kostritsa}, \citenamefont {Kothari}, \citenamefont {Kozlovskii}, \citenamefont {Kreikebaum}, \citenamefont {Kurilovich}, \citenamefont {Lacroix}, \citenamefont {Landhuis}, \citenamefont {Lange-Dei}, \citenamefont {Langley}, \citenamefont {Laptev}, \citenamefont {Lau}, \citenamefont {{Le Guevel}}, \citenamefont {Ledford}, \citenamefont {Lee}, \citenamefont {Lee}, \citenamefont {Lensky}, \citenamefont {Leon}, \citenamefont {Lester}, \citenamefont {Li}, \citenamefont {Li}, \citenamefont {Lill}, \citenamefont {Liu}, \citenamefont {Livingston}, \citenamefont {Locharla}, \citenamefont {Lucero}, \citenamefont {Lundahl}, \citenamefont {Lunt}, \citenamefont {Madhuk}, \citenamefont {Malone}, \citenamefont {Maloney}, \citenamefont {Mandr{\`{a}}}, \citenamefont {Manyika}, \citenamefont {Martin}, \citenamefont {Martin}, \citenamefont {Martin}, \citenamefont {Maxfield}, \citenamefont {McClean}, \citenamefont {McEwen}, \citenamefont {Meeks}, \citenamefont {Megrant}, \citenamefont {Mi}, \citenamefont {Miao}, \citenamefont
  {Mieszala}, \citenamefont {Molavi}, \citenamefont {Molina}, \citenamefont {Montazeri}, \citenamefont {Morvan}, \citenamefont {Movassagh}, \citenamefont {Mruczkiewicz}, \citenamefont {Naaman}, \citenamefont {Neeley}, \citenamefont {Neill}, \citenamefont {Nersisyan}, \citenamefont {Neven}, \citenamefont {Newman}, \citenamefont {Ng}, \citenamefont {Nguyen}, \citenamefont {Nguyen}, \citenamefont {Ni}, \citenamefont {Niu}, \citenamefont {O'Brien}, \citenamefont {Oliver}, \citenamefont {Opremcak}, \citenamefont {Ottosson}, \citenamefont {Petukhov}, \citenamefont {Pizzuto}, \citenamefont {Platt}, \citenamefont {Potter}, \citenamefont {Pritchard}, \citenamefont {Pryadko}, \citenamefont {Quintana}, \citenamefont {Ramachandran}, \citenamefont {Reagor}, \citenamefont {Redding}, \citenamefont {Rhodes}, \citenamefont {Roberts}, \citenamefont {Rosenberg}, \citenamefont {Rosenfeld}, \citenamefont {Roushan}, \citenamefont {Rubin}, \citenamefont {Saei}, \citenamefont {Sank}, \citenamefont {Sankaragomathi}, \citenamefont
  {Satzinger}, \citenamefont {Schurkus}, \citenamefont {Schuster}, \citenamefont {Senior}, \citenamefont {Shearn}, \citenamefont {Shorter}, \citenamefont {Shutty}, \citenamefont {Shvarts}, \citenamefont {Singh}, \citenamefont {Sivak}, \citenamefont {Skruzny}, \citenamefont {Small}, \citenamefont {Smelyanskiy}, \citenamefont {Smith}, \citenamefont {Somma}, \citenamefont {Springer}, \citenamefont {Sterling}, \citenamefont {Strain}, \citenamefont {Suchard}, \citenamefont {Szasz}, \citenamefont {Sztein}, \citenamefont {Thor}, \citenamefont {Torres}, \citenamefont {Torunbalci}, \citenamefont {Vaishnav}, \citenamefont {Vargas}, \citenamefont {Vdovichev}, \citenamefont {Vidal}, \citenamefont {Villalonga}, \citenamefont {Heidweiller}, \citenamefont {Waltman}, \citenamefont {Wang}, \citenamefont {Ware}, \citenamefont {Weber}, \citenamefont {Weidel}, \citenamefont {White}, \citenamefont {Wong}, \citenamefont {Woo}, \citenamefont {Xing}, \citenamefont {Yao}, \citenamefont {Yeh}, \citenamefont {Ying}, \citenamefont
  {Yoo}, \citenamefont {Yosri}, \citenamefont {Young}, \citenamefont {Zalcman}, \citenamefont {Zhang}, \citenamefont {Zhu},\ and\ \citenamefont {Zobrist}}]{collaborators2025quantum-3c7}%
  \BibitemOpen
  \bibfield  {author} {\bibinfo {author} {\bibfnamefont {R.}~\bibnamefont {Acharya}}, \bibinfo {author} {\bibfnamefont {D.~A.}\ \bibnamefont {Abanin}}, \bibinfo {author} {\bibfnamefont {L.}~\bibnamefont {Aghababaie-Beni}}, \bibinfo {author} {\bibfnamefont {I.}~\bibnamefont {Aleiner}}, \bibinfo {author} {\bibfnamefont {T.~I.}\ \bibnamefont {Andersen}}, \bibinfo {author} {\bibfnamefont {M.}~\bibnamefont {Ansmann}}, \bibinfo {author} {\bibfnamefont {F.}~\bibnamefont {Arute}}, \bibinfo {author} {\bibfnamefont {K.}~\bibnamefont {Arya}}, \bibinfo {author} {\bibfnamefont {A.}~\bibnamefont {Asfaw}}, \bibinfo {author} {\bibfnamefont {N.}~\bibnamefont {Astrakhantsev}}, \bibinfo {author} {\bibfnamefont {J.}~\bibnamefont {Atalaya}}, \bibinfo {author} {\bibfnamefont {R.}~\bibnamefont {Babbush}}, \bibinfo {author} {\bibfnamefont {D.}~\bibnamefont {Bacon}}, \bibinfo {author} {\bibfnamefont {B.}~\bibnamefont {Ballard}}, \bibinfo {author} {\bibfnamefont {J.~C.}\ \bibnamefont {Bardin}}, \bibinfo {author} {\bibfnamefont
  {J.}~\bibnamefont {Bausch}}, \bibinfo {author} {\bibfnamefont {A.}~\bibnamefont {Bengtsson}}, \bibinfo {author} {\bibfnamefont {A.}~\bibnamefont {Bilmes}}, \bibinfo {author} {\bibfnamefont {S.}~\bibnamefont {Blackwell}}, \bibinfo {author} {\bibfnamefont {S.}~\bibnamefont {Boixo}}, \bibinfo {author} {\bibfnamefont {G.}~\bibnamefont {Bortoli}}, \bibinfo {author} {\bibfnamefont {A.}~\bibnamefont {Bourassa}}, \bibinfo {author} {\bibfnamefont {J.}~\bibnamefont {Bovaird}}, \bibinfo {author} {\bibfnamefont {L.}~\bibnamefont {Brill}}, \bibinfo {author} {\bibfnamefont {M.}~\bibnamefont {Broughton}}, \bibinfo {author} {\bibfnamefont {D.~A.}\ \bibnamefont {Browne}}, \bibinfo {author} {\bibfnamefont {B.}~\bibnamefont {Buchea}}, \bibinfo {author} {\bibfnamefont {B.~B.}\ \bibnamefont {Buckley}}, \bibinfo {author} {\bibfnamefont {D.~A.}\ \bibnamefont {Buell}}, \bibinfo {author} {\bibfnamefont {T.}~\bibnamefont {Burger}}, \bibinfo {author} {\bibfnamefont {B.}~\bibnamefont {Burkett}}, \bibinfo {author} {\bibfnamefont
  {N.}~\bibnamefont {Bushnell}}, \bibinfo {author} {\bibfnamefont {A.}~\bibnamefont {Cabrera}}, \bibinfo {author} {\bibfnamefont {J.}~\bibnamefont {Campero}}, \bibinfo {author} {\bibfnamefont {H.-S.}\ \bibnamefont {Chang}}, \bibinfo {author} {\bibfnamefont {Y.}~\bibnamefont {Chen}}, \bibinfo {author} {\bibfnamefont {Z.}~\bibnamefont {Chen}}, \bibinfo {author} {\bibfnamefont {B.}~\bibnamefont {Chiaro}}, \bibinfo {author} {\bibfnamefont {D.}~\bibnamefont {Chik}}, \bibinfo {author} {\bibfnamefont {C.}~\bibnamefont {Chou}}, \bibinfo {author} {\bibfnamefont {J.}~\bibnamefont {Claes}}, \bibinfo {author} {\bibfnamefont {A.~Y.}\ \bibnamefont {Cleland}}, \bibinfo {author} {\bibfnamefont {J.}~\bibnamefont {Cogan}}, \bibinfo {author} {\bibfnamefont {R.}~\bibnamefont {Collins}}, \bibinfo {author} {\bibfnamefont {P.}~\bibnamefont {Conner}}, \bibinfo {author} {\bibfnamefont {W.}~\bibnamefont {Courtney}}, \bibinfo {author} {\bibfnamefont {A.~L.}\ \bibnamefont {Crook}}, \bibinfo {author} {\bibfnamefont {B.}~\bibnamefont
  {Curtin}}, \bibinfo {author} {\bibfnamefont {S.}~\bibnamefont {Das}}, \bibinfo {author} {\bibfnamefont {A.}~\bibnamefont {Davies}}, \bibinfo {author} {\bibfnamefont {L.}~\bibnamefont {{De Lorenzo}}}, \bibinfo {author} {\bibfnamefont {D.~M.}\ \bibnamefont {Debroy}}, \bibinfo {author} {\bibfnamefont {S.}~\bibnamefont {Demura}}, \bibinfo {author} {\bibfnamefont {M.}~\bibnamefont {Devoret}}, \bibinfo {author} {\bibfnamefont {A.}~\bibnamefont {{Di Paolo}}}, \bibinfo {author} {\bibfnamefont {P.}~\bibnamefont {Donohoe}}, \bibinfo {author} {\bibfnamefont {I.}~\bibnamefont {Drozdov}}, \bibinfo {author} {\bibfnamefont {A.}~\bibnamefont {Dunsworth}}, \bibinfo {author} {\bibfnamefont {C.}~\bibnamefont {Earle}}, \bibinfo {author} {\bibfnamefont {T.}~\bibnamefont {Edlich}}, \bibinfo {author} {\bibfnamefont {A.}~\bibnamefont {Eickbusch}}, \bibinfo {author} {\bibfnamefont {A.~M.}\ \bibnamefont {Elbag}}, \bibinfo {author} {\bibfnamefont {M.}~\bibnamefont {Elzouka}}, \bibinfo {author} {\bibfnamefont {C.}~\bibnamefont
  {Erickson}}, \bibinfo {author} {\bibfnamefont {L.}~\bibnamefont {Faoro}}, \bibinfo {author} {\bibfnamefont {E.}~\bibnamefont {Farhi}}, \bibinfo {author} {\bibfnamefont {V.~S.}\ \bibnamefont {Ferreira}}, \bibinfo {author} {\bibfnamefont {L.~F.}\ \bibnamefont {Burgos}}, \bibinfo {author} {\bibfnamefont {E.}~\bibnamefont {Forati}}, \bibinfo {author} {\bibfnamefont {A.~G.}\ \bibnamefont {Fowler}}, \bibinfo {author} {\bibfnamefont {B.}~\bibnamefont {Foxen}}, \bibinfo {author} {\bibfnamefont {S.}~\bibnamefont {Ganjam}}, \bibinfo {author} {\bibfnamefont {G.}~\bibnamefont {Garcia}}, \bibinfo {author} {\bibfnamefont {R.}~\bibnamefont {Gasca}}, \bibinfo {author} {\bibfnamefont {{\'{E}}.}~\bibnamefont {Genois}}, \bibinfo {author} {\bibfnamefont {W.}~\bibnamefont {Giang}}, \bibinfo {author} {\bibfnamefont {C.}~\bibnamefont {Gidney}}, \bibinfo {author} {\bibfnamefont {D.}~\bibnamefont {Gilboa}}, \bibinfo {author} {\bibfnamefont {R.}~\bibnamefont {Gosula}}, \bibinfo {author} {\bibfnamefont {A.~G.}\ \bibnamefont {Dau}},
  \bibinfo {author} {\bibfnamefont {D.}~\bibnamefont {Graumann}}, \bibinfo {author} {\bibfnamefont {A.}~\bibnamefont {Greene}}, \bibinfo {author} {\bibfnamefont {J.~A.}\ \bibnamefont {Gross}}, \bibinfo {author} {\bibfnamefont {S.}~\bibnamefont {Habegger}}, \bibinfo {author} {\bibfnamefont {J.}~\bibnamefont {Hall}}, \bibinfo {author} {\bibfnamefont {M.~C.}\ \bibnamefont {Hamilton}}, \bibinfo {author} {\bibfnamefont {M.}~\bibnamefont {Hansen}}, \bibinfo {author} {\bibfnamefont {M.~P.}\ \bibnamefont {Harrigan}}, \bibinfo {author} {\bibfnamefont {S.~D.}\ \bibnamefont {Harrington}}, \bibinfo {author} {\bibfnamefont {F.~J.~H.}\ \bibnamefont {Heras}}, \bibinfo {author} {\bibfnamefont {S.}~\bibnamefont {Heslin}}, \bibinfo {author} {\bibfnamefont {P.}~\bibnamefont {Heu}}, \bibinfo {author} {\bibfnamefont {O.}~\bibnamefont {Higgott}}, \bibinfo {author} {\bibfnamefont {G.}~\bibnamefont {Hill}}, \bibinfo {author} {\bibfnamefont {J.}~\bibnamefont {Hilton}}, \bibinfo {author} {\bibfnamefont {G.}~\bibnamefont {Holland}},
  \bibinfo {author} {\bibfnamefont {S.}~\bibnamefont {Hong}}, \bibinfo {author} {\bibfnamefont {H.-Y.}\ \bibnamefont {Huang}}, \bibinfo {author} {\bibfnamefont {A.}~\bibnamefont {Huff}}, \bibinfo {author} {\bibfnamefont {W.~J.}\ \bibnamefont {Huggins}}, \bibinfo {author} {\bibfnamefont {L.~B.}\ \bibnamefont {Ioffe}}, \bibinfo {author} {\bibfnamefont {S.~V.}\ \bibnamefont {Isakov}}, \bibinfo {author} {\bibfnamefont {J.}~\bibnamefont {Iveland}}, \bibinfo {author} {\bibfnamefont {E.}~\bibnamefont {Jeffrey}}, \bibinfo {author} {\bibfnamefont {Z.}~\bibnamefont {Jiang}}, \bibinfo {author} {\bibfnamefont {C.}~\bibnamefont {Jones}}, \bibinfo {author} {\bibfnamefont {S.}~\bibnamefont {Jordan}}, \bibinfo {author} {\bibfnamefont {C.}~\bibnamefont {Joshi}}, \bibinfo {author} {\bibfnamefont {P.}~\bibnamefont {Juhas}}, \bibinfo {author} {\bibfnamefont {D.}~\bibnamefont {Kafri}}, \bibinfo {author} {\bibfnamefont {H.}~\bibnamefont {Kang}}, \bibinfo {author} {\bibfnamefont {A.~H.}\ \bibnamefont {Karamlou}}, \bibinfo {author}
  {\bibfnamefont {K.}~\bibnamefont {Kechedzhi}}, \bibinfo {author} {\bibfnamefont {J.}~\bibnamefont {Kelly}}, \bibinfo {author} {\bibfnamefont {T.}~\bibnamefont {Khaire}}, \bibinfo {author} {\bibfnamefont {T.}~\bibnamefont {Khattar}}, \bibinfo {author} {\bibfnamefont {M.}~\bibnamefont {Khezri}}, \bibinfo {author} {\bibfnamefont {S.}~\bibnamefont {Kim}}, \bibinfo {author} {\bibfnamefont {P.~V.}\ \bibnamefont {Klimov}}, \bibinfo {author} {\bibfnamefont {A.~R.}\ \bibnamefont {Klots}}, \bibinfo {author} {\bibfnamefont {B.}~\bibnamefont {Kobrin}}, \bibinfo {author} {\bibfnamefont {P.}~\bibnamefont {Kohli}}, \bibinfo {author} {\bibfnamefont {A.~N.}\ \bibnamefont {Korotkov}}, \bibinfo {author} {\bibfnamefont {F.}~\bibnamefont {Kostritsa}}, \bibinfo {author} {\bibfnamefont {R.}~\bibnamefont {Kothari}}, \bibinfo {author} {\bibfnamefont {B.}~\bibnamefont {Kozlovskii}}, \bibinfo {author} {\bibfnamefont {J.~M.}\ \bibnamefont {Kreikebaum}}, \bibinfo {author} {\bibfnamefont {V.~D.}\ \bibnamefont {Kurilovich}}, \bibinfo
  {author} {\bibfnamefont {N.}~\bibnamefont {Lacroix}}, \bibinfo {author} {\bibfnamefont {D.}~\bibnamefont {Landhuis}}, \bibinfo {author} {\bibfnamefont {T.}~\bibnamefont {Lange-Dei}}, \bibinfo {author} {\bibfnamefont {B.~W.}\ \bibnamefont {Langley}}, \bibinfo {author} {\bibfnamefont {P.}~\bibnamefont {Laptev}}, \bibinfo {author} {\bibfnamefont {K.-M.}\ \bibnamefont {Lau}}, \bibinfo {author} {\bibfnamefont {L.}~\bibnamefont {{Le Guevel}}}, \bibinfo {author} {\bibfnamefont {J.}~\bibnamefont {Ledford}}, \bibinfo {author} {\bibfnamefont {J.}~\bibnamefont {Lee}}, \bibinfo {author} {\bibfnamefont {K.}~\bibnamefont {Lee}}, \bibinfo {author} {\bibfnamefont {Y.~D.}\ \bibnamefont {Lensky}}, \bibinfo {author} {\bibfnamefont {S.}~\bibnamefont {Leon}}, \bibinfo {author} {\bibfnamefont {B.~J.}\ \bibnamefont {Lester}}, \bibinfo {author} {\bibfnamefont {W.~Y.}\ \bibnamefont {Li}}, \bibinfo {author} {\bibfnamefont {Y.}~\bibnamefont {Li}}, \bibinfo {author} {\bibfnamefont {A.~T.}\ \bibnamefont {Lill}}, \bibinfo {author}
  {\bibfnamefont {W.}~\bibnamefont {Liu}}, \bibinfo {author} {\bibfnamefont {W.~P.}\ \bibnamefont {Livingston}}, \bibinfo {author} {\bibfnamefont {A.}~\bibnamefont {Locharla}}, \bibinfo {author} {\bibfnamefont {E.}~\bibnamefont {Lucero}}, \bibinfo {author} {\bibfnamefont {D.}~\bibnamefont {Lundahl}}, \bibinfo {author} {\bibfnamefont {A.}~\bibnamefont {Lunt}}, \bibinfo {author} {\bibfnamefont {S.}~\bibnamefont {Madhuk}}, \bibinfo {author} {\bibfnamefont {F.~D.}\ \bibnamefont {Malone}}, \bibinfo {author} {\bibfnamefont {A.}~\bibnamefont {Maloney}}, \bibinfo {author} {\bibfnamefont {S.}~\bibnamefont {Mandr{\`{a}}}}, \bibinfo {author} {\bibfnamefont {J.}~\bibnamefont {Manyika}}, \bibinfo {author} {\bibfnamefont {L.~S.}\ \bibnamefont {Martin}}, \bibinfo {author} {\bibfnamefont {O.}~\bibnamefont {Martin}}, \bibinfo {author} {\bibfnamefont {S.}~\bibnamefont {Martin}}, \bibinfo {author} {\bibfnamefont {C.}~\bibnamefont {Maxfield}}, \bibinfo {author} {\bibfnamefont {J.~R.}\ \bibnamefont {McClean}}, \bibinfo {author}
  {\bibfnamefont {M.}~\bibnamefont {McEwen}}, \bibinfo {author} {\bibfnamefont {S.}~\bibnamefont {Meeks}}, \bibinfo {author} {\bibfnamefont {A.}~\bibnamefont {Megrant}}, \bibinfo {author} {\bibfnamefont {X.}~\bibnamefont {Mi}}, \bibinfo {author} {\bibfnamefont {K.~C.}\ \bibnamefont {Miao}}, \bibinfo {author} {\bibfnamefont {A.}~\bibnamefont {Mieszala}}, \bibinfo {author} {\bibfnamefont {R.}~\bibnamefont {Molavi}}, \bibinfo {author} {\bibfnamefont {S.}~\bibnamefont {Molina}}, \bibinfo {author} {\bibfnamefont {S.}~\bibnamefont {Montazeri}}, \bibinfo {author} {\bibfnamefont {A.}~\bibnamefont {Morvan}}, \bibinfo {author} {\bibfnamefont {R.}~\bibnamefont {Movassagh}}, \bibinfo {author} {\bibfnamefont {W.}~\bibnamefont {Mruczkiewicz}}, \bibinfo {author} {\bibfnamefont {O.}~\bibnamefont {Naaman}}, \bibinfo {author} {\bibfnamefont {M.}~\bibnamefont {Neeley}}, \bibinfo {author} {\bibfnamefont {C.}~\bibnamefont {Neill}}, \bibinfo {author} {\bibfnamefont {A.}~\bibnamefont {Nersisyan}}, \bibinfo {author} {\bibfnamefont
  {H.}~\bibnamefont {Neven}}, \bibinfo {author} {\bibfnamefont {M.}~\bibnamefont {Newman}}, \bibinfo {author} {\bibfnamefont {J.~H.}\ \bibnamefont {Ng}}, \bibinfo {author} {\bibfnamefont {A.}~\bibnamefont {Nguyen}}, \bibinfo {author} {\bibfnamefont {M.}~\bibnamefont {Nguyen}}, \bibinfo {author} {\bibfnamefont {C.-H.}\ \bibnamefont {Ni}}, \bibinfo {author} {\bibfnamefont {M.~Y.}\ \bibnamefont {Niu}}, \bibinfo {author} {\bibfnamefont {T.~E.}\ \bibnamefont {O'Brien}}, \bibinfo {author} {\bibfnamefont {W.~D.}\ \bibnamefont {Oliver}}, \bibinfo {author} {\bibfnamefont {A.}~\bibnamefont {Opremcak}}, \bibinfo {author} {\bibfnamefont {K.}~\bibnamefont {Ottosson}}, \bibinfo {author} {\bibfnamefont {A.}~\bibnamefont {Petukhov}}, \bibinfo {author} {\bibfnamefont {A.}~\bibnamefont {Pizzuto}}, \bibinfo {author} {\bibfnamefont {J.}~\bibnamefont {Platt}}, \bibinfo {author} {\bibfnamefont {R.}~\bibnamefont {Potter}}, \bibinfo {author} {\bibfnamefont {O.}~\bibnamefont {Pritchard}}, \bibinfo {author} {\bibfnamefont {L.~P.}\
  \bibnamefont {Pryadko}}, \bibinfo {author} {\bibfnamefont {C.}~\bibnamefont {Quintana}}, \bibinfo {author} {\bibfnamefont {G.}~\bibnamefont {Ramachandran}}, \bibinfo {author} {\bibfnamefont {M.~J.}\ \bibnamefont {Reagor}}, \bibinfo {author} {\bibfnamefont {J.}~\bibnamefont {Redding}}, \bibinfo {author} {\bibfnamefont {D.~M.}\ \bibnamefont {Rhodes}}, \bibinfo {author} {\bibfnamefont {G.}~\bibnamefont {Roberts}}, \bibinfo {author} {\bibfnamefont {E.}~\bibnamefont {Rosenberg}}, \bibinfo {author} {\bibfnamefont {E.}~\bibnamefont {Rosenfeld}}, \bibinfo {author} {\bibfnamefont {P.}~\bibnamefont {Roushan}}, \bibinfo {author} {\bibfnamefont {N.~C.}\ \bibnamefont {Rubin}}, \bibinfo {author} {\bibfnamefont {N.}~\bibnamefont {Saei}}, \bibinfo {author} {\bibfnamefont {D.}~\bibnamefont {Sank}}, \bibinfo {author} {\bibfnamefont {K.}~\bibnamefont {Sankaragomathi}}, \bibinfo {author} {\bibfnamefont {K.~J.}\ \bibnamefont {Satzinger}}, \bibinfo {author} {\bibfnamefont {H.~F.}\ \bibnamefont {Schurkus}}, \bibinfo {author}
  {\bibfnamefont {C.}~\bibnamefont {Schuster}}, \bibinfo {author} {\bibfnamefont {A.~W.}\ \bibnamefont {Senior}}, \bibinfo {author} {\bibfnamefont {M.~J.}\ \bibnamefont {Shearn}}, \bibinfo {author} {\bibfnamefont {A.}~\bibnamefont {Shorter}}, \bibinfo {author} {\bibfnamefont {N.}~\bibnamefont {Shutty}}, \bibinfo {author} {\bibfnamefont {V.}~\bibnamefont {Shvarts}}, \bibinfo {author} {\bibfnamefont {S.}~\bibnamefont {Singh}}, \bibinfo {author} {\bibfnamefont {V.}~\bibnamefont {Sivak}}, \bibinfo {author} {\bibfnamefont {J.}~\bibnamefont {Skruzny}}, \bibinfo {author} {\bibfnamefont {S.}~\bibnamefont {Small}}, \bibinfo {author} {\bibfnamefont {V.}~\bibnamefont {Smelyanskiy}}, \bibinfo {author} {\bibfnamefont {W.~C.}\ \bibnamefont {Smith}}, \bibinfo {author} {\bibfnamefont {R.~D.}\ \bibnamefont {Somma}}, \bibinfo {author} {\bibfnamefont {S.}~\bibnamefont {Springer}}, \bibinfo {author} {\bibfnamefont {G.}~\bibnamefont {Sterling}}, \bibinfo {author} {\bibfnamefont {D.}~\bibnamefont {Strain}}, \bibinfo {author}
  {\bibfnamefont {J.}~\bibnamefont {Suchard}}, \bibinfo {author} {\bibfnamefont {A.}~\bibnamefont {Szasz}}, \bibinfo {author} {\bibfnamefont {A.}~\bibnamefont {Sztein}}, \bibinfo {author} {\bibfnamefont {D.}~\bibnamefont {Thor}}, \bibinfo {author} {\bibfnamefont {A.}~\bibnamefont {Torres}}, \bibinfo {author} {\bibfnamefont {M.~M.}\ \bibnamefont {Torunbalci}}, \bibinfo {author} {\bibfnamefont {A.}~\bibnamefont {Vaishnav}}, \bibinfo {author} {\bibfnamefont {J.}~\bibnamefont {Vargas}}, \bibinfo {author} {\bibfnamefont {S.}~\bibnamefont {Vdovichev}}, \bibinfo {author} {\bibfnamefont {G.}~\bibnamefont {Vidal}}, \bibinfo {author} {\bibfnamefont {B.}~\bibnamefont {Villalonga}}, \bibinfo {author} {\bibfnamefont {C.~V.}\ \bibnamefont {Heidweiller}}, \bibinfo {author} {\bibfnamefont {S.}~\bibnamefont {Waltman}}, \bibinfo {author} {\bibfnamefont {S.~X.}\ \bibnamefont {Wang}}, \bibinfo {author} {\bibfnamefont {B.}~\bibnamefont {Ware}}, \bibinfo {author} {\bibfnamefont {K.}~\bibnamefont {Weber}}, \bibinfo {author}
  {\bibfnamefont {T.}~\bibnamefont {Weidel}}, \bibinfo {author} {\bibfnamefont {T.}~\bibnamefont {White}}, \bibinfo {author} {\bibfnamefont {K.}~\bibnamefont {Wong}}, \bibinfo {author} {\bibfnamefont {B.~W.~K.}\ \bibnamefont {Woo}}, \bibinfo {author} {\bibfnamefont {C.}~\bibnamefont {Xing}}, \bibinfo {author} {\bibfnamefont {Z.~J.}\ \bibnamefont {Yao}}, \bibinfo {author} {\bibfnamefont {P.}~\bibnamefont {Yeh}}, \bibinfo {author} {\bibfnamefont {B.}~\bibnamefont {Ying}}, \bibinfo {author} {\bibfnamefont {J.}~\bibnamefont {Yoo}}, \bibinfo {author} {\bibfnamefont {N.}~\bibnamefont {Yosri}}, \bibinfo {author} {\bibfnamefont {G.}~\bibnamefont {Young}}, \bibinfo {author} {\bibfnamefont {A.}~\bibnamefont {Zalcman}}, \bibinfo {author} {\bibfnamefont {Y.}~\bibnamefont {Zhang}}, \bibinfo {author} {\bibfnamefont {N.}~\bibnamefont {Zhu}},\ and\ \bibinfo {author} {\bibfnamefont {N.}~\bibnamefont {Zobrist}},\ }\bibfield  {title} {\bibinfo {title} {{Quantum error correction below the surface code threshold}},\ }\href
  {https://doi.org/10.1038/s41586-024-08449-y} {\bibfield  {journal} {\bibinfo  {journal} {Nature}\ }\textbf {\bibinfo {volume} {638}},\ \bibinfo {pages} {920} (\bibinfo {year} {2025})}\BibitemShut {NoStop}%
\bibitem [{\citenamefont {Shor}(1996)}]{shor1996fault}%
  \BibitemOpen
  \bibfield  {author} {\bibinfo {author} {\bibfnamefont {P.~W.}\ \bibnamefont {Shor}},\ }\bibfield  {title} {\bibinfo {title} {Fault-tolerant quantum computation},\ }in\ \href@noop {} {\emph {\bibinfo {booktitle} {Proceedings of 37th conference on foundations of computer science}}}\ (\bibinfo {organization} {IEEE},\ \bibinfo {year} {1996})\ pp.\ \bibinfo {pages} {56--65}\BibitemShut {NoStop}%
\bibitem [{\citenamefont {Knill}\ and\ \citenamefont {Laflamme}(1996)}]{Knill1996}%
  \BibitemOpen
  \bibfield  {author} {\bibinfo {author} {\bibfnamefont {E.}~\bibnamefont {Knill}}\ and\ \bibinfo {author} {\bibfnamefont {R.}~\bibnamefont {Laflamme}},\ }\href {https://arxiv.org/abs/quant-ph/9608012} {\bibinfo {title} {Concatenated quantum codes}} (\bibinfo {year} {1996}),\ \Eprint {https://arxiv.org/abs/quant-ph/9608012} {arXiv:quant-ph/9608012 [quant-ph]} \BibitemShut {NoStop}%
\bibitem [{\citenamefont {Aharonov}\ and\ \citenamefont {Ben-Or}(1997)}]{aharonov1997}%
  \BibitemOpen
  \bibfield  {author} {\bibinfo {author} {\bibfnamefont {D.}~\bibnamefont {Aharonov}}\ and\ \bibinfo {author} {\bibfnamefont {M.}~\bibnamefont {Ben-Or}},\ }\bibfield  {title} {\bibinfo {title} {Fault-tolerant quantum computation with constant error},\ }in\ \href {https://doi.org/10.48550/arXiv.quant-ph/9611025} {\emph {\bibinfo {booktitle} {Proceedings of the twenty-ninth annual ACM symposium on Theory of computing}}}\ (\bibinfo {year} {1997})\ pp.\ \bibinfo {pages} {176--188}\BibitemShut {NoStop}%
\bibitem [{\citenamefont {Kitaev}(1997)}]{Kitaev1997}%
  \BibitemOpen
  \bibfield  {author} {\bibinfo {author} {\bibfnamefont {A.~Y.}\ \bibnamefont {Kitaev}},\ }\bibfield  {title} {\bibinfo {title} {{Quantum computations: algorithms and error correction}},\ }\href {https://doi.org/10.1070/RM1997v052n06ABEH002155} {\bibfield  {journal} {\bibinfo  {journal} {Russian Mathematical Surveys}\ }\textbf {\bibinfo {volume} {52}},\ \bibinfo {pages} {1191} (\bibinfo {year} {1997})}\BibitemShut {NoStop}%
\bibitem [{\citenamefont {Knill}\ \emph {et~al.}(1998)\citenamefont {Knill}, \citenamefont {Laflamme},\ and\ \citenamefont {Zurek}}]{Knill1998}%
  \BibitemOpen
  \bibfield  {author} {\bibinfo {author} {\bibfnamefont {E.}~\bibnamefont {Knill}}, \bibinfo {author} {\bibfnamefont {R.}~\bibnamefont {Laflamme}},\ and\ \bibinfo {author} {\bibfnamefont {W.~H.}\ \bibnamefont {Zurek}},\ }\bibfield  {title} {\bibinfo {title} {Resilient quantum computation},\ }\href {https://doi.org/10.1126/science.279.5349.342} {\bibfield  {journal} {\bibinfo  {journal} {Science}\ }\textbf {\bibinfo {volume} {279}},\ \bibinfo {pages} {342} (\bibinfo {year} {1998})}\BibitemShut {NoStop}%
\bibitem [{\citenamefont {Reichardt}(2005{\natexlab{a}})}]{reichardt2005}%
  \BibitemOpen
  \bibfield  {author} {\bibinfo {author} {\bibfnamefont {B.~W.}\ \bibnamefont {Reichardt}},\ }\href {https://arxiv.org/abs/quant-ph/0509203} {\bibinfo {title} {Fault-tolerance threshold for a distance-three quantum code}} (\bibinfo {year} {2005}{\natexlab{a}}),\ \Eprint {https://arxiv.org/abs/quant-ph/0509203} {arXiv:quant-ph/0509203 [quant-ph]} \BibitemShut {NoStop}%
\bibitem [{\citenamefont {Aliferis}\ \emph {et~al.}(2006)\citenamefont {Aliferis}, \citenamefont {Gottesman},\ and\ \citenamefont {Preskill}}]{Aliferis2006}%
  \BibitemOpen
  \bibfield  {author} {\bibinfo {author} {\bibfnamefont {P.}~\bibnamefont {Aliferis}}, \bibinfo {author} {\bibfnamefont {D.}~\bibnamefont {Gottesman}},\ and\ \bibinfo {author} {\bibfnamefont {J.}~\bibnamefont {Preskill}},\ }\bibfield  {title} {\bibinfo {title} {{Quantum accuracy threshold for concatenated distance-3 code}},\ }\href {https://doi.org/10.26421/QIC6.2-1} {\bibfield  {journal} {\bibinfo  {journal} {Quantum Information and Computation}\ }\textbf {\bibinfo {volume} {6}},\ \bibinfo {pages} {97} (\bibinfo {year} {2006})}\BibitemShut {NoStop}%
\bibitem [{\citenamefont {Poulin}(2006)}]{Poulin2006}%
  \BibitemOpen
  \bibfield  {author} {\bibinfo {author} {\bibfnamefont {D.}~\bibnamefont {Poulin}},\ }\bibfield  {title} {\bibinfo {title} {Optimal and efficient decoding of concatenated quantum block codes},\ }\href {https://doi.org/10.1103/PhysRevA.74.052333} {\bibfield  {journal} {\bibinfo  {journal} {Phys. Rev. A}\ }\textbf {\bibinfo {volume} {74}},\ \bibinfo {pages} {052333} (\bibinfo {year} {2006})}\BibitemShut {NoStop}%
\bibitem [{\citenamefont {Evans}\ and\ \citenamefont {Stephens}(2012)}]{Evans2012}%
  \BibitemOpen
  \bibfield  {author} {\bibinfo {author} {\bibfnamefont {Z.~W.~E.}\ \bibnamefont {Evans}}\ and\ \bibinfo {author} {\bibfnamefont {A.~M.}\ \bibnamefont {Stephens}},\ }\bibfield  {title} {\bibinfo {title} {{Optimal correction of concatenated fault-tolerant quantum codes}},\ }\href {https://doi.org/10.1007/s11128-011-0312-4} {\bibfield  {journal} {\bibinfo  {journal} {Quantum Information Processing}\ }\textbf {\bibinfo {volume} {11}},\ \bibinfo {pages} {1511} (\bibinfo {year} {2012})}\BibitemShut {NoStop}%
\bibitem [{\citenamefont {Yadavalli}\ and\ \citenamefont {Marvian}(2024)}]{Yadavalli2024}%
  \BibitemOpen
  \bibfield  {author} {\bibinfo {author} {\bibfnamefont {S.~A.}\ \bibnamefont {Yadavalli}}\ and\ \bibinfo {author} {\bibfnamefont {I.}~\bibnamefont {Marvian}},\ }\href {https://arxiv.org/abs/2306.14294} {\bibinfo {title} {Noisy quantum trees: Infinite protection without correction}} (\bibinfo {year} {2024}),\ \Eprint {https://arxiv.org/abs/2306.14294} {arXiv:2306.14294 [quant-ph]} \BibitemShut {NoStop}%
\bibitem [{\citenamefont {Pato}\ \emph {et~al.}(2024)\citenamefont {Pato}, \citenamefont {Tansuwannont},\ and\ \citenamefont {Brown}}]{Pato2024}%
  \BibitemOpen
  \bibfield  {author} {\bibinfo {author} {\bibfnamefont {B.}~\bibnamefont {Pato}}, \bibinfo {author} {\bibfnamefont {T.}~\bibnamefont {Tansuwannont}},\ and\ \bibinfo {author} {\bibfnamefont {K.~R.}\ \bibnamefont {Brown}},\ }\bibfield  {title} {\bibinfo {title} {Concatenated steane code with single-flag syndrome checks},\ }\href {https://doi.org/10.1103/PhysRevA.110.032411} {\bibfield  {journal} {\bibinfo  {journal} {Phys. Rev. A}\ }\textbf {\bibinfo {volume} {110}},\ \bibinfo {pages} {032411} (\bibinfo {year} {2024})}\BibitemShut {NoStop}%
\bibitem [{\citenamefont {Reichardt}(2004)}]{Reichardt2004}%
  \BibitemOpen
  \bibfield  {author} {\bibinfo {author} {\bibfnamefont {B.~W.}\ \bibnamefont {Reichardt}},\ }\href {http://arxiv.org/abs/quant-ph/0406025} {\bibinfo {title} {{Improved ancilla preparation scheme increases fault-tolerant threshold}}} (\bibinfo {year} {2004}),\ \Eprint {https://arxiv.org/abs/0406025} {arXiv:0406025 [quant-ph]} \BibitemShut {NoStop}%
\bibitem [{\citenamefont {Steane}(2004)}]{Steane2004}%
  \BibitemOpen
  \bibfield  {author} {\bibinfo {author} {\bibfnamefont {A.~M.}\ \bibnamefont {Steane}},\ }\href {http://arxiv.org/abs/quant-ph/0202036} {\bibinfo {title} {{Fast fault-tolerant filtering of quantum codewords}}} (\bibinfo {year} {2004}),\ \Eprint {https://arxiv.org/abs/0202036} {arXiv:0202036 [quant-ph]} \BibitemShut {NoStop}%
\bibitem [{\citenamefont {Paetznick}\ and\ \citenamefont {Reichardt}(2013{\natexlab{a}})}]{Paetznick2013}%
  \BibitemOpen
  \bibfield  {author} {\bibinfo {author} {\bibfnamefont {A.}~\bibnamefont {Paetznick}}\ and\ \bibinfo {author} {\bibfnamefont {B.~W.}\ \bibnamefont {Reichardt}},\ }\bibfield  {title} {\bibinfo {title} {{Fault-tolerant ancilla preparation and noise threshold lower bounds for the 23-qubit Golay code}},\ }\href {https://doi.org/10.26421/QIC12.11-12-10} {\bibfield  {journal} {\bibinfo  {journal} {Quantum Information and Computation}\ }\textbf {\bibinfo {volume} {12}},\ \bibinfo {pages} {1034} (\bibinfo {year} {2013}{\natexlab{a}})}\BibitemShut {NoStop}%
\bibitem [{\citenamefont {Steane}(2003)}]{Steane2003}%
  \BibitemOpen
  \bibfield  {author} {\bibinfo {author} {\bibfnamefont {A.~M.}\ \bibnamefont {Steane}},\ }\bibfield  {title} {\bibinfo {title} {{Overhead and noise threshold of fault-tolerant quantum error correction}},\ }\href {https://doi.org/10.1103/PhysRevA.68.042322} {\bibfield  {journal} {\bibinfo  {journal} {Physical Review A}\ }\textbf {\bibinfo {volume} {68}},\ \bibinfo {pages} {042322} (\bibinfo {year} {2003})}\BibitemShut {NoStop}%
\bibitem [{\citenamefont {Knill}(2005)}]{Knill2005}%
  \BibitemOpen
  \bibfield  {author} {\bibinfo {author} {\bibfnamefont {E.}~\bibnamefont {Knill}},\ }\bibfield  {title} {\bibinfo {title} {{Quantum computing with realistically noisy devices}},\ }\href {https://doi.org/10.1038/nature03350} {\bibfield  {journal} {\bibinfo  {journal} {Nature}\ }\textbf {\bibinfo {volume} {434}},\ \bibinfo {pages} {39} (\bibinfo {year} {2005})}\BibitemShut {NoStop}%
\bibitem [{\citenamefont {Cross}\ \emph {et~al.}(2009)\citenamefont {Cross}, \citenamefont {DiVincenzo},\ and\ \citenamefont {Terhal}}]{Cross2009}%
  \BibitemOpen
  \bibfield  {author} {\bibinfo {author} {\bibfnamefont {A.}~\bibnamefont {Cross}}, \bibinfo {author} {\bibfnamefont {D.}~\bibnamefont {DiVincenzo}},\ and\ \bibinfo {author} {\bibfnamefont {B.}~\bibnamefont {Terhal}},\ }\bibfield  {title} {\bibinfo {title} {{A comparative code study for quantum fault tolerance}},\ }\href {https://doi.org/10.26421/QIC9.7-8-1} {\bibfield  {journal} {\bibinfo  {journal} {Quantum Information and Computation}\ }\textbf {\bibinfo {volume} {9}},\ \bibinfo {pages} {541} (\bibinfo {year} {2009})}\BibitemShut {NoStop}%
\bibitem [{\citenamefont {Suchara}\ \emph {et~al.}(2013)\citenamefont {Suchara}, \citenamefont {Kubiatowicz}, \citenamefont {Faruque}, \citenamefont {Chong}, \citenamefont {Lai},\ and\ \citenamefont {Paz}}]{Suchara2013}%
  \BibitemOpen
  \bibfield  {author} {\bibinfo {author} {\bibfnamefont {M.}~\bibnamefont {Suchara}}, \bibinfo {author} {\bibfnamefont {J.}~\bibnamefont {Kubiatowicz}}, \bibinfo {author} {\bibfnamefont {A.}~\bibnamefont {Faruque}}, \bibinfo {author} {\bibfnamefont {F.~T.}\ \bibnamefont {Chong}}, \bibinfo {author} {\bibfnamefont {C.-Y.}\ \bibnamefont {Lai}},\ and\ \bibinfo {author} {\bibfnamefont {G.}~\bibnamefont {Paz}},\ }\bibfield  {title} {\bibinfo {title} {{QuRE: The Quantum Resource Estimator toolbox}},\ }in\ \href {https://doi.org/10.1109/ICCD.2013.6657074} {\emph {\bibinfo {booktitle} {2013 IEEE 31st International Conference on Computer Design (ICCD)}}}\ (\bibinfo  {publisher} {IEEE},\ \bibinfo {year} {2013})\ pp.\ \bibinfo {pages} {419--426}\BibitemShut {NoStop}%
\bibitem [{\citenamefont {Chamberland}\ \emph {et~al.}(2017)\citenamefont {Chamberland}, \citenamefont {Jochym-O'Connor},\ and\ \citenamefont {Laflamme}}]{Chamberland2017}%
  \BibitemOpen
  \bibfield  {author} {\bibinfo {author} {\bibfnamefont {C.}~\bibnamefont {Chamberland}}, \bibinfo {author} {\bibfnamefont {T.}~\bibnamefont {Jochym-O'Connor}},\ and\ \bibinfo {author} {\bibfnamefont {R.}~\bibnamefont {Laflamme}},\ }\bibfield  {title} {\bibinfo {title} {{Overhead analysis of universal concatenated quantum codes}},\ }\href {https://doi.org/10.1103/PhysRevA.95.022313} {\bibfield  {journal} {\bibinfo  {journal} {Physical Review A}\ }\textbf {\bibinfo {volume} {95}},\ \bibinfo {pages} {1} (\bibinfo {year} {2017})}\BibitemShut {NoStop}%
\bibitem [{\citenamefont {Jochym-O'Connor}\ and\ \citenamefont {Laflamme}(2014)}]{Jochym-OConnor2014}%
  \BibitemOpen
  \bibfield  {author} {\bibinfo {author} {\bibfnamefont {T.}~\bibnamefont {Jochym-O'Connor}}\ and\ \bibinfo {author} {\bibfnamefont {R.}~\bibnamefont {Laflamme}},\ }\bibfield  {title} {\bibinfo {title} {{Using Concatenated Quantum Codes for Universal Fault-Tolerant Quantum Gates}},\ }\href {https://doi.org/10.1103/PhysRevLett.112.010505} {\bibfield  {journal} {\bibinfo  {journal} {Physical Review Letters}\ }\textbf {\bibinfo {volume} {112}},\ \bibinfo {pages} {010505} (\bibinfo {year} {2014})}\BibitemShut {NoStop}%
\bibitem [{\citenamefont {Chamberland}\ \emph {et~al.}(2016)\citenamefont {Chamberland}, \citenamefont {Jochym-O'Connor},\ and\ \citenamefont {Laflamme}}]{Chamberland2016}%
  \BibitemOpen
  \bibfield  {author} {\bibinfo {author} {\bibfnamefont {C.}~\bibnamefont {Chamberland}}, \bibinfo {author} {\bibfnamefont {T.}~\bibnamefont {Jochym-O'Connor}},\ and\ \bibinfo {author} {\bibfnamefont {R.}~\bibnamefont {Laflamme}},\ }\bibfield  {title} {\bibinfo {title} {{Thresholds for Universal Concatenated Quantum Codes}},\ }\href {https://doi.org/10.1103/PhysRevLett.117.010501} {\bibfield  {journal} {\bibinfo  {journal} {Physical Review Letters}\ }\textbf {\bibinfo {volume} {117}},\ \bibinfo {pages} {010501} (\bibinfo {year} {2016})}\BibitemShut {NoStop}%
\bibitem [{\citenamefont {Yoder}\ \emph {et~al.}(2016)\citenamefont {Yoder}, \citenamefont {Takagi},\ and\ \citenamefont {Chuang}}]{Yoder2016}%
  \BibitemOpen
  \bibfield  {author} {\bibinfo {author} {\bibfnamefont {T.~J.}\ \bibnamefont {Yoder}}, \bibinfo {author} {\bibfnamefont {R.}~\bibnamefont {Takagi}},\ and\ \bibinfo {author} {\bibfnamefont {I.~L.}\ \bibnamefont {Chuang}},\ }\bibfield  {title} {\bibinfo {title} {{Universal Fault-Tolerant Gates on Concatenated Stabilizer Codes}},\ }\href {https://doi.org/10.1103/PhysRevX.6.031039} {\bibfield  {journal} {\bibinfo  {journal} {Physical Review X}\ }\textbf {\bibinfo {volume} {6}},\ \bibinfo {pages} {031039} (\bibinfo {year} {2016})}\BibitemShut {NoStop}%
\bibitem [{\citenamefont {Nikahd}\ \emph {et~al.}(2017)\citenamefont {Nikahd}, \citenamefont {Sedighi},\ and\ \citenamefont {{Saheb Zamani}}}]{Nikahd2017}%
  \BibitemOpen
  \bibfield  {author} {\bibinfo {author} {\bibfnamefont {E.}~\bibnamefont {Nikahd}}, \bibinfo {author} {\bibfnamefont {M.}~\bibnamefont {Sedighi}},\ and\ \bibinfo {author} {\bibfnamefont {M.}~\bibnamefont {{Saheb Zamani}}},\ }\bibfield  {title} {\bibinfo {title} {{Nonuniform code concatenation for universal fault-tolerant quantum computing}},\ }\href {https://doi.org/10.1103/PhysRevA.96.032337} {\bibfield  {journal} {\bibinfo  {journal} {Physical Review A}\ }\textbf {\bibinfo {volume} {96}},\ \bibinfo {pages} {032337} (\bibinfo {year} {2017})}\BibitemShut {NoStop}%
\bibitem [{\citenamefont {Lin}\ and\ \citenamefont {Yang}(2020)}]{Lin2020}%
  \BibitemOpen
  \bibfield  {author} {\bibinfo {author} {\bibfnamefont {C.}~\bibnamefont {Lin}}\ and\ \bibinfo {author} {\bibfnamefont {G.}~\bibnamefont {Yang}},\ }\bibfield  {title} {\bibinfo {title} {{Concatenated pieceable fault-tolerant scheme for universal quantum computation}},\ }\href {https://doi.org/10.1103/PhysRevA.102.052415} {\bibfield  {journal} {\bibinfo  {journal} {Physical Review A}\ }\textbf {\bibinfo {volume} {102}},\ \bibinfo {pages} {052415} (\bibinfo {year} {2020})}\BibitemShut {NoStop}%
\bibitem [{\citenamefont {Yamasaki}\ and\ \citenamefont {Koashi}(2024)}]{yamasaki2024time-efficient-b61}%
  \BibitemOpen
  \bibfield  {author} {\bibinfo {author} {\bibfnamefont {H.}~\bibnamefont {Yamasaki}}\ and\ \bibinfo {author} {\bibfnamefont {M.}~\bibnamefont {Koashi}},\ }\bibfield  {title} {\bibinfo {title} {Time-efficient constant-space-overhead fault-tolerant quantum computation},\ }\href {https://doi.org/10.1038/s41567-023-02325-8} {\bibfield  {journal} {\bibinfo  {journal} {Nature Physics}\ }\textbf {\bibinfo {volume} {20}},\ \bibinfo {pages} {247} (\bibinfo {year} {2024})}\BibitemShut {NoStop}%
\bibitem [{\citenamefont {Yoshida}\ \emph {et~al.}(2025)\citenamefont {Yoshida}, \citenamefont {Tamiya},\ and\ \citenamefont {Yamasaki}}]{yoshida2024concatenate-611}%
  \BibitemOpen
  \bibfield  {author} {\bibinfo {author} {\bibfnamefont {S.}~\bibnamefont {Yoshida}}, \bibinfo {author} {\bibfnamefont {S.}~\bibnamefont {Tamiya}},\ and\ \bibinfo {author} {\bibfnamefont {H.}~\bibnamefont {Yamasaki}},\ }\bibfield  {title} {\bibinfo {title} {{Concatenate codes, save qubits}},\ }\href {https://doi.org/10.1038/s41534-025-01035-8} {\bibfield  {journal} {\bibinfo  {journal} {npj Quantum Information}\ }\textbf {\bibinfo {volume} {11}},\ \bibinfo {pages} {88} (\bibinfo {year} {2025})}\BibitemShut {NoStop}%
\bibitem [{\citenamefont {Fowler}\ \emph {et~al.}(2012)\citenamefont {Fowler}, \citenamefont {Mariantoni}, \citenamefont {Martinis},\ and\ \citenamefont {Cleland}}]{Fowler2012fi}%
  \BibitemOpen
  \bibfield  {author} {\bibinfo {author} {\bibfnamefont {A.~G.}\ \bibnamefont {Fowler}}, \bibinfo {author} {\bibfnamefont {M.}~\bibnamefont {Mariantoni}}, \bibinfo {author} {\bibfnamefont {J.~M.}\ \bibnamefont {Martinis}},\ and\ \bibinfo {author} {\bibfnamefont {A.~N.}\ \bibnamefont {Cleland}},\ }\bibfield  {title} {\bibinfo {title} {Surface codes: Towards practical large-scale quantum computation},\ }\href {https://doi.org/10.1103/physreva.86.032324} {\bibfield  {journal} {\bibinfo  {journal} {Physical Review A}\ }\textbf {\bibinfo {volume} {86}},\ \bibinfo {pages} {032324} (\bibinfo {year} {2012})}\BibitemShut {NoStop}%
\bibitem [{\citenamefont {Gottesman}(2013)}]{Gottesman2013ug}%
  \BibitemOpen
  \bibfield  {author} {\bibinfo {author} {\bibfnamefont {D.}~\bibnamefont {Gottesman}},\ }\href {https://arxiv.org/abs/1310.2984} {\bibinfo {title} {Fault-tolerant quantum computation with constant overhead}} (\bibinfo {year} {2013})\BibitemShut {NoStop}%
\bibitem [{\citenamefont {Bravyi}\ \emph {et~al.}(2024)\citenamefont {Bravyi}, \citenamefont {Cross}, \citenamefont {Gambetta}, \citenamefont {Maslov}, \citenamefont {Rall},\ and\ \citenamefont {Yoder}}]{bravyi2024high-threshold-623}%
  \BibitemOpen
  \bibfield  {author} {\bibinfo {author} {\bibfnamefont {S.}~\bibnamefont {Bravyi}}, \bibinfo {author} {\bibfnamefont {A.~W.}\ \bibnamefont {Cross}}, \bibinfo {author} {\bibfnamefont {J.~M.}\ \bibnamefont {Gambetta}}, \bibinfo {author} {\bibfnamefont {D.}~\bibnamefont {Maslov}}, \bibinfo {author} {\bibfnamefont {P.}~\bibnamefont {Rall}},\ and\ \bibinfo {author} {\bibfnamefont {T.~J.}\ \bibnamefont {Yoder}},\ }\bibfield  {title} {\bibinfo {title} {High-threshold and low-overhead fault-tolerant quantum memory},\ }\href {https://doi.org/10.1038/s41586-024-07107-7} {\bibfield  {journal} {\bibinfo  {journal} {Nature}\ }\textbf {\bibinfo {volume} {627}},\ \bibinfo {pages} {778} (\bibinfo {year} {2024})}\BibitemShut {NoStop}%
\bibitem [{\citenamefont {Sommers}\ \emph {et~al.}(2025)\citenamefont {Sommers}, \citenamefont {Huse},\ and\ \citenamefont {Gullans}}]{Sommers2024tree}%
  \BibitemOpen
  \bibfield  {author} {\bibinfo {author} {\bibfnamefont {G.~M.}\ \bibnamefont {Sommers}}, \bibinfo {author} {\bibfnamefont {D.~A.}\ \bibnamefont {Huse}},\ and\ \bibinfo {author} {\bibfnamefont {M.~J.}\ \bibnamefont {Gullans}},\ }\bibfield  {title} {\bibinfo {title} {Dynamically generated concatenated codes and their phase diagrams},\ }\href {https://doi.org/10.1103/PhysRevResearch.7.023086} {\bibfield  {journal} {\bibinfo  {journal} {Phys. Rev. Res.}\ }\textbf {\bibinfo {volume} {7}},\ \bibinfo {pages} {023086} (\bibinfo {year} {2025})}\BibitemShut {NoStop}%
\bibitem [{\citenamefont {Moses}\ \emph {et~al.}(2023)\citenamefont {Moses}, \citenamefont {Baldwin}, \citenamefont {Allman}, \citenamefont {Ancona}, \citenamefont {Ascarrunz}, \citenamefont {Barnes}, \citenamefont {Bartolotta}, \citenamefont {Bjork}, \citenamefont {Blanchard}, \citenamefont {Bohn}, \citenamefont {Bohnet}, \citenamefont {Brown}, \citenamefont {Burdick}, \citenamefont {Burton}, \citenamefont {Campbell}, \citenamefont {Campora}, \citenamefont {Carron}, \citenamefont {Chambers}, \citenamefont {Chan}, \citenamefont {Chen}, \citenamefont {Chernoguzov}, \citenamefont {Chertkov}, \citenamefont {Colina}, \citenamefont {Curtis}, \citenamefont {Daniel}, \citenamefont {DeCross}, \citenamefont {Deen}, \citenamefont {Delaney}, \citenamefont {Dreiling}, \citenamefont {Ertsgaard}, \citenamefont {Esposito}, \citenamefont {Estey}, \citenamefont {Fabrikant}, \citenamefont {Figgatt}, \citenamefont {Foltz}, \citenamefont {Foss-Feig}, \citenamefont {Francois}, \citenamefont {Gaebler}, \citenamefont {Gatterman},
  \citenamefont {Gilbreth}, \citenamefont {Giles}, \citenamefont {Glynn}, \citenamefont {Hall}, \citenamefont {Hankin}, \citenamefont {Hansen}, \citenamefont {Hayes}, \citenamefont {Higashi}, \citenamefont {Hoffman}, \citenamefont {Horning}, \citenamefont {Hout}, \citenamefont {Jacobs}, \citenamefont {Johansen}, \citenamefont {Jones}, \citenamefont {Karcz}, \citenamefont {Klein}, \citenamefont {Lauria}, \citenamefont {Lee}, \citenamefont {Liefer}, \citenamefont {Lu}, \citenamefont {Lucchetti}, \citenamefont {Lytle}, \citenamefont {Malm}, \citenamefont {Matheny}, \citenamefont {Mathewson}, \citenamefont {Mayer}, \citenamefont {Miller}, \citenamefont {Mills}, \citenamefont {Neyenhuis}, \citenamefont {Nugent}, \citenamefont {Olson}, \citenamefont {Parks}, \citenamefont {Price}, \citenamefont {Price}, \citenamefont {Pugh}, \citenamefont {Ransford}, \citenamefont {Reed}, \citenamefont {Roman}, \citenamefont {Rowe}, \citenamefont {Ryan-Anderson}, \citenamefont {Sanders}, \citenamefont {Sedlacek}, \citenamefont
  {Shevchuk}, \citenamefont {Siegfried}, \citenamefont {Skripka}, \citenamefont {Spaun}, \citenamefont {Sprenkle}, \citenamefont {Stutz}, \citenamefont {Swallows}, \citenamefont {Tobey}, \citenamefont {Tran}, \citenamefont {Tran}, \citenamefont {Vogt}, \citenamefont {Volin}, \citenamefont {Walker}, \citenamefont {Zolot},\ and\ \citenamefont {Pino}}]{quantinuum2023}%
  \BibitemOpen
  \bibfield  {author} {\bibinfo {author} {\bibfnamefont {S.~A.}\ \bibnamefont {Moses}}, \bibinfo {author} {\bibfnamefont {C.~H.}\ \bibnamefont {Baldwin}}, \bibinfo {author} {\bibfnamefont {M.~S.}\ \bibnamefont {Allman}}, \bibinfo {author} {\bibfnamefont {R.}~\bibnamefont {Ancona}}, \bibinfo {author} {\bibfnamefont {L.}~\bibnamefont {Ascarrunz}}, \bibinfo {author} {\bibfnamefont {C.}~\bibnamefont {Barnes}}, \bibinfo {author} {\bibfnamefont {J.}~\bibnamefont {Bartolotta}}, \bibinfo {author} {\bibfnamefont {B.}~\bibnamefont {Bjork}}, \bibinfo {author} {\bibfnamefont {P.}~\bibnamefont {Blanchard}}, \bibinfo {author} {\bibfnamefont {M.}~\bibnamefont {Bohn}}, \bibinfo {author} {\bibfnamefont {J.~G.}\ \bibnamefont {Bohnet}}, \bibinfo {author} {\bibfnamefont {N.~C.}\ \bibnamefont {Brown}}, \bibinfo {author} {\bibfnamefont {N.~Q.}\ \bibnamefont {Burdick}}, \bibinfo {author} {\bibfnamefont {W.~C.}\ \bibnamefont {Burton}}, \bibinfo {author} {\bibfnamefont {S.~L.}\ \bibnamefont {Campbell}}, \bibinfo {author}
  {\bibfnamefont {J.~P.}\ \bibnamefont {Campora}}, \bibinfo {author} {\bibfnamefont {C.}~\bibnamefont {Carron}}, \bibinfo {author} {\bibfnamefont {J.}~\bibnamefont {Chambers}}, \bibinfo {author} {\bibfnamefont {J.~W.}\ \bibnamefont {Chan}}, \bibinfo {author} {\bibfnamefont {Y.~H.}\ \bibnamefont {Chen}}, \bibinfo {author} {\bibfnamefont {A.}~\bibnamefont {Chernoguzov}}, \bibinfo {author} {\bibfnamefont {E.}~\bibnamefont {Chertkov}}, \bibinfo {author} {\bibfnamefont {J.}~\bibnamefont {Colina}}, \bibinfo {author} {\bibfnamefont {J.~P.}\ \bibnamefont {Curtis}}, \bibinfo {author} {\bibfnamefont {R.}~\bibnamefont {Daniel}}, \bibinfo {author} {\bibfnamefont {M.}~\bibnamefont {DeCross}}, \bibinfo {author} {\bibfnamefont {D.}~\bibnamefont {Deen}}, \bibinfo {author} {\bibfnamefont {C.}~\bibnamefont {Delaney}}, \bibinfo {author} {\bibfnamefont {J.~M.}\ \bibnamefont {Dreiling}}, \bibinfo {author} {\bibfnamefont {C.~T.}\ \bibnamefont {Ertsgaard}}, \bibinfo {author} {\bibfnamefont {J.}~\bibnamefont {Esposito}}, \bibinfo
  {author} {\bibfnamefont {B.}~\bibnamefont {Estey}}, \bibinfo {author} {\bibfnamefont {M.}~\bibnamefont {Fabrikant}}, \bibinfo {author} {\bibfnamefont {C.}~\bibnamefont {Figgatt}}, \bibinfo {author} {\bibfnamefont {C.}~\bibnamefont {Foltz}}, \bibinfo {author} {\bibfnamefont {M.}~\bibnamefont {Foss-Feig}}, \bibinfo {author} {\bibfnamefont {D.}~\bibnamefont {Francois}}, \bibinfo {author} {\bibfnamefont {J.~P.}\ \bibnamefont {Gaebler}}, \bibinfo {author} {\bibfnamefont {T.~M.}\ \bibnamefont {Gatterman}}, \bibinfo {author} {\bibfnamefont {C.~N.}\ \bibnamefont {Gilbreth}}, \bibinfo {author} {\bibfnamefont {J.}~\bibnamefont {Giles}}, \bibinfo {author} {\bibfnamefont {E.}~\bibnamefont {Glynn}}, \bibinfo {author} {\bibfnamefont {A.}~\bibnamefont {Hall}}, \bibinfo {author} {\bibfnamefont {A.~M.}\ \bibnamefont {Hankin}}, \bibinfo {author} {\bibfnamefont {A.}~\bibnamefont {Hansen}}, \bibinfo {author} {\bibfnamefont {D.}~\bibnamefont {Hayes}}, \bibinfo {author} {\bibfnamefont {B.}~\bibnamefont {Higashi}}, \bibinfo
  {author} {\bibfnamefont {I.~M.}\ \bibnamefont {Hoffman}}, \bibinfo {author} {\bibfnamefont {B.}~\bibnamefont {Horning}}, \bibinfo {author} {\bibfnamefont {J.~J.}\ \bibnamefont {Hout}}, \bibinfo {author} {\bibfnamefont {R.}~\bibnamefont {Jacobs}}, \bibinfo {author} {\bibfnamefont {J.}~\bibnamefont {Johansen}}, \bibinfo {author} {\bibfnamefont {L.}~\bibnamefont {Jones}}, \bibinfo {author} {\bibfnamefont {J.}~\bibnamefont {Karcz}}, \bibinfo {author} {\bibfnamefont {T.}~\bibnamefont {Klein}}, \bibinfo {author} {\bibfnamefont {P.}~\bibnamefont {Lauria}}, \bibinfo {author} {\bibfnamefont {P.}~\bibnamefont {Lee}}, \bibinfo {author} {\bibfnamefont {D.}~\bibnamefont {Liefer}}, \bibinfo {author} {\bibfnamefont {S.~T.}\ \bibnamefont {Lu}}, \bibinfo {author} {\bibfnamefont {D.}~\bibnamefont {Lucchetti}}, \bibinfo {author} {\bibfnamefont {C.}~\bibnamefont {Lytle}}, \bibinfo {author} {\bibfnamefont {A.}~\bibnamefont {Malm}}, \bibinfo {author} {\bibfnamefont {M.}~\bibnamefont {Matheny}}, \bibinfo {author} {\bibfnamefont
  {B.}~\bibnamefont {Mathewson}}, \bibinfo {author} {\bibfnamefont {K.}~\bibnamefont {Mayer}}, \bibinfo {author} {\bibfnamefont {D.~B.}\ \bibnamefont {Miller}}, \bibinfo {author} {\bibfnamefont {M.}~\bibnamefont {Mills}}, \bibinfo {author} {\bibfnamefont {B.}~\bibnamefont {Neyenhuis}}, \bibinfo {author} {\bibfnamefont {L.}~\bibnamefont {Nugent}}, \bibinfo {author} {\bibfnamefont {S.}~\bibnamefont {Olson}}, \bibinfo {author} {\bibfnamefont {J.}~\bibnamefont {Parks}}, \bibinfo {author} {\bibfnamefont {G.~N.}\ \bibnamefont {Price}}, \bibinfo {author} {\bibfnamefont {Z.}~\bibnamefont {Price}}, \bibinfo {author} {\bibfnamefont {M.}~\bibnamefont {Pugh}}, \bibinfo {author} {\bibfnamefont {A.}~\bibnamefont {Ransford}}, \bibinfo {author} {\bibfnamefont {A.~P.}\ \bibnamefont {Reed}}, \bibinfo {author} {\bibfnamefont {C.}~\bibnamefont {Roman}}, \bibinfo {author} {\bibfnamefont {M.}~\bibnamefont {Rowe}}, \bibinfo {author} {\bibfnamefont {C.}~\bibnamefont {Ryan-Anderson}}, \bibinfo {author} {\bibfnamefont
  {S.}~\bibnamefont {Sanders}}, \bibinfo {author} {\bibfnamefont {J.}~\bibnamefont {Sedlacek}}, \bibinfo {author} {\bibfnamefont {P.}~\bibnamefont {Shevchuk}}, \bibinfo {author} {\bibfnamefont {P.}~\bibnamefont {Siegfried}}, \bibinfo {author} {\bibfnamefont {T.}~\bibnamefont {Skripka}}, \bibinfo {author} {\bibfnamefont {B.}~\bibnamefont {Spaun}}, \bibinfo {author} {\bibfnamefont {R.~T.}\ \bibnamefont {Sprenkle}}, \bibinfo {author} {\bibfnamefont {R.~P.}\ \bibnamefont {Stutz}}, \bibinfo {author} {\bibfnamefont {M.}~\bibnamefont {Swallows}}, \bibinfo {author} {\bibfnamefont {R.~I.}\ \bibnamefont {Tobey}}, \bibinfo {author} {\bibfnamefont {A.}~\bibnamefont {Tran}}, \bibinfo {author} {\bibfnamefont {T.}~\bibnamefont {Tran}}, \bibinfo {author} {\bibfnamefont {E.}~\bibnamefont {Vogt}}, \bibinfo {author} {\bibfnamefont {C.}~\bibnamefont {Volin}}, \bibinfo {author} {\bibfnamefont {J.}~\bibnamefont {Walker}}, \bibinfo {author} {\bibfnamefont {A.~M.}\ \bibnamefont {Zolot}},\ and\ \bibinfo {author} {\bibfnamefont
  {J.~M.}\ \bibnamefont {Pino}},\ }\bibfield  {title} {\bibinfo {title} {A race-track trapped-ion quantum processor},\ }\href {https://doi.org/10.1103/PhysRevX.13.041052} {\bibfield  {journal} {\bibinfo  {journal} {Phys. Rev. X}\ }\textbf {\bibinfo {volume} {13}},\ \bibinfo {pages} {041052} (\bibinfo {year} {2023})}\BibitemShut {NoStop}%
\bibitem [{\citenamefont {Steane}(1997)}]{Steane1997}%
  \BibitemOpen
  \bibfield  {author} {\bibinfo {author} {\bibfnamefont {A.~M.}\ \bibnamefont {Steane}},\ }\bibfield  {title} {\bibinfo {title} {{Active Stabilization, Quantum Computation, and Quantum State Synthesis}},\ }\href {https://doi.org/10.1103/PhysRevLett.78.2252} {\bibfield  {journal} {\bibinfo  {journal} {Physical Review Letters}\ }\textbf {\bibinfo {volume} {78}},\ \bibinfo {pages} {2252} (\bibinfo {year} {1997})}\BibitemShut {NoStop}%
\bibitem [{\citenamefont {Bravyi}\ and\ \citenamefont {Kitaev}(2005)}]{Bravyi2005}%
  \BibitemOpen
  \bibfield  {author} {\bibinfo {author} {\bibfnamefont {S.}~\bibnamefont {Bravyi}}\ and\ \bibinfo {author} {\bibfnamefont {A.}~\bibnamefont {Kitaev}},\ }\bibfield  {title} {\bibinfo {title} {{Universal quantum computation with ideal Clifford gates and noisy ancillas}},\ }\href {https://doi.org/10.1103/PhysRevA.71.022316} {\bibfield  {journal} {\bibinfo  {journal} {Physical Review A}\ }\textbf {\bibinfo {volume} {71}},\ \bibinfo {pages} {022316} (\bibinfo {year} {2005})}\BibitemShut {NoStop}%
\bibitem [{sup()}]{supp-ref}%
  \BibitemOpen
  \href@noop {} {\bibinfo {title} {{See Supplemental Material for additional details on the spacetime code, variations on the state preparation circuit, optimal and approximate decoding via tensor networks, {implementation on Quantinuum System H2, the quantum memory test, and code switching.}}}}\BibitemShut {Stop}%
\bibitem [{\citenamefont {Cao}\ and\ \citenamefont {Lackey}(2022)}]{Cao2022lego}%
  \BibitemOpen
  \bibfield  {author} {\bibinfo {author} {\bibfnamefont {C.}~\bibnamefont {Cao}}\ and\ \bibinfo {author} {\bibfnamefont {B.}~\bibnamefont {Lackey}},\ }\bibfield  {title} {\bibinfo {title} {{Quantum Lego: Building Quantum Error Correction Codes from Tensor Networks}},\ }\href {https://doi.org/10.1103/PRXQuantum.3.020332} {\bibfield  {journal} {\bibinfo  {journal} {PRX Quantum}\ }\textbf {\bibinfo {volume} {3}},\ \bibinfo {pages} {020332} (\bibinfo {year} {2022})}\BibitemShut {NoStop}%
\bibitem [{\citenamefont {Cao}\ and\ \citenamefont {Lackey}(2024)}]{Cao2024}%
  \BibitemOpen
  \bibfield  {author} {\bibinfo {author} {\bibfnamefont {C.}~\bibnamefont {Cao}}\ and\ \bibinfo {author} {\bibfnamefont {B.}~\bibnamefont {Lackey}},\ }\bibfield  {title} {\bibinfo {title} {{Quantum Weight Enumerators and Tensor Networks}},\ }\href {https://doi.org/10.1109/TIT.2023.3340503} {\bibfield  {journal} {\bibinfo  {journal} {IEEE Transactions on Information Theory}\ }\textbf {\bibinfo {volume} {70}},\ \bibinfo {pages} {3512} (\bibinfo {year} {2024})}\BibitemShut {NoStop}%
\bibitem [{\citenamefont {Cao}\ \emph {et~al.}(2024)\citenamefont {Cao}, \citenamefont {Gullans}, \citenamefont {Lackey},\ and\ \citenamefont {Wang}}]{Cao2024expansion}%
  \BibitemOpen
  \bibfield  {author} {\bibinfo {author} {\bibfnamefont {C.}~\bibnamefont {Cao}}, \bibinfo {author} {\bibfnamefont {M.~J.}\ \bibnamefont {Gullans}}, \bibinfo {author} {\bibfnamefont {B.}~\bibnamefont {Lackey}},\ and\ \bibinfo {author} {\bibfnamefont {Z.}~\bibnamefont {Wang}},\ }\bibfield  {title} {\bibinfo {title} {Quantum lego expansion pack: Enumerators from tensor networks},\ }\href {https://doi.org/10.1103/PRXQuantum.5.030313} {\bibfield  {journal} {\bibinfo  {journal} {PRX Quantum}\ }\textbf {\bibinfo {volume} {5}},\ \bibinfo {pages} {030313} (\bibinfo {year} {2024})}\BibitemShut {NoStop}%
\bibitem [{\citenamefont {Shor}(1995)}]{Shor1995}%
  \BibitemOpen
  \bibfield  {author} {\bibinfo {author} {\bibfnamefont {P.~W.}\ \bibnamefont {Shor}},\ }\bibfield  {title} {\bibinfo {title} {Scheme for reducing decoherence in quantum computer memory},\ }\href {https://doi.org/10.1103/PhysRevA.52.R2493} {\bibfield  {journal} {\bibinfo  {journal} {Phys. Rev. A}\ }\textbf {\bibinfo {volume} {52}},\ \bibinfo {pages} {R2493} (\bibinfo {year} {1995})}\BibitemShut {NoStop}%
\bibitem [{\citenamefont {Nguyen}\ \emph {et~al.}(2021)\citenamefont {Nguyen}, \citenamefont {Li}, \citenamefont {Green}, \citenamefont {{Huerta Alderete}}, \citenamefont {Zhu}, \citenamefont {Zhu}, \citenamefont {Brown},\ and\ \citenamefont {Linke}}]{Nguyen2021}%
  \BibitemOpen
  \bibfield  {author} {\bibinfo {author} {\bibfnamefont {N.~H.}\ \bibnamefont {Nguyen}}, \bibinfo {author} {\bibfnamefont {M.}~\bibnamefont {Li}}, \bibinfo {author} {\bibfnamefont {A.~M.}\ \bibnamefont {Green}}, \bibinfo {author} {\bibfnamefont {C.}~\bibnamefont {{Huerta Alderete}}}, \bibinfo {author} {\bibfnamefont {Y.}~\bibnamefont {Zhu}}, \bibinfo {author} {\bibfnamefont {D.}~\bibnamefont {Zhu}}, \bibinfo {author} {\bibfnamefont {K.~R.}\ \bibnamefont {Brown}},\ and\ \bibinfo {author} {\bibfnamefont {N.~M.}\ \bibnamefont {Linke}},\ }\bibfield  {title} {\bibinfo {title} {{Demonstration of Shor Encoding on a Trapped-Ion Quantum Computer}},\ }\href {https://doi.org/10.1103/PhysRevApplied.16.024057} {\bibfield  {journal} {\bibinfo  {journal} {Physical Review Applied}\ }\textbf {\bibinfo {volume} {16}},\ \bibinfo {pages} {024057} (\bibinfo {year} {2021})}\BibitemShut {NoStop}%
\bibitem [{zx()}]{zx}%
  \BibitemOpen
  \href@noop {} {}\bibinfo {note} {\gs{The way the circuit is drawn may look familiar to practitioners of ZX calculus~\cite{Vandewetering2020}, and indeed if we absorb the stabilizer inputs into the gates, the corresponding ZX diagram is a binary tree composed of alternating layers of Z and X spiders, as discussed in Ref.~\cite{Sommers2024tree}. Here we leave the inputs explicit as they can also suffer errors in our noise model.}}\BibitemShut {Stop}%
\bibitem [{\citenamefont {Calderbank}\ and\ \citenamefont {Shor}(1996)}]{Calderbank1996}%
  \BibitemOpen
  \bibfield  {author} {\bibinfo {author} {\bibfnamefont {A.~R.}\ \bibnamefont {Calderbank}}\ and\ \bibinfo {author} {\bibfnamefont {P.~W.}\ \bibnamefont {Shor}},\ }\bibfield  {title} {\bibinfo {title} {{Good quantum error-correcting codes exist}},\ }\href {https://doi.org/10.1103/PhysRevA.54.1098} {\bibfield  {journal} {\bibinfo  {journal} {Physical Review A}\ }\textbf {\bibinfo {volume} {54}},\ \bibinfo {pages} {1098} (\bibinfo {year} {1996})}\BibitemShut {NoStop}%
\bibitem [{\citenamefont {Steane}(1996)}]{Steane1996}%
  \BibitemOpen
  \bibfield  {author} {\bibinfo {author} {\bibfnamefont {A.}~\bibnamefont {Steane}},\ }\bibfield  {title} {\bibinfo {title} {{Multiple-particle interference and quantum error correction}},\ }\href {https://doi.org/10.1098/rspa.1996.0136} {\bibfield  {journal} {\bibinfo  {journal} {Proceedings of the Royal Society of London. Series A: Mathematical, Physical and Engineering Sciences}\ }\textbf {\bibinfo {volume} {452}},\ \bibinfo {pages} {2551} (\bibinfo {year} {1996})}\BibitemShut {NoStop}%
\bibitem [{roo()}]{root}%
  \BibitemOpen
  \href@noop {} {\bibinfo {title} {This stabilizer will henceforth be referred to as the ``root stabilizer'' to distinguish it from the stabilizer inputs that enter each node on the right side; under the encoding circuit, it evolves to the operator $\overline{P}$.}}\BibitemShut {Stop}%
\bibitem [{\citenamefont {Leighton}(1992)}]{Leighton1992}%
  \BibitemOpen
  \bibfield  {author} {\bibinfo {author} {\bibfnamefont {F.~T.}\ \bibnamefont {Leighton}},\ }\href@noop {} {\emph {\bibinfo {title} {Introduction to parallel algorithms and architectures: arrays, trees, hypercubes}}}\ (\bibinfo  {publisher} {Morgan Kaufmann Publishers Inc.},\ \bibinfo {address} {San Francisco, CA, USA},\ \bibinfo {year} {1992})\BibitemShut {NoStop}%
\bibitem [{\citenamefont {Cooley}\ and\ \citenamefont {Tukey}(1965)}]{cooley1965algorithm}%
  \BibitemOpen
  \bibfield  {author} {\bibinfo {author} {\bibfnamefont {J.~W.}\ \bibnamefont {Cooley}}\ and\ \bibinfo {author} {\bibfnamefont {J.~W.}\ \bibnamefont {Tukey}},\ }\bibfield  {title} {\bibinfo {title} {An algorithm for the machine calculation of complex fourier series},\ }\href@noop {} {\bibfield  {journal} {\bibinfo  {journal} {Mathematics of computation}\ }\textbf {\bibinfo {volume} {19}},\ \bibinfo {pages} {297} (\bibinfo {year} {1965})}\BibitemShut {NoStop}%
\bibitem [{\citenamefont {Weinstein}(1969)}]{weinstein1969}%
  \BibitemOpen
  \bibfield  {author} {\bibinfo {author} {\bibfnamefont {C.~J.}\ \bibnamefont {Weinstein}},\ }\href@noop {} {\emph {\bibinfo {title} {Quantization effects in digital filters}}},\ \bibinfo {type} {Tech. Rep.}\ (\bibinfo {year} {1969})\BibitemShut {NoStop}%
\bibitem [{\citenamefont {Ferris}\ and\ \citenamefont {Poulin}(2014)}]{Ferris2014}%
  \BibitemOpen
  \bibfield  {author} {\bibinfo {author} {\bibfnamefont {A.~J.}\ \bibnamefont {Ferris}}\ and\ \bibinfo {author} {\bibfnamefont {D.}~\bibnamefont {Poulin}},\ }\bibfield  {title} {\bibinfo {title} {{Tensor Networks and Quantum Error Correction}},\ }\href {https://doi.org/10.1103/PhysRevLett.113.030501} {\bibfield  {journal} {\bibinfo  {journal} {Physical Review Letters}\ }\textbf {\bibinfo {volume} {113}},\ \bibinfo {pages} {030501} (\bibinfo {year} {2014})}\BibitemShut {NoStop}%
\bibitem [{\citenamefont {Delfosse}\ and\ \citenamefont {Paetznick}(2023)}]{Delfosse2023}%
  \BibitemOpen
  \bibfield  {author} {\bibinfo {author} {\bibfnamefont {N.}~\bibnamefont {Delfosse}}\ and\ \bibinfo {author} {\bibfnamefont {A.}~\bibnamefont {Paetznick}},\ }\href {http://arxiv.org/abs/2304.05943} {\bibinfo {title} {{Spacetime codes of Clifford circuits}}} (\bibinfo {year} {2023}),\ \Eprint {https://arxiv.org/abs/2304.05943} {arXiv:2304.05943} \BibitemShut {NoStop}%
\bibitem [{err()}]{errors}%
  \BibitemOpen
  \href@noop {} {\bibinfo {title} {{Each measurement error occurs immediately after a two-qubit gates, so the ancilla legs exiting a gate have two error opportunities. A Pauli error immediately preceding measurement can be interpreted as a bit flip error on the classical measurement outcome.}}}\BibitemShut {Stop}%
\bibitem [{\citenamefont {Bacon}\ \emph {et~al.}(2017)\citenamefont {Bacon}, \citenamefont {Flammia}, \citenamefont {Harrow},\ and\ \citenamefont {Shi}}]{Bacon2017}%
  \BibitemOpen
  \bibfield  {author} {\bibinfo {author} {\bibfnamefont {D.}~\bibnamefont {Bacon}}, \bibinfo {author} {\bibfnamefont {S.~T.}\ \bibnamefont {Flammia}}, \bibinfo {author} {\bibfnamefont {A.~W.}\ \bibnamefont {Harrow}},\ and\ \bibinfo {author} {\bibfnamefont {J.}~\bibnamefont {Shi}},\ }\bibfield  {title} {\bibinfo {title} {{Sparse quantum codes from quantum circuits}},\ }in\ \href {https://doi.org/10.1109/TIT.2017.2663199} {\emph {\bibinfo {booktitle} {IEEE Transactions on Information Theory}}},\ Vol.~\bibinfo {volume} {63}\ (\bibinfo {year} {2017})\ pp.\ \bibinfo {pages} {2464--2479},\ \Eprint {https://arxiv.org/abs/1411.3334} {arXiv:1411.3334} \BibitemShut {NoStop}%
\bibitem [{\citenamefont {Gottesman}(2022)}]{Gottesman2022}%
  \BibitemOpen
  \bibfield  {author} {\bibinfo {author} {\bibfnamefont {D.}~\bibnamefont {Gottesman}},\ }\href {http://arxiv.org/abs/2210.15844} {\bibinfo {title} {{Opportunities and Challenges in Fault-Tolerant Quantum Computation}}} (\bibinfo {year} {2022}),\ \Eprint {https://arxiv.org/abs/2210.15844} {arXiv:2210.15844} \BibitemShut {NoStop}%
\bibitem [{spa()}]{spacetime}%
  \BibitemOpen
  \href@noop {} {\bibinfo {title} {{The precise number of spacetime qubits for this error model $(5T + \lfloor (T+1)/2 \rfloor - 2)2^{T-1}$ for $\lstate{0}$, $(5T + \lfloor (T+1)/2 \rfloor - 2)2^{T-1}$ for $\lstate{+}$. For simplicity, the scaling collapse in Fig. 2a(ii) drops the subleading factors and simply defines $N(T) = T 2^T$. This convention does not affect the inferred exponent $\nu$}}}\BibitemShut {NoStop}%
\bibitem [{\citenamefont {Bravyi}\ \emph {et~al.}(2018)\citenamefont {Bravyi}, \citenamefont {Englbrecht}, \citenamefont {K{\"{o}}nig},\ and\ \citenamefont {Peard}}]{Bravyi2018}%
  \BibitemOpen
  \bibfield  {author} {\bibinfo {author} {\bibfnamefont {S.}~\bibnamefont {Bravyi}}, \bibinfo {author} {\bibfnamefont {M.}~\bibnamefont {Englbrecht}}, \bibinfo {author} {\bibfnamefont {R.}~\bibnamefont {K{\"{o}}nig}},\ and\ \bibinfo {author} {\bibfnamefont {N.}~\bibnamefont {Peard}},\ }\bibfield  {title} {\bibinfo {title} {{Correcting coherent errors with surface codes}},\ }\href {https://doi.org/10.1038/s41534-018-0106-y} {\bibfield  {journal} {\bibinfo  {journal} {npj Quantum Information}\ }\textbf {\bibinfo {volume} {4}},\ \bibinfo {pages} {55} (\bibinfo {year} {2018})}\BibitemShut {NoStop}%
\bibitem [{\citenamefont {Venn}\ \emph {et~al.}(2023)\citenamefont {Venn}, \citenamefont {Behrends},\ and\ \citenamefont {B{\'{e}}ri}}]{Venn2023}%
  \BibitemOpen
  \bibfield  {author} {\bibinfo {author} {\bibfnamefont {F.}~\bibnamefont {Venn}}, \bibinfo {author} {\bibfnamefont {J.}~\bibnamefont {Behrends}},\ and\ \bibinfo {author} {\bibfnamefont {B.}~\bibnamefont {B{\'{e}}ri}},\ }\bibfield  {title} {\bibinfo {title} {{Coherent-Error Threshold for Surface Codes from Majorana Delocalization}},\ }\href {https://doi.org/10.1103/PhysRevLett.131.060603} {\bibfield  {journal} {\bibinfo  {journal} {Physical Review Letters}\ }\textbf {\bibinfo {volume} {131}},\ \bibinfo {pages} {060603} (\bibinfo {year} {2023})}\BibitemShut {NoStop}%
\bibitem [{\citenamefont {Audenaert}\ and\ \citenamefont {Plenio}(2006)}]{Audenaert2006}%
  \BibitemOpen
  \bibfield  {author} {\bibinfo {author} {\bibfnamefont {K.~M.~R.}\ \bibnamefont {Audenaert}}\ and\ \bibinfo {author} {\bibfnamefont {M.~B.}\ \bibnamefont {Plenio}},\ }\bibfield  {title} {\bibinfo {title} {{When are correlations quantum?—verification and quantification of entanglement by simple measurements}},\ }\href {https://doi.org/10.1088/1367-2630/8/11/266} {\bibfield  {journal} {\bibinfo  {journal} {New Journal of Physics}\ }\textbf {\bibinfo {volume} {8}},\ \bibinfo {pages} {266} (\bibinfo {year} {2006})}\BibitemShut {NoStop}%
\bibitem [{\citenamefont {Stephens}\ \emph {et~al.}(2008)\citenamefont {Stephens}, \citenamefont {Evans}, \citenamefont {Devitt},\ and\ \citenamefont {Hollenberg}}]{Stephens2008}%
  \BibitemOpen
  \bibfield  {author} {\bibinfo {author} {\bibfnamefont {A.~M.}\ \bibnamefont {Stephens}}, \bibinfo {author} {\bibfnamefont {Z.~W.}\ \bibnamefont {Evans}}, \bibinfo {author} {\bibfnamefont {S.~J.}\ \bibnamefont {Devitt}},\ and\ \bibinfo {author} {\bibfnamefont {L.~C.}\ \bibnamefont {Hollenberg}},\ }\bibfield  {title} {\bibinfo {title} {{Asymmetric quantum error correction via code conversion}},\ }\href {https://doi.org/10.1103/PhysRevA.77.062335} {\bibfield  {journal} {\bibinfo  {journal} {Physical Review A - Atomic, Molecular, and Optical Physics}\ }\textbf {\bibinfo {volume} {77}},\ \bibinfo {pages} {2} (\bibinfo {year} {2008})}\BibitemShut {NoStop}%
\bibitem [{\citenamefont {Paetznick}\ and\ \citenamefont {Reichardt}(2013{\natexlab{b}})}]{Paetznick2013switch}%
  \BibitemOpen
  \bibfield  {author} {\bibinfo {author} {\bibfnamefont {A.}~\bibnamefont {Paetznick}}\ and\ \bibinfo {author} {\bibfnamefont {B.~W.}\ \bibnamefont {Reichardt}},\ }\bibfield  {title} {\bibinfo {title} {{Universal Fault-Tolerant Quantum Computation with Only Transversal Gates and Error Correction}},\ }\href {https://doi.org/10.1103/PhysRevLett.111.090505} {\bibfield  {journal} {\bibinfo  {journal} {Physical Review Letters}\ }\textbf {\bibinfo {volume} {111}},\ \bibinfo {pages} {090505} (\bibinfo {year} {2013}{\natexlab{b}})}\BibitemShut {NoStop}%
\bibitem [{\citenamefont {Anderson}\ \emph {et~al.}(2014)\citenamefont {Anderson}, \citenamefont {Duclos-Cianci},\ and\ \citenamefont {Poulin}}]{Anderson2014}%
  \BibitemOpen
  \bibfield  {author} {\bibinfo {author} {\bibfnamefont {J.~T.}\ \bibnamefont {Anderson}}, \bibinfo {author} {\bibfnamefont {G.}~\bibnamefont {Duclos-Cianci}},\ and\ \bibinfo {author} {\bibfnamefont {D.}~\bibnamefont {Poulin}},\ }\bibfield  {title} {\bibinfo {title} {{Fault-Tolerant Conversion between the Steane and Reed-Muller Quantum Codes}},\ }\href {https://doi.org/10.1103/PhysRevLett.113.080501} {\bibfield  {journal} {\bibinfo  {journal} {Physical Review Letters}\ }\textbf {\bibinfo {volume} {113}},\ \bibinfo {pages} {080501} (\bibinfo {year} {2014})}\BibitemShut {NoStop}%
\bibitem [{\citenamefont {Kubica}\ and\ \citenamefont {Beverland}(2015)}]{Kubica2015}%
  \BibitemOpen
  \bibfield  {author} {\bibinfo {author} {\bibfnamefont {A.}~\bibnamefont {Kubica}}\ and\ \bibinfo {author} {\bibfnamefont {M.~E.}\ \bibnamefont {Beverland}},\ }\bibfield  {title} {\bibinfo {title} {Universal transversal gates with color codes: A simplified approach},\ }\href {https://doi.org/10.1103/PhysRevA.91.032330} {\bibfield  {journal} {\bibinfo  {journal} {Phys. Rev. A}\ }\textbf {\bibinfo {volume} {91}},\ \bibinfo {pages} {032330} (\bibinfo {year} {2015})}\BibitemShut {NoStop}%
\bibitem [{\citenamefont {Bombín}(2016)}]{Bombin2016}%
  \BibitemOpen
  \bibfield  {author} {\bibinfo {author} {\bibfnamefont {H.}~\bibnamefont {Bombín}},\ }\bibfield  {title} {\bibinfo {title} {Dimensional jump in quantum error correction},\ }\href {https://doi.org/10.1088/1367-2630/18/4/043038} {\bibfield  {journal} {\bibinfo  {journal} {New Journal of Physics}\ }\textbf {\bibinfo {volume} {18}},\ \bibinfo {pages} {043038} (\bibinfo {year} {2016})}\BibitemShut {NoStop}%
\bibitem [{\citenamefont {Vuillot}\ \emph {et~al.}(2019)\citenamefont {Vuillot}, \citenamefont {Lao}, \citenamefont {Criger}, \citenamefont {{Garc{\'{i}}a Almud{\'{e}}ver}}, \citenamefont {Bertels},\ and\ \citenamefont {Terhal}}]{Vuillot2019}%
  \BibitemOpen
  \bibfield  {author} {\bibinfo {author} {\bibfnamefont {C.}~\bibnamefont {Vuillot}}, \bibinfo {author} {\bibfnamefont {L.}~\bibnamefont {Lao}}, \bibinfo {author} {\bibfnamefont {B.}~\bibnamefont {Criger}}, \bibinfo {author} {\bibfnamefont {C.}~\bibnamefont {{Garc{\'{i}}a Almud{\'{e}}ver}}}, \bibinfo {author} {\bibfnamefont {K.}~\bibnamefont {Bertels}},\ and\ \bibinfo {author} {\bibfnamefont {B.~M.}\ \bibnamefont {Terhal}},\ }\bibfield  {title} {\bibinfo {title} {{Code deformation and lattice surgery are gauge fixing}},\ }\href {https://doi.org/10.1088/1367-2630/ab0199} {\bibfield  {journal} {\bibinfo  {journal} {New Journal of Physics}\ }\textbf {\bibinfo {volume} {21}},\ \bibinfo {pages} {033028} (\bibinfo {year} {2019})}\BibitemShut {NoStop}%
\bibitem [{\citenamefont {Pogorelov}\ \emph {et~al.}(2025)\citenamefont {Pogorelov}, \citenamefont {Butt}, \citenamefont {Postler}, \citenamefont {Marciniak}, \citenamefont {Schindler}, \citenamefont {M{\"u}ller},\ and\ \citenamefont {Monz}}]{pogorelov2025}%
  \BibitemOpen
  \bibfield  {author} {\bibinfo {author} {\bibfnamefont {I.}~\bibnamefont {Pogorelov}}, \bibinfo {author} {\bibfnamefont {F.}~\bibnamefont {Butt}}, \bibinfo {author} {\bibfnamefont {L.}~\bibnamefont {Postler}}, \bibinfo {author} {\bibfnamefont {C.~D.}\ \bibnamefont {Marciniak}}, \bibinfo {author} {\bibfnamefont {P.}~\bibnamefont {Schindler}}, \bibinfo {author} {\bibfnamefont {M.}~\bibnamefont {M{\"u}ller}},\ and\ \bibinfo {author} {\bibfnamefont {T.}~\bibnamefont {Monz}},\ }\bibfield  {title} {\bibinfo {title} {Experimental fault-tolerant code switching},\ }\href@noop {} {\bibfield  {journal} {\bibinfo  {journal} {Nature Physics}\ }\textbf {\bibinfo {volume} {21}},\ \bibinfo {pages} {298} (\bibinfo {year} {2025})}\BibitemShut {NoStop}%
\bibitem [{\citenamefont {Gottesman}(raft)}]{Gottesman2024}%
  \BibitemOpen
  \bibfield  {author} {\bibinfo {author} {\bibfnamefont {D.}~\bibnamefont {Gottesman}},\ }\href {https://www.cs.umd.edu/class/spring2024/cmsc858G/} {\emph {\bibinfo {title} {{Surviving as a Quantum Computer in a Classical World}}}}\ (\bibinfo {year} {2024 draft})\BibitemShut {NoStop}%
\bibitem [{T()}]{T}%
  \BibitemOpen
  \href@noop {} {\bibinfo {title} {{In} some works, including ref.~\cite{Bravyi2005}, the state we call $\ket{T}$ is instead called the ``hadamard magic state'' $\ket{H}$, while $\ket{T}$ is reserved for an eigenstate of the $t$ gate; elsewhere, e.g. in ref.~\cite{Fowler2012fi}, it is denoted $\ket{A}$. we use $\ket{T}$ (as in, e.g., ref.~\cite{Gidney2024}), to emphasize that it is a resource state for the $t$ gate.}}\BibitemShut {Stop}%
\bibitem [{thr()}]{threshold}%
  \BibitemOpen
  \href@noop {} {\bibinfo {title} {{A protocol subsequently developed in Ref.~\cite{Reichardt2005distillation} achieves an optimal distillation threshold of $(1-1/\sqrt{2})/2 = 0.85355...$, but has inferior subthreshold performance to the scheme of Ref.~\cite{Bravyi2005}. We propose using the distillation scheme of Ref.~\cite{Bravyi2005} since the fixed-point logical fidelity on the singletree already exceeds its threshold in the entire coding phase.}}}\BibitemShut {Stop}%
\bibitem [{eve()}]{even}%
  \BibitemOpen
  \href@noop {} {\bibinfo {title} {{Eagle-eyed readers will note that this figure uses the opposite encoding order (\GSC{e} rather than \GSC{o}) from the rest of the text. We made this choice due to an odd-even effect in the fixed-point logical fidelity, which results in inferior performance of \GSC{o} for this basis of logical $T$ state at even depths. See~\autoref{app:T} of the Supplemental Material for further details.}}}\BibitemShut {Stop}%
\bibitem [{\citenamefont {Gidney}\ \emph {et~al.}(2024)\citenamefont {Gidney}, \citenamefont {Shutty},\ and\ \citenamefont {Jones}}]{Gidney2024}%
  \BibitemOpen
  \bibfield  {author} {\bibinfo {author} {\bibfnamefont {C.}~\bibnamefont {Gidney}}, \bibinfo {author} {\bibfnamefont {N.}~\bibnamefont {Shutty}},\ and\ \bibinfo {author} {\bibfnamefont {C.}~\bibnamefont {Jones}},\ }\href {https://arxiv.org/abs/2409.17595} {\bibinfo {title} {Magic state cultivation: growing {T} states as cheap as {CNOT} gates}} (\bibinfo {year} {2024}),\ \Eprint {https://arxiv.org/abs/2409.17595} {arXiv:2409.17595 [quant-ph]} \BibitemShut {NoStop}%
\bibitem [{\citenamefont {Svore}\ \emph {et~al.}(2005)\citenamefont {Svore}, \citenamefont {Terhal},\ and\ \citenamefont {DiVincenzo}}]{Svore2005}%
  \BibitemOpen
  \bibfield  {author} {\bibinfo {author} {\bibfnamefont {K.~M.}\ \bibnamefont {Svore}}, \bibinfo {author} {\bibfnamefont {B.~M.}\ \bibnamefont {Terhal}},\ and\ \bibinfo {author} {\bibfnamefont {D.~P.}\ \bibnamefont {DiVincenzo}},\ }\bibfield  {title} {\bibinfo {title} {{Local fault-tolerant quantum computation}},\ }\href {https://doi.org/10.1103/PhysRevA.72.022317} {\bibfield  {journal} {\bibinfo  {journal} {Physical Review A}\ }\textbf {\bibinfo {volume} {72}},\ \bibinfo {pages} {022317} (\bibinfo {year} {2005})}\BibitemShut {NoStop}%
\bibitem [{\citenamefont {Svore}\ \emph {et~al.}(2006)\citenamefont {Svore}, \citenamefont {Cross}, \citenamefont {Chuang},\ and\ \citenamefont {Aho}}]{Svore2006}%
  \BibitemOpen
  \bibfield  {author} {\bibinfo {author} {\bibfnamefont {K.}~\bibnamefont {Svore}}, \bibinfo {author} {\bibfnamefont {A.}~\bibnamefont {Cross}}, \bibinfo {author} {\bibfnamefont {I.}~\bibnamefont {Chuang}},\ and\ \bibinfo {author} {\bibfnamefont {A.}~\bibnamefont {Aho}},\ }\bibfield  {title} {\bibinfo {title} {{A flow-map model for analyzing pseudothresholds in fault-tolerant quantum computing}},\ }\href {https://doi.org/10.26421/QIC6.3-1} {\bibfield  {journal} {\bibinfo  {journal} {Quantum Information and Computation}\ }\textbf {\bibinfo {volume} {6}},\ \bibinfo {pages} {193} (\bibinfo {year} {2006})}\BibitemShut {NoStop}%
\bibitem [{\citenamefont {Wu}\ \emph {et~al.}(2022)\citenamefont {Wu}, \citenamefont {Kolkowitz}, \citenamefont {Puri},\ and\ \citenamefont {Thompson}}]{Wu2022}%
  \BibitemOpen
  \bibfield  {author} {\bibinfo {author} {\bibfnamefont {Y.}~\bibnamefont {Wu}}, \bibinfo {author} {\bibfnamefont {S.}~\bibnamefont {Kolkowitz}}, \bibinfo {author} {\bibfnamefont {S.}~\bibnamefont {Puri}},\ and\ \bibinfo {author} {\bibfnamefont {J.~D.}\ \bibnamefont {Thompson}},\ }\bibfield  {title} {\bibinfo {title} {{Erasure conversion for fault-tolerant quantum computing in alkaline earth Rydberg atom arrays}},\ }\href {https://doi.org/10.1038/s41467-022-32094-6} {\bibfield  {journal} {\bibinfo  {journal} {Nature Communications}\ }\textbf {\bibinfo {volume} {13}},\ \bibinfo {pages} {1} (\bibinfo {year} {2022})}\BibitemShut {NoStop}%
\bibitem [{\citenamefont {Sahay}\ \emph {et~al.}(2023)\citenamefont {Sahay}, \citenamefont {Jin}, \citenamefont {Claes}, \citenamefont {Thompson},\ and\ \citenamefont {Puri}}]{Sahay2023}%
  \BibitemOpen
  \bibfield  {author} {\bibinfo {author} {\bibfnamefont {K.}~\bibnamefont {Sahay}}, \bibinfo {author} {\bibfnamefont {J.}~\bibnamefont {Jin}}, \bibinfo {author} {\bibfnamefont {J.}~\bibnamefont {Claes}}, \bibinfo {author} {\bibfnamefont {J.~D.}\ \bibnamefont {Thompson}},\ and\ \bibinfo {author} {\bibfnamefont {S.}~\bibnamefont {Puri}},\ }\bibfield  {title} {\bibinfo {title} {{High-Threshold Codes for Neutral-Atom Qubits with Biased Erasure Errors}},\ }\href {https://doi.org/10.1103/PhysRevX.13.041013} {\bibfield  {journal} {\bibinfo  {journal} {Physical Review X}\ }\textbf {\bibinfo {volume} {13}},\ \bibinfo {pages} {041013} (\bibinfo {year} {2023})}\BibitemShut {NoStop}%
\bibitem [{\citenamefont {Kang}\ \emph {et~al.}(2023)\citenamefont {Kang}, \citenamefont {Campbell},\ and\ \citenamefont {Brown}}]{Kang2023}%
  \BibitemOpen
  \bibfield  {author} {\bibinfo {author} {\bibfnamefont {M.}~\bibnamefont {Kang}}, \bibinfo {author} {\bibfnamefont {W.~C.}\ \bibnamefont {Campbell}},\ and\ \bibinfo {author} {\bibfnamefont {K.~R.}\ \bibnamefont {Brown}},\ }\bibfield  {title} {\bibinfo {title} {Quantum error correction with metastable states of trapped ions using erasure conversion},\ }\href {https://doi.org/10.1103/PRXQuantum.4.020358} {\bibfield  {journal} {\bibinfo  {journal} {PRX Quantum}\ }\textbf {\bibinfo {volume} {4}},\ \bibinfo {pages} {020358} (\bibinfo {year} {2023})}\BibitemShut {NoStop}%
\bibitem [{\citenamefont {Kubica}\ \emph {et~al.}(2023)\citenamefont {Kubica}, \citenamefont {Haim}, \citenamefont {Vaknin}, \citenamefont {Levine}, \citenamefont {Brand\~ao},\ and\ \citenamefont {Retzker}}]{Kubica2023}%
  \BibitemOpen
  \bibfield  {author} {\bibinfo {author} {\bibfnamefont {A.}~\bibnamefont {Kubica}}, \bibinfo {author} {\bibfnamefont {A.}~\bibnamefont {Haim}}, \bibinfo {author} {\bibfnamefont {Y.}~\bibnamefont {Vaknin}}, \bibinfo {author} {\bibfnamefont {H.}~\bibnamefont {Levine}}, \bibinfo {author} {\bibfnamefont {F.}~\bibnamefont {Brand\~ao}},\ and\ \bibinfo {author} {\bibfnamefont {A.}~\bibnamefont {Retzker}},\ }\bibfield  {title} {\bibinfo {title} {Erasure qubits: Overcoming the ${T}_{1}$ limit in superconducting circuits},\ }\href {https://doi.org/10.1103/PhysRevX.13.041022} {\bibfield  {journal} {\bibinfo  {journal} {Phys. Rev. X}\ }\textbf {\bibinfo {volume} {13}},\ \bibinfo {pages} {041022} (\bibinfo {year} {2023})}\BibitemShut {NoStop}%
\bibitem [{\citenamefont {Teoh}\ \emph {et~al.}(2023)\citenamefont {Teoh}, \citenamefont {Winkel}, \citenamefont {Babla}, \citenamefont {Chapman}, \citenamefont {Claes}, \citenamefont {de~Graaf}, \citenamefont {Garmon}, \citenamefont {Kalfus}, \citenamefont {Lu}, \citenamefont {Maiti}, \citenamefont {Sahay}, \citenamefont {Thakur}, \citenamefont {Tsunoda}, \citenamefont {Xue}, \citenamefont {Frunzio}, \citenamefont {Girvin}, \citenamefont {Puri},\ and\ \citenamefont {Schoelkopf}}]{Teoh2023}%
  \BibitemOpen
  \bibfield  {author} {\bibinfo {author} {\bibfnamefont {J.~D.}\ \bibnamefont {Teoh}}, \bibinfo {author} {\bibfnamefont {P.}~\bibnamefont {Winkel}}, \bibinfo {author} {\bibfnamefont {H.~K.}\ \bibnamefont {Babla}}, \bibinfo {author} {\bibfnamefont {B.~J.}\ \bibnamefont {Chapman}}, \bibinfo {author} {\bibfnamefont {J.}~\bibnamefont {Claes}}, \bibinfo {author} {\bibfnamefont {S.~J.}\ \bibnamefont {de~Graaf}}, \bibinfo {author} {\bibfnamefont {J.~W.~O.}\ \bibnamefont {Garmon}}, \bibinfo {author} {\bibfnamefont {W.~D.}\ \bibnamefont {Kalfus}}, \bibinfo {author} {\bibfnamefont {Y.}~\bibnamefont {Lu}}, \bibinfo {author} {\bibfnamefont {A.}~\bibnamefont {Maiti}}, \bibinfo {author} {\bibfnamefont {K.}~\bibnamefont {Sahay}}, \bibinfo {author} {\bibfnamefont {N.}~\bibnamefont {Thakur}}, \bibinfo {author} {\bibfnamefont {T.}~\bibnamefont {Tsunoda}}, \bibinfo {author} {\bibfnamefont {S.~H.}\ \bibnamefont {Xue}}, \bibinfo {author} {\bibfnamefont {L.}~\bibnamefont {Frunzio}}, \bibinfo {author} {\bibfnamefont {S.~M.}\
  \bibnamefont {Girvin}}, \bibinfo {author} {\bibfnamefont {S.}~\bibnamefont {Puri}},\ and\ \bibinfo {author} {\bibfnamefont {R.~J.}\ \bibnamefont {Schoelkopf}},\ }\bibfield  {title} {\bibinfo {title} {Dual-rail encoding with superconducting cavities},\ }\href {https://doi.org/10.1073/pnas.2221736120} {\bibfield  {journal} {\bibinfo  {journal} {Proceedings of the National Academy of Sciences}\ }\textbf {\bibinfo {volume} {120}},\ \bibinfo {pages} {e2221736120} (\bibinfo {year} {2023})}\BibitemShut {NoStop}%
\bibitem [{\citenamefont {Scholl}\ \emph {et~al.}(2023)\citenamefont {Scholl}, \citenamefont {Shaw}, \citenamefont {Tsai}, \citenamefont {Finkelstein}, \citenamefont {Choi},\ and\ \citenamefont {Endres}}]{Scholl2023}%
  \BibitemOpen
  \bibfield  {author} {\bibinfo {author} {\bibfnamefont {P.}~\bibnamefont {Scholl}}, \bibinfo {author} {\bibfnamefont {A.~L.}\ \bibnamefont {Shaw}}, \bibinfo {author} {\bibfnamefont {R.~B.-s.}\ \bibnamefont {Tsai}}, \bibinfo {author} {\bibfnamefont {R.}~\bibnamefont {Finkelstein}}, \bibinfo {author} {\bibfnamefont {J.}~\bibnamefont {Choi}},\ and\ \bibinfo {author} {\bibfnamefont {M.}~\bibnamefont {Endres}},\ }\bibfield  {title} {\bibinfo {title} {{Erasure conversion in a high-fidelity Rydberg quantum simulator}},\ }\href {https://doi.org/10.1038/s41586-023-06516-4} {\bibfield  {journal} {\bibinfo  {journal} {Nature}\ }\textbf {\bibinfo {volume} {622}},\ \bibinfo {pages} {273} (\bibinfo {year} {2023})}\BibitemShut {NoStop}%
\bibitem [{\citenamefont {Ma}\ \emph {et~al.}(2023)\citenamefont {Ma}, \citenamefont {Liu}, \citenamefont {Peng}, \citenamefont {Zhang}, \citenamefont {Jandura}, \citenamefont {Claes}, \citenamefont {Burgers}, \citenamefont {Pupillo}, \citenamefont {Puri},\ and\ \citenamefont {Thompson}}]{Ma2023}%
  \BibitemOpen
  \bibfield  {author} {\bibinfo {author} {\bibfnamefont {S.}~\bibnamefont {Ma}}, \bibinfo {author} {\bibfnamefont {G.}~\bibnamefont {Liu}}, \bibinfo {author} {\bibfnamefont {P.}~\bibnamefont {Peng}}, \bibinfo {author} {\bibfnamefont {B.}~\bibnamefont {Zhang}}, \bibinfo {author} {\bibfnamefont {S.}~\bibnamefont {Jandura}}, \bibinfo {author} {\bibfnamefont {J.}~\bibnamefont {Claes}}, \bibinfo {author} {\bibfnamefont {A.~P.}\ \bibnamefont {Burgers}}, \bibinfo {author} {\bibfnamefont {G.}~\bibnamefont {Pupillo}}, \bibinfo {author} {\bibfnamefont {S.}~\bibnamefont {Puri}},\ and\ \bibinfo {author} {\bibfnamefont {J.~D.}\ \bibnamefont {Thompson}},\ }\bibfield  {title} {\bibinfo {title} {{High-fidelity gates and mid-circuit erasure conversion in an atomic qubit}},\ }\href {https://doi.org/10.1038/s41586-023-06438-1} {\bibfield  {journal} {\bibinfo  {journal} {Nature}\ }\textbf {\bibinfo {volume} {622}},\ \bibinfo {pages} {279} (\bibinfo {year} {2023})}\BibitemShut {NoStop}%
\bibitem [{\citenamefont {Levine}\ \emph {et~al.}(2024)\citenamefont {Levine}, \citenamefont {Haim}, \citenamefont {Hung}, \citenamefont {Alidoust}, \citenamefont {Kalaee}, \citenamefont {DeLorenzo}, \citenamefont {Wollack}, \citenamefont {Arrangoiz-Arriola}, \citenamefont {Khalajhedayati}, \citenamefont {Sanil}, \citenamefont {Moradinejad}, \citenamefont {Vaknin}, \citenamefont {Kubica}, \citenamefont {Hover}, \citenamefont {Aghaeimeibodi}, \citenamefont {Alcid}, \citenamefont {Baek}, \citenamefont {Barnett}, \citenamefont {Bawdekar}, \citenamefont {Bienias}, \citenamefont {Carson}, \citenamefont {Chen}, \citenamefont {Chen}, \citenamefont {Chinkezian}, \citenamefont {Chisholm}, \citenamefont {Clifford}, \citenamefont {Cosmic}, \citenamefont {Crisosto}, \citenamefont {Dalzell}, \citenamefont {Davis}, \citenamefont {D'Ewart}, \citenamefont {Diez}, \citenamefont {D'Souza}, \citenamefont {Dumitrescu}, \citenamefont {Elkhouly}, \citenamefont {Fang}, \citenamefont {Fang}, \citenamefont {Flammia}, \citenamefont
  {Fling}, \citenamefont {Garcia}, \citenamefont {Gharzai}, \citenamefont {Gorshkov}, \citenamefont {Gray}, \citenamefont {Grimberg}, \citenamefont {Grimsmo}, \citenamefont {Hann}, \citenamefont {He}, \citenamefont {Heidel}, \citenamefont {Howell}, \citenamefont {Hunt}, \citenamefont {Iverson}, \citenamefont {Jarrige}, \citenamefont {Jiang}, \citenamefont {Jones}, \citenamefont {Karabalin}, \citenamefont {Karalekas}, \citenamefont {Keller}, \citenamefont {Lasi}, \citenamefont {Lee}, \citenamefont {Ly}, \citenamefont {MacCabe}, \citenamefont {Mahuli}, \citenamefont {Marcaud}, \citenamefont {Matheny}, \citenamefont {McArdle}, \citenamefont {McCabe}, \citenamefont {Merton}, \citenamefont {Miles}, \citenamefont {Milsted}, \citenamefont {Mishra}, \citenamefont {Moncelsi}, \citenamefont {Naghiloo}, \citenamefont {Noh}, \citenamefont {Oblepias}, \citenamefont {Ortuno}, \citenamefont {Owens}, \citenamefont {Pagdilao}, \citenamefont {Panduro}, \citenamefont {Paquette}, \citenamefont {Patel}, \citenamefont {Peairs},
  \citenamefont {Perello}, \citenamefont {Peterson}, \citenamefont {Ponte}, \citenamefont {Putterman}, \citenamefont {Refael}, \citenamefont {Reinhold}, \citenamefont {Resnick}, \citenamefont {Reyna}, \citenamefont {Rodriguez}, \citenamefont {Rose}, \citenamefont {Rubin}, \citenamefont {Runyan}, \citenamefont {Ryan}, \citenamefont {Sahmoud}, \citenamefont {Scaffidi}, \citenamefont {Shah}, \citenamefont {Siavoshi}, \citenamefont {Sivarajah}, \citenamefont {Skogland}, \citenamefont {Su}, \citenamefont {Swenson}, \citenamefont {Sylvia}, \citenamefont {Teo}, \citenamefont {Tomada}, \citenamefont {Torlai}, \citenamefont {Wistrom}, \citenamefont {Zhang}, \citenamefont {Zuk}, \citenamefont {Clerk}, \citenamefont {Brand\~ao}, \citenamefont {Retzker},\ and\ \citenamefont {Painter}}]{Levine2024}%
  \BibitemOpen
  \bibfield  {author} {\bibinfo {author} {\bibfnamefont {H.}~\bibnamefont {Levine}}, \bibinfo {author} {\bibfnamefont {A.}~\bibnamefont {Haim}}, \bibinfo {author} {\bibfnamefont {J.~S.~C.}\ \bibnamefont {Hung}}, \bibinfo {author} {\bibfnamefont {N.}~\bibnamefont {Alidoust}}, \bibinfo {author} {\bibfnamefont {M.}~\bibnamefont {Kalaee}}, \bibinfo {author} {\bibfnamefont {L.}~\bibnamefont {DeLorenzo}}, \bibinfo {author} {\bibfnamefont {E.~A.}\ \bibnamefont {Wollack}}, \bibinfo {author} {\bibfnamefont {P.}~\bibnamefont {Arrangoiz-Arriola}}, \bibinfo {author} {\bibfnamefont {A.}~\bibnamefont {Khalajhedayati}}, \bibinfo {author} {\bibfnamefont {R.}~\bibnamefont {Sanil}}, \bibinfo {author} {\bibfnamefont {H.}~\bibnamefont {Moradinejad}}, \bibinfo {author} {\bibfnamefont {Y.}~\bibnamefont {Vaknin}}, \bibinfo {author} {\bibfnamefont {A.}~\bibnamefont {Kubica}}, \bibinfo {author} {\bibfnamefont {D.}~\bibnamefont {Hover}}, \bibinfo {author} {\bibfnamefont {S.}~\bibnamefont {Aghaeimeibodi}}, \bibinfo {author}
  {\bibfnamefont {J.~A.}\ \bibnamefont {Alcid}}, \bibinfo {author} {\bibfnamefont {C.}~\bibnamefont {Baek}}, \bibinfo {author} {\bibfnamefont {J.}~\bibnamefont {Barnett}}, \bibinfo {author} {\bibfnamefont {K.}~\bibnamefont {Bawdekar}}, \bibinfo {author} {\bibfnamefont {P.}~\bibnamefont {Bienias}}, \bibinfo {author} {\bibfnamefont {H.~A.}\ \bibnamefont {Carson}}, \bibinfo {author} {\bibfnamefont {C.}~\bibnamefont {Chen}}, \bibinfo {author} {\bibfnamefont {L.}~\bibnamefont {Chen}}, \bibinfo {author} {\bibfnamefont {H.}~\bibnamefont {Chinkezian}}, \bibinfo {author} {\bibfnamefont {E.~M.}\ \bibnamefont {Chisholm}}, \bibinfo {author} {\bibfnamefont {A.}~\bibnamefont {Clifford}}, \bibinfo {author} {\bibfnamefont {R.}~\bibnamefont {Cosmic}}, \bibinfo {author} {\bibfnamefont {N.}~\bibnamefont {Crisosto}}, \bibinfo {author} {\bibfnamefont {A.~M.}\ \bibnamefont {Dalzell}}, \bibinfo {author} {\bibfnamefont {E.}~\bibnamefont {Davis}}, \bibinfo {author} {\bibfnamefont {J.~M.}\ \bibnamefont {D'Ewart}}, \bibinfo {author}
  {\bibfnamefont {S.}~\bibnamefont {Diez}}, \bibinfo {author} {\bibfnamefont {N.}~\bibnamefont {D'Souza}}, \bibinfo {author} {\bibfnamefont {P.~T.}\ \bibnamefont {Dumitrescu}}, \bibinfo {author} {\bibfnamefont {E.}~\bibnamefont {Elkhouly}}, \bibinfo {author} {\bibfnamefont {M.~T.}\ \bibnamefont {Fang}}, \bibinfo {author} {\bibfnamefont {Y.}~\bibnamefont {Fang}}, \bibinfo {author} {\bibfnamefont {S.}~\bibnamefont {Flammia}}, \bibinfo {author} {\bibfnamefont {M.~J.}\ \bibnamefont {Fling}}, \bibinfo {author} {\bibfnamefont {G.}~\bibnamefont {Garcia}}, \bibinfo {author} {\bibfnamefont {M.~K.}\ \bibnamefont {Gharzai}}, \bibinfo {author} {\bibfnamefont {A.~V.}\ \bibnamefont {Gorshkov}}, \bibinfo {author} {\bibfnamefont {M.~J.}\ \bibnamefont {Gray}}, \bibinfo {author} {\bibfnamefont {S.}~\bibnamefont {Grimberg}}, \bibinfo {author} {\bibfnamefont {A.~L.}\ \bibnamefont {Grimsmo}}, \bibinfo {author} {\bibfnamefont {C.~T.}\ \bibnamefont {Hann}}, \bibinfo {author} {\bibfnamefont {Y.}~\bibnamefont {He}}, \bibinfo {author}
  {\bibfnamefont {S.}~\bibnamefont {Heidel}}, \bibinfo {author} {\bibfnamefont {S.}~\bibnamefont {Howell}}, \bibinfo {author} {\bibfnamefont {M.}~\bibnamefont {Hunt}}, \bibinfo {author} {\bibfnamefont {J.}~\bibnamefont {Iverson}}, \bibinfo {author} {\bibfnamefont {I.}~\bibnamefont {Jarrige}}, \bibinfo {author} {\bibfnamefont {L.}~\bibnamefont {Jiang}}, \bibinfo {author} {\bibfnamefont {W.~M.}\ \bibnamefont {Jones}}, \bibinfo {author} {\bibfnamefont {R.}~\bibnamefont {Karabalin}}, \bibinfo {author} {\bibfnamefont {P.~J.}\ \bibnamefont {Karalekas}}, \bibinfo {author} {\bibfnamefont {A.~J.}\ \bibnamefont {Keller}}, \bibinfo {author} {\bibfnamefont {D.}~\bibnamefont {Lasi}}, \bibinfo {author} {\bibfnamefont {M.}~\bibnamefont {Lee}}, \bibinfo {author} {\bibfnamefont {V.}~\bibnamefont {Ly}}, \bibinfo {author} {\bibfnamefont {G.}~\bibnamefont {MacCabe}}, \bibinfo {author} {\bibfnamefont {N.}~\bibnamefont {Mahuli}}, \bibinfo {author} {\bibfnamefont {G.}~\bibnamefont {Marcaud}}, \bibinfo {author} {\bibfnamefont
  {M.~H.}\ \bibnamefont {Matheny}}, \bibinfo {author} {\bibfnamefont {S.}~\bibnamefont {McArdle}}, \bibinfo {author} {\bibfnamefont {G.}~\bibnamefont {McCabe}}, \bibinfo {author} {\bibfnamefont {G.}~\bibnamefont {Merton}}, \bibinfo {author} {\bibfnamefont {C.}~\bibnamefont {Miles}}, \bibinfo {author} {\bibfnamefont {A.}~\bibnamefont {Milsted}}, \bibinfo {author} {\bibfnamefont {A.}~\bibnamefont {Mishra}}, \bibinfo {author} {\bibfnamefont {L.}~\bibnamefont {Moncelsi}}, \bibinfo {author} {\bibfnamefont {M.}~\bibnamefont {Naghiloo}}, \bibinfo {author} {\bibfnamefont {K.}~\bibnamefont {Noh}}, \bibinfo {author} {\bibfnamefont {E.}~\bibnamefont {Oblepias}}, \bibinfo {author} {\bibfnamefont {G.}~\bibnamefont {Ortuno}}, \bibinfo {author} {\bibfnamefont {J.~C.}\ \bibnamefont {Owens}}, \bibinfo {author} {\bibfnamefont {J.}~\bibnamefont {Pagdilao}}, \bibinfo {author} {\bibfnamefont {A.}~\bibnamefont {Panduro}}, \bibinfo {author} {\bibfnamefont {J.-P.}\ \bibnamefont {Paquette}}, \bibinfo {author} {\bibfnamefont {R.~N.}\
  \bibnamefont {Patel}}, \bibinfo {author} {\bibfnamefont {G.}~\bibnamefont {Peairs}}, \bibinfo {author} {\bibfnamefont {D.~J.}\ \bibnamefont {Perello}}, \bibinfo {author} {\bibfnamefont {E.~C.}\ \bibnamefont {Peterson}}, \bibinfo {author} {\bibfnamefont {S.}~\bibnamefont {Ponte}}, \bibinfo {author} {\bibfnamefont {H.}~\bibnamefont {Putterman}}, \bibinfo {author} {\bibfnamefont {G.}~\bibnamefont {Refael}}, \bibinfo {author} {\bibfnamefont {P.}~\bibnamefont {Reinhold}}, \bibinfo {author} {\bibfnamefont {R.}~\bibnamefont {Resnick}}, \bibinfo {author} {\bibfnamefont {O.~A.}\ \bibnamefont {Reyna}}, \bibinfo {author} {\bibfnamefont {R.}~\bibnamefont {Rodriguez}}, \bibinfo {author} {\bibfnamefont {J.}~\bibnamefont {Rose}}, \bibinfo {author} {\bibfnamefont {A.~H.}\ \bibnamefont {Rubin}}, \bibinfo {author} {\bibfnamefont {M.}~\bibnamefont {Runyan}}, \bibinfo {author} {\bibfnamefont {C.~A.}\ \bibnamefont {Ryan}}, \bibinfo {author} {\bibfnamefont {A.}~\bibnamefont {Sahmoud}}, \bibinfo {author} {\bibfnamefont
  {T.}~\bibnamefont {Scaffidi}}, \bibinfo {author} {\bibfnamefont {B.}~\bibnamefont {Shah}}, \bibinfo {author} {\bibfnamefont {S.}~\bibnamefont {Siavoshi}}, \bibinfo {author} {\bibfnamefont {P.}~\bibnamefont {Sivarajah}}, \bibinfo {author} {\bibfnamefont {T.}~\bibnamefont {Skogland}}, \bibinfo {author} {\bibfnamefont {C.-J.}\ \bibnamefont {Su}}, \bibinfo {author} {\bibfnamefont {L.~J.}\ \bibnamefont {Swenson}}, \bibinfo {author} {\bibfnamefont {J.}~\bibnamefont {Sylvia}}, \bibinfo {author} {\bibfnamefont {S.~M.}\ \bibnamefont {Teo}}, \bibinfo {author} {\bibfnamefont {A.}~\bibnamefont {Tomada}}, \bibinfo {author} {\bibfnamefont {G.}~\bibnamefont {Torlai}}, \bibinfo {author} {\bibfnamefont {M.}~\bibnamefont {Wistrom}}, \bibinfo {author} {\bibfnamefont {K.}~\bibnamefont {Zhang}}, \bibinfo {author} {\bibfnamefont {I.}~\bibnamefont {Zuk}}, \bibinfo {author} {\bibfnamefont {A.~A.}\ \bibnamefont {Clerk}}, \bibinfo {author} {\bibfnamefont {F.~G. S.~L.}\ \bibnamefont {Brand\~ao}}, \bibinfo {author} {\bibfnamefont
  {A.}~\bibnamefont {Retzker}},\ and\ \bibinfo {author} {\bibfnamefont {O.}~\bibnamefont {Painter}},\ }\bibfield  {title} {\bibinfo {title} {Demonstrating a long-coherence dual-rail erasure qubit using tunable transmons},\ }\href {https://doi.org/10.1103/PhysRevX.14.011051} {\bibfield  {journal} {\bibinfo  {journal} {Phys. Rev. X}\ }\textbf {\bibinfo {volume} {14}},\ \bibinfo {pages} {011051} (\bibinfo {year} {2024})}\BibitemShut {NoStop}%
\bibitem [{\citenamefont {Chou}\ \emph {et~al.}(2024)\citenamefont {Chou}, \citenamefont {Shemma}, \citenamefont {McCarrick}, \citenamefont {Chien}, \citenamefont {Teoh}, \citenamefont {Winkel}, \citenamefont {Anderson}, \citenamefont {Chen}, \citenamefont {Curtis}, \citenamefont {de~Graaf}, \citenamefont {Garmon}, \citenamefont {Gudlewski}, \citenamefont {Kalfus}, \citenamefont {Keen}, \citenamefont {Khedkar}, \citenamefont {Lei}, \citenamefont {Liu}, \citenamefont {Lu}, \citenamefont {Lu}, \citenamefont {Maiti}, \citenamefont {Mastalli-Kelly}, \citenamefont {Mehta}, \citenamefont {Mundhada}, \citenamefont {Narla}, \citenamefont {Noh}, \citenamefont {Tsunoda}, \citenamefont {Xue}, \citenamefont {Yuan}, \citenamefont {Frunzio}, \citenamefont {Aumentado}, \citenamefont {Puri}, \citenamefont {Girvin}, \citenamefont {Moseley},\ and\ \citenamefont {Schoelkopf}}]{Chou2024}%
  \BibitemOpen
  \bibfield  {author} {\bibinfo {author} {\bibfnamefont {K.~S.}\ \bibnamefont {Chou}}, \bibinfo {author} {\bibfnamefont {T.}~\bibnamefont {Shemma}}, \bibinfo {author} {\bibfnamefont {H.}~\bibnamefont {McCarrick}}, \bibinfo {author} {\bibfnamefont {T.-C.}\ \bibnamefont {Chien}}, \bibinfo {author} {\bibfnamefont {J.~D.}\ \bibnamefont {Teoh}}, \bibinfo {author} {\bibfnamefont {P.}~\bibnamefont {Winkel}}, \bibinfo {author} {\bibfnamefont {A.}~\bibnamefont {Anderson}}, \bibinfo {author} {\bibfnamefont {J.}~\bibnamefont {Chen}}, \bibinfo {author} {\bibfnamefont {J.~C.}\ \bibnamefont {Curtis}}, \bibinfo {author} {\bibfnamefont {S.~J.}\ \bibnamefont {de~Graaf}}, \bibinfo {author} {\bibfnamefont {J.~W.~O.}\ \bibnamefont {Garmon}}, \bibinfo {author} {\bibfnamefont {B.}~\bibnamefont {Gudlewski}}, \bibinfo {author} {\bibfnamefont {W.~D.}\ \bibnamefont {Kalfus}}, \bibinfo {author} {\bibfnamefont {T.}~\bibnamefont {Keen}}, \bibinfo {author} {\bibfnamefont {N.}~\bibnamefont {Khedkar}}, \bibinfo {author} {\bibfnamefont
  {C.~U.}\ \bibnamefont {Lei}}, \bibinfo {author} {\bibfnamefont {G.}~\bibnamefont {Liu}}, \bibinfo {author} {\bibfnamefont {P.}~\bibnamefont {Lu}}, \bibinfo {author} {\bibfnamefont {Y.}~\bibnamefont {Lu}}, \bibinfo {author} {\bibfnamefont {A.}~\bibnamefont {Maiti}}, \bibinfo {author} {\bibfnamefont {L.}~\bibnamefont {Mastalli-Kelly}}, \bibinfo {author} {\bibfnamefont {N.}~\bibnamefont {Mehta}}, \bibinfo {author} {\bibfnamefont {S.~O.}\ \bibnamefont {Mundhada}}, \bibinfo {author} {\bibfnamefont {A.}~\bibnamefont {Narla}}, \bibinfo {author} {\bibfnamefont {T.}~\bibnamefont {Noh}}, \bibinfo {author} {\bibfnamefont {T.}~\bibnamefont {Tsunoda}}, \bibinfo {author} {\bibfnamefont {S.~H.}\ \bibnamefont {Xue}}, \bibinfo {author} {\bibfnamefont {J.~O.}\ \bibnamefont {Yuan}}, \bibinfo {author} {\bibfnamefont {L.}~\bibnamefont {Frunzio}}, \bibinfo {author} {\bibfnamefont {J.}~\bibnamefont {Aumentado}}, \bibinfo {author} {\bibfnamefont {S.}~\bibnamefont {Puri}}, \bibinfo {author} {\bibfnamefont {S.~M.}\ \bibnamefont
  {Girvin}}, \bibinfo {author} {\bibfnamefont {S.~H.}\ \bibnamefont {Moseley}},\ and\ \bibinfo {author} {\bibfnamefont {R.~J.}\ \bibnamefont {Schoelkopf}},\ }\bibfield  {title} {\bibinfo {title} {{A superconducting dual-rail cavity qubit with erasure-detected logical measurements}},\ }\href {https://doi.org/10.1038/s41567-024-02539-4} {\bibfield  {journal} {\bibinfo  {journal} {Nature Physics}\ }\textbf {\bibinfo {volume} {20}},\ \bibinfo {pages} {1454} (\bibinfo {year} {2024})}\BibitemShut {NoStop}%
\bibitem [{\citenamefont {Koottandavida}\ \emph {et~al.}(2024)\citenamefont {Koottandavida}, \citenamefont {Tsioutsios}, \citenamefont {Kargioti}, \citenamefont {Smith}, \citenamefont {Joshi}, \citenamefont {Dai}, \citenamefont {Teoh}, \citenamefont {Curtis}, \citenamefont {Frunzio}, \citenamefont {Schoelkopf},\ and\ \citenamefont {Devoret}}]{Koottandavida2024}%
  \BibitemOpen
  \bibfield  {author} {\bibinfo {author} {\bibfnamefont {A.}~\bibnamefont {Koottandavida}}, \bibinfo {author} {\bibfnamefont {I.}~\bibnamefont {Tsioutsios}}, \bibinfo {author} {\bibfnamefont {A.}~\bibnamefont {Kargioti}}, \bibinfo {author} {\bibfnamefont {C.~R.}\ \bibnamefont {Smith}}, \bibinfo {author} {\bibfnamefont {V.~R.}\ \bibnamefont {Joshi}}, \bibinfo {author} {\bibfnamefont {W.}~\bibnamefont {Dai}}, \bibinfo {author} {\bibfnamefont {J.~D.}\ \bibnamefont {Teoh}}, \bibinfo {author} {\bibfnamefont {J.~C.}\ \bibnamefont {Curtis}}, \bibinfo {author} {\bibfnamefont {L.}~\bibnamefont {Frunzio}}, \bibinfo {author} {\bibfnamefont {R.~J.}\ \bibnamefont {Schoelkopf}},\ and\ \bibinfo {author} {\bibfnamefont {M.~H.}\ \bibnamefont {Devoret}},\ }\bibfield  {title} {\bibinfo {title} {{Erasure Detection of a Dual-Rail Qubit Encoded in a Double-Post Superconducting Cavity}},\ }\href {https://doi.org/10.1103/PhysRevLett.132.180601} {\bibfield  {journal} {\bibinfo  {journal} {Physical Review Letters}\ }\textbf {\bibinfo
  {volume} {132}},\ \bibinfo {pages} {180601} (\bibinfo {year} {2024})}\BibitemShut {NoStop}%
\bibitem [{\citenamefont {Holland}\ \emph {et~al.}(2024)\citenamefont {Holland}, \citenamefont {Lu}, \citenamefont {Li}, \citenamefont {Welsh},\ and\ \citenamefont {Cheuk}}]{Holland2024}%
  \BibitemOpen
  \bibfield  {author} {\bibinfo {author} {\bibfnamefont {C.~M.}\ \bibnamefont {Holland}}, \bibinfo {author} {\bibfnamefont {Y.}~\bibnamefont {Lu}}, \bibinfo {author} {\bibfnamefont {S.~J.}\ \bibnamefont {Li}}, \bibinfo {author} {\bibfnamefont {C.~L.}\ \bibnamefont {Welsh}},\ and\ \bibinfo {author} {\bibfnamefont {L.~W.}\ \bibnamefont {Cheuk}},\ }\href {https://arxiv.org/abs/2406.02391} {\bibinfo {title} {Demonstration of erasure conversion in a molecular tweezer array}} (\bibinfo {year} {2024}),\ \Eprint {https://arxiv.org/abs/2406.02391} {arXiv:2406.02391 [quant-ph]} \BibitemShut {NoStop}%
\bibitem [{\citenamefont {Zhang}\ \emph {et~al.}(2025)\citenamefont {Zhang}, \citenamefont {Liu}, \citenamefont {Bornet}, \citenamefont {Horvath}, \citenamefont {Peng}, \citenamefont {Ma}, \citenamefont {Huang}, \citenamefont {Puri},\ and\ \citenamefont {Thompson}}]{Zhang2025}%
  \BibitemOpen
  \bibfield  {author} {\bibinfo {author} {\bibfnamefont {B.}~\bibnamefont {Zhang}}, \bibinfo {author} {\bibfnamefont {G.}~\bibnamefont {Liu}}, \bibinfo {author} {\bibfnamefont {G.}~\bibnamefont {Bornet}}, \bibinfo {author} {\bibfnamefont {S.~P.}\ \bibnamefont {Horvath}}, \bibinfo {author} {\bibfnamefont {P.}~\bibnamefont {Peng}}, \bibinfo {author} {\bibfnamefont {S.}~\bibnamefont {Ma}}, \bibinfo {author} {\bibfnamefont {S.}~\bibnamefont {Huang}}, \bibinfo {author} {\bibfnamefont {S.}~\bibnamefont {Puri}},\ and\ \bibinfo {author} {\bibfnamefont {J.~D.}\ \bibnamefont {Thompson}},\ }\href {https://arxiv.org/abs/2506.13724} {\bibinfo {title} {Leveraging erasure errors in logical qubits with metastable $^{171}$yb atoms}} (\bibinfo {year} {2025}),\ \Eprint {https://arxiv.org/abs/2506.13724} {arXiv:2506.13724 [quant-ph]} \BibitemShut {NoStop}%
\bibitem [{\citenamefont {Singleton}(1969)}]{1162042}%
  \BibitemOpen
  \bibfield  {author} {\bibinfo {author} {\bibfnamefont {R.}~\bibnamefont {Singleton}},\ }\bibfield  {title} {\bibinfo {title} {An algorithm for computing the mixed radix fast fourier transform},\ }\href {https://doi.org/10.1109/TAU.1969.1162042} {\bibfield  {journal} {\bibinfo  {journal} {IEEE Transactions on Audio and Electroacoustics}\ }\textbf {\bibinfo {volume} {17}},\ \bibinfo {pages} {93} (\bibinfo {year} {1969})}\BibitemShut {NoStop}%
\bibitem [{Qua()}]{QuantumClifford}%
  \BibitemOpen
  \href {https://doi.org/10.5281/zenodo.5208167} {\bibinfo {title} {{QuantumClifford.jl}}},\ \bibinfo {note} {computer code}\BibitemShut {NoStop}%
\bibitem [{myc()}]{mycode}%
  \BibitemOpen
  \href {https://github.com/gsommers/multitree/} {\bibinfo {title} {https://github.com/gsommers/multitree/}}\BibitemShut {NoStop}%
\bibitem [{cha()}]{channels}%
  \BibitemOpen
  \href@noop {} {\bibinfo {title} {{Note that without the classical register, $\mathcal{E}_{X,e}$ would be equivalent to $\mathcal{N}_{X,e/2}$, while for $\theta=\pi/2$, the Pauli twirl of $\mathcal{C}_{X,q,\pi/2}$ would match $\mathcal{E}_{X,q}$}}}\BibitemShut {NoStop}%
\bibitem [{\citenamefont {van~de Wetering}(2020)}]{Vandewetering2020}%
  \BibitemOpen
  \bibfield  {author} {\bibinfo {author} {\bibfnamefont {J.}~\bibnamefont {van~de Wetering}},\ }\href {http://arxiv.org/abs/2012.13966} {\bibinfo {title} {{ZX-calculus for the working quantum computer scientist}}} (\bibinfo {year} {2020}),\ \Eprint {https://arxiv.org/abs/2012.13966} {arXiv:2012.13966} \BibitemShut {NoStop}%
\bibitem [{\citenamefont {Reichardt}(2005{\natexlab{b}})}]{Reichardt2005distillation}%
  \BibitemOpen
  \bibfield  {author} {\bibinfo {author} {\bibfnamefont {B.~W.}\ \bibnamefont {Reichardt}},\ }\bibfield  {title} {\bibinfo {title} {{Quantum Universality from Magic States Distillation Applied to CSS Codes}},\ }\href {https://doi.org/10.1007/s11128-005-7654-8} {\bibfield  {journal} {\bibinfo  {journal} {Quantum Information Processing}\ }\textbf {\bibinfo {volume} {4}},\ \bibinfo {pages} {251} (\bibinfo {year} {2005}{\natexlab{b}})}\BibitemShut {NoStop}%
\bibitem [{\citenamefont {Poulin}(2005)}]{Poulin2005}%
  \BibitemOpen
  \bibfield  {author} {\bibinfo {author} {\bibfnamefont {D.}~\bibnamefont {Poulin}},\ }\bibfield  {title} {\bibinfo {title} {{Stabilizer Formalism for Operator Quantum Error Correction}},\ }\href {https://doi.org/10.1103/PhysRevLett.95.230504} {\bibfield  {journal} {\bibinfo  {journal} {Physical Review Letters}\ }\textbf {\bibinfo {volume} {95}},\ \bibinfo {pages} {230504} (\bibinfo {year} {2005})}\BibitemShut {NoStop}%
\bibitem [{\citenamefont {Albert}(2023)}]{Albert2023}%
  \BibitemOpen
  \bibfield  {author} {\bibinfo {author} {\bibfnamefont {V.~V.}\ \bibnamefont {Albert}},\ }\href {https://boulderschool.yale.edu/sites/default/files/files/BSS_notes_-_VVA.pdf} {\bibinfo {title} {2023 boulder lecture notes: Modern quantum error correction with 4 qubits}} (\bibinfo {year} {2023})\BibitemShut {NoStop}%
\bibitem [{\citenamefont {Gullans}\ \emph {et~al.}(2021)\citenamefont {Gullans}, \citenamefont {Krastanov}, \citenamefont {Huse}, \citenamefont {Jiang},\ and\ \citenamefont {Flammia}}]{Gullans21}%
  \BibitemOpen
  \bibfield  {author} {\bibinfo {author} {\bibfnamefont {M.~J.}\ \bibnamefont {Gullans}}, \bibinfo {author} {\bibfnamefont {S.}~\bibnamefont {Krastanov}}, \bibinfo {author} {\bibfnamefont {D.~A.}\ \bibnamefont {Huse}}, \bibinfo {author} {\bibfnamefont {L.}~\bibnamefont {Jiang}},\ and\ \bibinfo {author} {\bibfnamefont {S.~T.}\ \bibnamefont {Flammia}},\ }\bibfield  {title} {\bibinfo {title} {Quantum coding with low-depth random circuits},\ }\href {https://doi.org/10.1103/PhysRevX.11.031066} {\bibfield  {journal} {\bibinfo  {journal} {Phys. Rev. X}\ }\textbf {\bibinfo {volume} {11}},\ \bibinfo {pages} {031066} (\bibinfo {year} {2021})}\BibitemShut {NoStop}%
\bibitem [{\citenamefont {Kukliansky}\ and\ \citenamefont {Lackey}(2024)}]{Kukliansky2024}%
  \BibitemOpen
  \bibfield  {author} {\bibinfo {author} {\bibfnamefont {A.}~\bibnamefont {Kukliansky}}\ and\ \bibinfo {author} {\bibfnamefont {B.}~\bibnamefont {Lackey}},\ }\href {https://arxiv.org/abs/2405.19643} {\bibinfo {title} {Quantum circuit tensors and enumerators with applications to quantum fault tolerance}} (\bibinfo {year} {2024}),\ \Eprint {https://arxiv.org/abs/2405.19643} {arXiv:2405.19643 [quant-ph]} \BibitemShut {NoStop}%
\bibitem [{oei(2025)}]{oeis}%
  \BibitemOpen
  \href {https://oeis.org/A000695} {\bibinfo {title} {{The On-Line Encyclopedia of Integer Sequences, Sequence A000695}}} (\bibinfo {year} {2025})\BibitemShut {NoStop}%
\bibitem [{\citenamefont {Bacon}(2006)}]{Bacon2006}%
  \BibitemOpen
  \bibfield  {author} {\bibinfo {author} {\bibfnamefont {D.}~\bibnamefont {Bacon}},\ }\bibfield  {title} {\bibinfo {title} {{Operator quantum error-correcting subsystems for self-correcting quantum memories}},\ }\href {https://doi.org/10.1103/PhysRevA.73.012340} {\bibfield  {journal} {\bibinfo  {journal} {Physical Review A}\ }\textbf {\bibinfo {volume} {73}},\ \bibinfo {pages} {012340} (\bibinfo {year} {2006})}\BibitemShut {NoStop}%
\end{thebibliography}
\end{document}